\documentclass[12pt]{article}
\usepackage{amsmath}
\usepackage{graphicx}
\usepackage{enumerate}
\usepackage{natbib}
\usepackage{url} 

\usepackage{amssymb}
\usepackage{graphicx,overpic,amsmath}
\usepackage{bm}
\usepackage{upgreek}
\usepackage{url}
\usepackage{color}
\usepackage{verbatim}

\newcommand{\blind}{0}

\newtheorem{theorem}{Theorem}[section]

\newtheorem{remark}{Remark}[section]
\def\proof {\noindent{\bf Proof.}\quad}

\DeclareMathOperator*{\argmin}{arg\,min}

\def \bfm#1{\mbox{\boldmath$#1$}}
\def \es2{$E(s^2)$}

  \def \I {{\bfm I}}
\def \X {{\bfm X}} \def \x {{\bfm x}}   
 
  \def \A {{\bf A}}
  \def \Y {{\bfm Y}}  
\def \R {{\bfm R}}    \def \L {{\bfm L}}

\def \1{{\bf 1}} \def \0{{\bf 0}} \def \A{{\bf A}}

\def \Beta {{\bfm \beta}} \def \s2{{\sigma^2}} \def \Mu{{\bfm \mu}} 

\def \fl {{\noindent}}

\begin{document}

\def\spacingset#1{\renewcommand{\baselinestretch}%
{#1}\small\normalsize} \spacingset{1}


\if0\blind
{
  \title{\bf BOLT-SSI: A Statistical Approach to Screening Interaction Effects for Ultra-High Dimensional Data}
  \author{Min Zhou
  	\hspace{.2cm}\\
     BNU-HKBU United International College\\   
    Mingwei Dai \\
    Southwestern University of Finance and Economics\\
     Yuan Yao \\
  Victoria University of Wellington\\
     Jin Liu\\
   Duke-NUS Medical School\\   
   Can Yang \\
   The Hong Kong University of Science and Technology\\
   and  Heng Peng\\
    Hong Kong Baptist University\\
}

\date{}

  \maketitle
} \fi

\if1\blind
{
  \bigskip
  \bigskip
  \bigskip
  \begin{center}
    {\LARGE\bf BOLT-SSI: A Statistical Approach to Screening Interaction Effects for Ultra-High Dimensional Data}
\end{center}
  \medskip
} \fi

\begin{abstract}
Detecting interaction effects among predictors on the response variable is a crucial
step in various applications. In this paper, we first propose a simple method for sure screening interactions (SSI). Although its computation complexity is $O(p^2n)$, SSI works well for problems of moderate dimensionality (e.g., $p=10^3\sim10^4$), without the heredity assumption.  To ultra-high dimensional problems (e.g., $p = 10^6$),  motivated by discretization associated Boolean representation and operations and the contingency table for discrete variables, we propose a fast algorithm, named ``BOLT-SSI''. The statistical theory has been established for SSI and BOLT-SSI, guaranteeing their sure screening property. The performance of SSI and BOLT-SSI are evaluated by comprehensive simulation and real case studies. Numerical results demonstrate that SSI and BOLT-SSI can often outperform their competitors in terms of computational efficiency and statistical accuracy.  The proposed method can be applied for fully detecting interactions with more than 300,000 predictors. Based on this study, we believe that there is a great need to rethink the relationship between statistical accuracy and computational efficiency. We have shown that the computational performance of a statistical method can often be greatly improved by exploring the advantages of computational architecture with a tolerable loss of statistical accuracy.
\end{abstract}

\noindent%

\spacingset{1.15} 
\section{Introduction}
\label{sec:intro}

The recent two decades are the golden age for the development of statistical science on high dimensional problems.  A large number of innovative algorithms have been proposed to address the computational challenges in statistical inference for high dimensional problems. Despite a fruitful achievement in statistical science, there still exists a gap between the established statistical theory and computational performance of developed algorithms.  On one hand, many statistical models can deal with the high dimensional problems under some theoretically  mild conditions, but their computational cost can be too expensive to be affordable  when  dimensionality becomes extremely large. On the other hand,  to address many real problems, many algorithms are not developed in a principled way, leading to computational results without statistical guarantees. As argued by \cite{chandrasekaran2013computational}, there is a great need to rethink the relationship between statistical accuracy and computational efficiency.

To bridge the gap, most statistical literatures focus on reducing the theoretical complexity of an algorithm, or simply using parallel computing to speed it up, without paying not enough attention to  taking advantage of the computational architecture. In fact, the computational performance of statistical models can often be greatly improved by designing new data structures or using hardware acceleration (e.g., graphical processing units for training deep neural networks).  In this paper, we use the interaction detection problem in high dimensional models as an example, to  demonstrate that it is possible to design statistically guaranteed algorithms to overcome seemingly unaffordable computational cost by taking advantage of the computational architecture.

\subsection{Related work for interaction effect detection}
The word ``interaction'', in Oxford English Dictionary, is illustrated  as the reciprocal action, or influence of persons or things on each other. It is one kind of relationship among two or more objects, which have mutual influence upon one another.
There is a long history of investigating the interaction effects in many different scientific fields. For example, in physical chemistry, the main topics are interactions between atoms and molecules. A simple example in the real-world is that neither of carbon and steel has much effect on the strength, but a combination of them has substantial effects.
In medicine and pharmacology, the interaction effects of multiple drugs have been widely observed (\cite{lees2004principles}). In genomics, gene-gene interactions and gene-environment interactions have been widely studied by bio-medical researchers since the seminal work of  \cite{bateson1909mendel}. In recent years, increasing interest has been focusing on detecting gene-gene interactions from genome-wide association studies (GWAS) (\cite{cordell2009detecting}).

In this paper, we investigate the interaction effects from a statistical perspective, where the interaction effect is characterized by the statistical departure from the additive effects of two or more factors (see \cite{fisher1918correlation,cox1984interaction}). In the framework of high dimensional regression, it is common to use  products of  explanatory variables to study  interaction effects of explanatory variables on response variables. Consider three explanatory variables  $X_i$, $X_j$ and $X_k$, their two-way interaction terms are  $X_jX_k$, $X_iX_j$ and $X_iX_k$. By including these interaction terms, the standard linear regression model becomes
\begin{equation}\label{inter1}
Y=\beta_0+\sum_{i=1}^p\beta_iX_i+\sum_{1\leq j<k\leq p}\beta_{jk}X_{j}X_k+\varepsilon.
\end{equation}
where $Y$ is the response variable, $\beta_0$ is the intercept term, $\beta_i$ is the  coefficient of main effect term $X_i$, $\beta_{jk}$ is the coefficient of interaction term $X_jX_k$, and $\varepsilon$ is the independent error. For the high  dimensional data, the number of variables $p$ can be much larger than the sample size  $n$.
Clearly, the number of parameters to be determined would be $p+p(p-1)/2$ if all two-way interaction terms are included. For example, in GWAS, there are millions of genotyped genetic variants, i.e., $p\approx10^6$. The number of interaction terms goes up to an astronomical number at the order of $10^{12}$. The computational cost of detecting interaction effects in such a scale becomes seemingly un-affordable,
making the theoretical guarantees  with mild conditions (e.g. sparsity assumptions) useless.

To reduce the computational cost, methods developed recently often make two types of heredity assumptions: the strong heredity assumption means that the interaction effect is important only if its both parent are significant, while the weak heredity assumption illustrates that the interaction term is important only if at least one of its parent is included in the model.
To name a few, \cite{choi2010variable} extended the LASSO method and identified the significant interaction terms in the linear model  and generalized linear models under the strong heredity assumption.  \cite{choi2010variable} proved that their method possessed the oracle property  (\cite{fan2001variable} and \cite{fan2004nonconcave}), that is, it performed well as if the true model was known in advance.  The algorithm hierNet was developed by \cite{bien2013lasso} to select the interactions, which added   a set of convex constraints to LASSO in the linear model and  constructed the sparse interaction model with the strong or weak heredity assumptions. For the linear model, \cite{hao2014interaction} also proposed two algorithms iFORT and iFORM, and identified the interaction effects in a greedy fashion under the heredity assumption. \cite{hao2018model}  further improved interaction detection by proposing a regularization algorithm under marginality principle (RAMP).
To deviate from  these heredity assumptions for interaction  detection,    \cite{fan2016interaction} suggested a flexible sure screening procedure, called the interaction pursuit (IP), in ultra-high dimensional linear interaction models, applying a strong assumption on the joint normality between the response and predictor variables. The idea of the IP method is  to select the  ``active interaction variables" by screening significant predictor variables with the strong Pearson correlation between $X_j^2$ and $Y^2$ firstly,  and then detect the interaction effects among those identified active interaction variables.  \cite{kong2017interaction} extended IP to the ultra-high dimensional linear interaction model with multiple responses by  identifying the active interactive variables using the distance correlation  with $X_j^2$ and the multiple response $\Y^2$, where $\Y=(Y_1,\ldots,Y_q)$ be  a $q$-dimensional vector of responses and $\Y^2=(Y_1^2,\ldots,Y_q^2)$.

However, the heredity assumption may not be satisfied in practice due to the existence of pure interaction effects.
In human genetics, a number of gene-gene interaction effects have been detected in the absence of their main effects (\cite{cordell2009detecting} and \cite{wan2010boost}).
This motivates new methods to detect interactions without any heredity assumptions.
Recently, a new algorithm $xyz$ based on random projection was introduced by \cite{thanei2018xyz}  to screen interaction effects. This algorithm does not rely on the heredity assumption, thus it can detect interaction effects in the absence of the corresponding main effects. Based on our empirical observations, however, its performance in the real applications is not entirely  satisfactory because its accuracy of detecting interaction effects largely depends on the number of random projections. Yet, computationally efficient algorithms with statistically guaranteed performance for interaction detection are still lacking.

\subsection{Our Contribution}

Our contribution is to develop a computationally efficient and statistically guaranteed method for interaction detection in high dimensional problems:
\begin{itemize}
	\item[a.] We propose a new sure screening procedure (SSI) based on the increment of log-likelihood function to fully detect significant interactions  for the high dimensional generalized linear models.  Furthermore, in order to reduce the computational burden, we  take the advantages of computer architecture such as parallel techniques and Boolean operations to construct more computationally efficient algorithm BOLT-SSI, and make  available the detection for interaction effects in a large-scale data set. For example, for the data set Northern Finland Birth Cohort (NFBC) with $n=5,123$ individuals and $p=319,147$ SNPs, the number of interactions is about $5\times 10^{10}$. BOLT-SSI can quickly screen all these interactions with a short time. The details can be seen in  section 6.
	
	\item [b.]  Moreover, we investigate the sure screening properties of  SSI and BOLT-SSI  from theoretical insights, and show that our computationally efficient methods are statistically guaranteed. We provide implementations of both the core SSI algorithm and its extension BOLT-SSI in the R package BOLT-SSI, available on the authors' website
	({\url{h ttps://github.com/daviddaigithub/BOLTSSIRR}}).
	
	\item [c.] More importantly, our work is a practical attempt to integrate the advantages of well-designed  computer architecture and  statistically rigorous methodology. We take it as an example to promote the application of  computational structure in the statistical modeling and practice, especially in the era of ``Big Data''. We hope this example motivates  more combination of statistical methods and computational techniques, greatly improving the computational performance of statistical methods.
\end{itemize}

The rest of this paper is organized as follows. In Sections 2 and 3, we propose the sure screening algorithms  SSI and BOLT-SSI for detecting interactions in ultra-high dimensional generalized linear regression model, where we briefly introduce the Boolean representation and operations.  The theoretical properties of sure screening for the proposed methods are investigated in Section 4. In Section 5, we examine the finite sample performance of SSI and BOLT-SSI in comparison to alternative methods, RAMP, $xyz$-algorithm, and IP, through simulation studies.   In Section 6, three real data sets are used to demonstrate the utility of our approaches. Our findings and conclusions are summarized in Section 7. The details of the proof are given in the Appendix.

\section{Sure Screening Methods for Interaction in GLM}
\subsection{Generalized linear models(GLM) with Two-way Interaction}

Assume that given the predictor vector $\x$, the conditional distribution of the random variable $Y$ belongs to an exponential family, whose  probability density function has the canonical form
\begin{equation}\label{GLIM}
f_{Y|\x}(y|\x)=\exp\{y\theta(\x)-b(\theta(\x))+c(y)\}
\end{equation}
where $b(\cdot)$ and $c(\cdot)$ are some known functions and $\theta(\x)$ is a canonical natural parameter. Here we ignore the dispersion parameter $\phi$ in (\ref{GLIM}), since we only concentrate on the estimation of mean regression function. It is well known that the distributions in the exponential family include the Binomial, Gaussian, Gamma, Inverse-Gaussian and Poisson distributions.

We consider the following generalized linear model with two-way interactions:
\begin{equation}\label{model}
E(Y|\X)=b'(\theta(\X))=g^{-1}\left(\beta_0+\sum_{i=1}^p\beta_iX_i+\sum_{i<j}\beta_{ij}X_iX_j\right)
\end{equation}
for the canonical  link function $g^{-1}(\cdot)=b'$ with $$\theta(\X)=\beta_0+\sum_{i=1}^p\beta_iX_i+\sum_{i<j}\beta_{ij}X_iX_j\hat{=}
\beta_0+\sum_{i=1}^p\beta_iX_i+\sum_{i<j}\beta_{ij}X_{ij}.$$
where $\X=(\X_{\mathcal{C}}^T,\X_{\mathcal{I}}^T)^T$ with $\X_{\mathcal{C}}=(X_0,X_1,X_2,X_3,\ldots,X_p)^T$  and $\X_{\mathcal{I}}=(X_{12},X_{13},\ldots,X_{(p-1)p})^T$.  For simplicity, we assume that  $X_0=1$ and each of the other predictor variables is standardized with zero mean and unit variance. The corresponding sets of coefficient are
$$\Beta_{\mathcal{C}}=(\beta_0,\beta_{1},\beta_{2},\ldots,\beta_{p})^T\in \mathbb{R}^{p},\ \ \ \mbox{and}\ \
\Beta_{\mathcal{I}}=(\beta_{12},\beta_{13},\ldots,\beta_{(p-1)p})^T\in \mathbb{R}^{q},$$
where $q=\binom{p}{2}=p(p-1)/2.$

In the ultra-high dimensional regression model, we usually assume that there is a sparse structure in  the underlying  model. It means that only a few of predictor variables or features are significantly correlated with response $Y$. Hence for the above model with two-way interactions, we assume there are only a small number of interactions contributing to the response $Y$. Denote that the true parameter $\Beta^\star=({\Beta_{\mathcal{C}}^\star}^T, {\Beta_{\mathcal{I}}^\star}^T)^T$, where $\Beta_{\mathcal{C}}^\star=(\beta_0^\star,\beta_{1}^\star,\beta_{2}^\star,\ldots,\beta_{p}^\star)^T\in \mathbb{R}^{p+1}$ for main effects, and
$\Beta_{\mathcal{I}}^\star=(\beta_{12}^\star,\beta_{13}^\star,\ldots,\beta_{(p-1)p}^\star)^T\in \mathbb{R}^{q}$ with $q=\binom{p}{2}=p(p-1)/2$ for interactions.

Let
$$\mathcal{N}_\star=\{(i,j): \beta_{ij}^\star\neq0,\ 1\leq i <j\leq p\},$$ and denote that  $s_n=|\mathcal{N}_\star|$, then the non-sparsity size  $s_n$ is a relative small number compared to the dimension $p$ of the model.

\subsection{SSI for two-way interaction in GLM}
The model (\ref{model}) can be simply rewritten  as an ordinary generalized linear regression model form
\begin{equation}\label{model2}
E(Y|\X)=b'(\theta(\X))=g^{-1}(\X^T\Beta).
\end{equation}
\cite{fan2009ultrahigh} suggested to select the important variables by sorting the marginal likelihood, and  \cite{fan2010sure} pointed out that such  technique can be considered as the marginal likelihood ratio screening, which builds on the difference between two marginal log-likelihood functions.   If we regard the interaction variable $X_{ij}$ the same as other main effects from predictor variables $X_i, X_j$, by considering the marginal likelihood of $(X_{ij}, Y)$,  we could directly apply the sure screening techniques of  \cite{fan2009ultrahigh} and \cite{fan2010sure} to detect the significant interaction effects. But such a direct screening method ignores the main effects of $X_i$ and $X_j$, as argued by \cite{jaccard1990detection}, it often leads to  false discoveries for the pure significant interaction effects. Hence we consider the following sure screening procedure to detect pure interaction effects in the model (\ref{model}).

Denote that the random samples $\{(\X^{(k)},Y^{(k)}, k=1,\ldots, n\}$ are i.i.d. from the model  (\ref{model}) with the canonical link. Let $\X_{ij}=(1,X_i,X_j,X_{ij})^T$ and $\X_{i,j}=(1,X_i,X_j)^T$. And their coefficients are expressed as $\Beta_{ij}=(\beta_{ij0},\beta_i,\beta_j,\beta_{ij})^T$ and $\Beta_{i,j}=(\beta_{i,j0},\beta_{i,},\beta_{j,})^T$, respectively.

The first step of  the Sure Screening procedure to detect the Interaction effects (SSI) is to calculate the maximum marginal likelihood estimator $\hat{\Beta}_{ij}^M$ by the minimizer of the marginal regression
$$\hat{\Beta}_{ij}^M =\argmin_{\Beta_{ij}}{\mathbb{P}}_n\{l(\X_{ij}^T\Beta_{ij},Y)\}$$
where $l(\theta,Y)=b(\theta)-\theta Y-c(Y)$ and ${\mathbb{P}}_nf(\X,Y)=n^{-1}\sum_{k=1}^nf(\X^{(k)}_i,Y^{(k)}_i)$ is the empirical measure.
Similarly, we can  calculate the maximum marginal likelihood estimator $\hat{\Beta}_{i, j}^M$ without the interaction effect by the minimizer of the marginal regression
$$\hat{\Beta}_{i,j}^M
=\argmin_{\Beta_{i,j}}{\mathbb{P}}_n\{l(\X_{i,j}^T\Beta_{i,j},Y)\}\,.$$

Correspondingly, let the population version of the above minimizers of the marginal regressions be
$$\Beta_{ij}^M
=\argmin_{\Beta_{ij}}\mathrm{E}\{l(\X_{ij}^T\Beta_{ij},Y)\}$$
and
$$\Beta_{i,j}^M
=\argmin_{\Beta_{i,j}}\mathrm{E}\{l(\X_{i,j}^T\Beta_{i,j},Y)\}\,.$$

In fact, the coefficient $\beta_{ij}^M$ can measure the importance of the interaction terms from population insight. Though the real joint regression parameter $\beta_{ij}^\star$ would not be the same as the marginal regression coefficient $\beta_{ij}^M$,  we could still expect that, under mild conditions, $|\beta_{ij}^M|$ or the increment of the marginal log-likelihood function
$$L_{ij}^\star=\mathrm{E}\{l(\X_{i,j}^T\Beta_{i,j}^M,Y)-l(\X_{ij}^T\Beta_{ij}^M,Y)\}$$
is large,  if and only if $|\beta_{ij}^\star|$ is some large.

Hence the second step of the SSI procedure is to calculate the increment of the empirical maximum marginal likelihood function,  $$L_{ij,n}={\mathbb{P}}_n\{l(\X_{i,j}^T\hat{\Beta}_{i,j}^M,Y)-l(\X_{ij}^T\hat{\Beta}_{ij}^M,Y)\}$$
and $\L_n=(L_{12,n},\ldots,L_{(p-1)p,n})^T \in \mathbb{R}^{q}$ . Then  $L_{ij,n}$ measures the strength of the interaction $X_{ij}$ in the marginal model from the empirical version.  The larger $L_{ij,n}$, similar to $L^\ast_{ij}$, the more the interaction $X_{ij}$ contributes to the response $Y$.

The final step of the SSI procedure is to sort the vector $\L_n$ in a decreasing order and given threshold value $\gamma_n$,  select the following interaction effect variables
$$\widehat{\mathcal{N}}_{\gamma_n}=\{(i,j): L_{ij,n}\geq\gamma_n,\ 1\leq i<j\leq p\},$$
as the final candidates of the significant pure interaction effects.

Under regularized conditions and similar as the classical approach, it is not difficult to show that SSI has the so-called ``sure screening properties". So here we delegate those investigations of SSI properties to our supplement file. From practical insight, the proposed SSI procedure's computational complexity is in the order of $O(p^2n)$. When $p$ is of moderate size ($10^3-10^4$), SSI can quickly screen all interaction terms. It can be further accelerated by parallel computing because all the interaction terms can be evaluated independently.

\section{BOLT-SSI}
Despite the simplicity of SSI, it can not be scaled up to handle the case that dimensionality $p$ is very large, e.g., $p = 10^6$. To such a scenario, as other methods, we could impose similar uncheckable heredity assumptions to shrink the screening space of SSI to detect the interaction effects. But for such an approach, some significant interaction effects could never be discovered. Hence, even though we could have enough large observational samples, the method's efficiency could still be worst. The other approach is to use a rough but fast algorithm or calculation method to approximate and accelerate SSI's speed to deal with ultra-high dimensional scenarios. Though from theoretical insight, it would not decrease the original SSI algorithm's complexity and has to sacrifice SSI stability; such an approach would not lose much information about the data and miss essential discoveries. Especially, as the number of observations is large enough, such an approach's statistical efficiency could be satisfied by the requirement of real applications as our experience. It is the other kind of trade-off between statistical efficiency and computational efficiency.

In this paper, utilizing the computer's computational architecture, we follow the second approach and present a computationally efficient algorithm named ``BOLT-SSI'' to detect interactions in ultra-high dimensional problems. The BOLT-SSI algorithm is motivated by the following fact: when $X_j$, $X_k$ and $Y$ all are discrete variables, the interaction effects of $X_j$ and $X_k$ on $Y$ measured by logistic regression can be exactly calculated based on a few numbers in the contingency table of $X_j$, $X_k$ and $Y$. These numbers can be efficiently obtained by designing a new data structure and its associated operations, i.e., Boolean representation and Boolean operations. To handle continuous variables, we propose discretization first and then use the above strategy for screening. This section describes the details of BOLT-SSI algorithm and establishes statistical theory to guarantee its performance in the next section.

\subsection{Equivalence between the logistic models and log-linear models}
When all predictors and the response are categorical variables, we usually take the logistic model (for binary response)  or baseline-category logit models (for the response with several categories) to fit the data set. Actually, the logistic regression models or baseline-category logit models  have their corresponding log-linear regression models for the  contingency table when the predictor and the response are categorical (See \cite{agresti2011categorical}, Chapter 9 Section 9.5). Based on this equivalence, the significance of interaction effects can be measured by the increment of the corresponding log-linear regression models.

Assume that we consider the following two logistic models with main effects and full model, respectively:
\begin{equation}\label{logit1}
\text{logit}(P(Y=1|X,Z))=\beta_0+\beta_i^X+\beta_j^Z
\end{equation}
and
\begin{equation}\label{logit2}
\text{logit}(P(Y=1|X,Z))=\beta_0+\beta_i^X+\beta_j^Z+\beta_{ij}^{XZ}.
\end{equation}

Denote that $\widehat{l}_M$  and $\widehat{l}_F$ be the sample version of the negative maximum log-likelihood for the logistic regression models with main effects (\ref{logit1}) and full model (\ref{logit2}), respectively. The increment of the log-likelihood function is defined as   $\widehat{l}_M-\widehat{l}_F$. The corresponding log-linear  regression models can be expressed as
\begin{equation}\label{loglinear1}
\log(\mu_{ijk})=\lambda+\lambda_i^X+\lambda_j^Z+\lambda_k^Y+\lambda_{ij}^{XZ}+\lambda_{ik}^{XY}+\lambda_{jk}^{ZY}
\end{equation}
and
\begin{equation}\label{loglinear2}
\log(\mu_{ijk})=\lambda+\lambda_i^X+\lambda_j^Z+\lambda_k^Y+\lambda_{ij}^{XZ}+\lambda_{ik}^{XY}+\lambda_{jk}^{ZY}+\lambda_{ijk}^{XZY}.
\end{equation}
Let  $\widehat{l}_H$ and $\widehat{l}_S$  be the  sample version of the negative maximum log-likelihood for the homogeneous association regression model (\ref{loglinear1}) and the saturated model (\ref{loglinear2}), respectively. $\widehat{l}_H-\widehat{l}_S$ is the corresponding  increment of log-likelihood function. Thus, we can take advantage of $\widehat{l}_H-\widehat{l}_S$ to screen the interaction terms instead of using $\widehat{l}_M-\widehat{l}_F$.

Now we want to obtain the difference $\widehat{l}_H-\widehat{l}_S$. Suppose that we have one three-way ($I\times J\times K$) table with cell counts $\{n_{ijk}\}$ of random variables $X$, $Z$ and $Y$. The kernel of the log-likelihood function for this contingency table is
$$L(\Mu)=\sum_{ijk}n_{ijk}\log(\mu_{ijk})-\sum_{ijk}\mu_{ijk}.$$
Denote that $\pi_{i++}=\sum_{jk}\pi_{ijk}$ is the marginal probability of $X=i$ and $n_{i++}=\sum_{jk}n_{ijk}$ is the number of samples with $X=i$, $\pi_{ij+}=\sum_k\pi_{ijk}$ is the marginal probability of $X=i$ and $Z=j$ and $n_{ij+}=\sum_kn_{ijk}$ is the corresponding count. Similarly, $\pi_{+j+}=\sum_{ik}\pi_{ijk}$, $\pi_{++k}=\sum_{ij}\pi_{ijk}$, $\pi_{i+k}=\sum_j\pi_{ijk}$, $n_{i+k}=\sum_jn_{ijk}$, $\pi_{+jk}=\sum_i\pi_{ijk}$, $n_{+j+}=\sum_{ik}n_{ijk}$, $n_{++k}=\sum_{ij}n_{ijk}$, $n_{+jk}=\sum_in_{ijk}$.

For the saturated model (\ref{loglinear2}), we know that $\widehat{\mu}_{ijk}=n_{ijk}$ and directly get the estimation $\widehat{l}_S=\sum_{ijk}n_{ijk}\log(n_{ijk})-\sum_{ijk}n_{ijk}$.  For the  homogeneous association regression model (\ref{loglinear1}), the iterative proportional fitting (IPF) algorithm \cite{deming1940least} is used to calculate the estimate of $u_{ijk}$ efficiently. Three steps are included in the first cycle of the IPF algorithm:
$$\mu_{ijk}^{(1)}=\mu_{ijk}^{(0)}\frac{n_{ij+}}{\mu_{ij+}^{(0)}},\ \ \mu_{ijk}^{(2)}=\mu_{ijk}^{(1)}\frac{n_{i+k}}{\mu_{i+k}^{(1)}},\ \ \mu_{ijk}^{(3)}=\mu_{ijk}^{(2)}\frac{n_{+jk}}{\mu_{+jk}^{(2)}},$$
where  $\mu_{ij+}=\sum_{k}\mu_{ijk}$, , $\mu_{i+k}=\sum_{j}\mu_{ijk}$, $\mu_{+jk}=\sum_{i}\mu_{ijk}$.
This cycle does not stop until the process  converges and the convergence  property has been proved by \cite{fienberg1970iterative} and \cite{haberman1974analysis}. We count the number $n_{ijk}$ by using the Boolean representation, thus the contingency table for $X$ and $Z$ given $Y$ can be quickly constructed in a fast manner. In this way, the estimation $\widehat{l}_H$ will be obtained.

Consequently, we can take advantage of this equivalence to efficiently estimate the corresponding increment of log-likelihood function by the IPF algorithm when the predictors and the response are qualitative.  If some variables are continuous, we can discretize them and the details can be seen in the next section. In section 4, we show that our algorithm is still statistically guaranteed after discretization.

\subsection{Discretization}
In the case that  some of the predictors and/or response are continuous, we suggest discretizing them simply binned by equal width or frequency.
Considering the variation of random observations,  it would be more reasonable to use the equal-frequency method by quantiles to split the domain of variables to several intervals. The number of intervals is called ``arity'' in the discretization context (See \cite{liu2002discretization}). Assume that the arity is denoted by $l$, and then $l-1$ is the maximum number of cut-points of the continuous features.

For more detail, we follow the assumption of \cite{fan2010sure}, and consider variable or feature selection of the generalized linear model:
\begin{equation}\label{m2}
Y=b'(\X^T\Beta)+\varepsilon.
\end{equation}
where $\X=(X_1,\ X_2,\ \ldots,\ X_p)^T$ is a $p\times 1$ random vector,  $\Beta=\{\beta_1,\ \beta_2,\ \ldots,\ \beta_p\}$ is the parameter vector, $Y$ is the response, $b'(\cdot)$ is the canonical link function,  and assume that
$$\mathcal{M}_\star=\{1\leq k \leq p:\ \beta_k\neq0\}$$
is the set of indexes of nonzero parameter. Define the marginal log-likelihood increment
$$L_{k}^\star=\mbox{E}\{l(\beta_0^M,Y)-l(\X_k^T\Beta_k^M,Y)\},\ \ \ \ k=1,2,\ldots,p$$
where $\beta_0^M=\argmin_{\beta_0}\mbox{E}l(\beta_0,Y)$, $\X_k^T=\{1,X_k\}$, $\Beta_{k}^M=\{\beta_{k,0},\beta_k^M\}^T$ and  $$\Beta_{k}^M=\argmin_{\Beta_k}\mbox{E}l(\X_k^T\Beta_k,Y).$$
Furthermore, $\mbox{E}(Y)=\mbox{E}(X_k)=0$  and $\mbox{E}(Y^2)=\mbox{E}(X_k^2)=1$, $k=1,2,\ldots, p$. Let $\rho_k=\mbox{Corr}(Y,X_k)$ and $(Y_1,X_{1k})$, $(Y_2, X_{2k})$ be the independent copies of $(Y, X_k)$.\\
\indent Assume that $S^{X_k}$ and $S^Y$ are the support sets of variables $X_k$ and $Y$, respectively. Denote that $\{P_i^{X_k}\}_{i=1}^{l}$ and $\{P_j^{Y}\}_{j=1}^{m}$ are  partitions of their  supports, which means that
$$\bigcup_{i=1}^lP_i^{X_k}=S^{X_k}\ \ \ \mbox{and}\ \ \ P_{i_1}^{X_k}\bigcap P_{i_2}^{X_k}=\emptyset \ \ \mbox{for}\ i_1\neq i_2;$$
and
$$\bigcup_{j=1}^mP_j^{Y}=S^{Y}\ \ \ \mbox{and}\ \ \ P_{j_1}^{Y}\bigcap P_{j_2}^{Y}=\emptyset \ \ \mbox{for}\ j_1\neq j_2;$$
where $l$ and $m$ are two positive constants. Here, the $l-$quantiles and $m-$quantiles are considered as the break points for the partitions of variables $X_k$ and $Y$. Define
$$\widetilde{X}_k=\left\{\begin{array}{cc}
0,\ & X_k\in P_1^{X_k} \\
1,\ & X_k\in P_2^{X_k} \\
\vdots\ & \vdots \\
l-1,\ & X_k\in P_l^{X_k}
\end{array}
\right. \ \ \ \mbox{and}\ \ \  \widetilde{Y}=\left\{\begin{array}{cc}
0,\ & Y\in P_1^Y \\
1,\ & Y\in P_2^Y\\
\vdots & \vdots\\
m-1,\ & Y\in P_m^Y
\end{array}
\right.,$$
and then variables $X_k$ and $Y$ are discretized  to two categorical variables $\widetilde{X}_k$ and $\widetilde{Y}$, respectively. Furthermore, denote that $\widetilde{X}_{k_i}=I(X_k\in P_i^{X_k})$, $1\leq i\leq l$ and $\widetilde{Y}_{j}=I(Y\in P_j^{Y})$, $1\leq j \leq m$, where $I(\cdot)$ is the indicator function.
After discretization, we have the new increment of log-likelihood function as
$$\widetilde{L}_k^\star=\mbox{E}\{l(\widetilde{\beta}_0^M,\widetilde{Y})-l(\widetilde{\X}_k^T\widetilde{\Beta}_k^M,\widetilde{Y})\},\ \ \ \ k=1,2,\ldots,p.$$

 Now consider the discretization for the marginal model with the interaction effect. Assume that $S^{X_i}$, $S^{X_j}$ and $S^Y$ are the support sets of variables $X_i$, $X_j$ and $Y$, respectively. Denote that $\{P_s^{X_i}\}_{s=1}^{l_1}$, $\{P_t^{X_j}\}_{t=1}^{l_2}$and $\{P_k^{Y}\}_{k=1}^{m}$ are  partitions of their  supports, which means that
$$\bigcup_{s=1}^lP_s^{X_i}=S^{X_i}\ \ \ \mbox{and}\ \ \ P_{s_1}^{X_i}\bigcap P_{s_2}^{X_i}=\emptyset \ \ \mbox{for}\ s_1\neq s_2;$$
$$\bigcup_{t=1}^lP_t^{X_j}=S^{X_j}\ \ \ \mbox{and}\ \ \ P_{t_1}^{X_j}\bigcap P_{t_2}^{X_j}=\emptyset \ \ \mbox{for}\ t_1\neq t_2;$$
and
$$\bigcup_{k=1}^mP_k^{Y}=S^{Y}\ \ \ \mbox{and}\ \ \ P_{k_1}^{Y}\bigcap P_{k_2}^{Y}=\emptyset \ \ \mbox{for}\ k_1\neq k_2;$$
where $l_1$, $l_2$ and $m$ are  positive constants. Here, we still consider the $l_1-$quantiles, $l_2-$quantiles and $m-$quantiles as the break points for the partitions of variables $X_i$, $X_j$ and $Y$, respectively. Define
$$\widetilde{X}_i=\left\{\begin{array}{cc}
0,\ & X_i\in P_1^{X_i} \\
1,\ & X_i\in P_2^{X_i}\\
\vdots\ & \vdots \\
l_1-1,\ & X_i\in P_{l_1}^{X_i}
\end{array}
\right. \ \ \ \mbox{and}\ \ \  \widetilde{X}_j=\left\{\begin{array}{cc}
0,\ & X_j\in P_1^{X_j} \\
1,\ & X_j\in P_2^{X_j}\\
\vdots\ & \vdots \\
l_2-1,\ & X_j\in P_{l_2}^{X_j}
\end{array}
\right..$$
Furthermore, denote that
$$\widetilde{X}^{ij}=\left\{\begin{array}{cc}
0,\ & X_i\in P_1^{X_i}\ \  \mbox{and}\ \  X_j\in P_1^{X_j} \\
1,\ & X_i\in P_1^{X_i}\ \  \mbox{and}\ \  X_j\in P_2^{X_j} \\
\vdots\ & \vdots \\
l_1*l_2-1,\ & X_i\in P_{l_1}^{X_i}\ \  \mbox{and}\ \  X_j\in P_{l_2}^{X_j}
\end{array}
\right. $$
And also, we define the discretized response $\widetilde{Y}$,
$$\widetilde{Y}=\left\{\begin{array}{cc}
0,\ & Y\in P_1^Y \\
1,\ & Y\in P_2^Y\\
\vdots & \vdots\\
m-1,\ & Y\in P_m^Y
\end{array}
\right..$$
Hence, we have the new categorical predictor $\widetilde{X}_i$, $\widetilde{X}_j$ and response $\widetilde{Y}$, respectively.  And also, we get the new interaction variable $\widetilde{X}^{ij}$. Furthermore, denote that $$\widetilde{X}_{st}^{ij}=I\left(\left\{X_i\in P_s^{X_i}\right\}\ \bigcap\ \left\{X_j\in P_t^{X_j}\right\}\right),\ \ \ 1\leq s\leq l_1,\ \ 1\leq t\leq l_2$$
and $\widetilde{Y}_{j}=I(Y\in P_j^{Y})$, $1\leq j \leq m$, where $I(\cdot)$ is the indicator function.
After discretization,  the new increment of log-likelihood function in population version is defined as
$$\widetilde{L}_{ij}^\star=E\{l(\widetilde{\X}_{i,j}^T\widetilde{\Beta}_{i,j}^M,\widetilde{Y})
-l(\widetilde{\X}_{ij}^T\widetilde{\Beta}_{ij}^M,\widetilde{Y})\},\ \ \ 1\leq i<j\leq p.$$

\begin{remark}
	Actually, there is a trade-off between the arity $l$ and the accuracy of screening procedures. Higher arity would lead to  a more accurate sure screening. However, when the sample size of data is large enough, the relatively small arity $l$ could also guarantee the accuracy of the screening procedure from our theoretical investigation and numerical studies. Hence though large $l_i$ for different continuous features $X_i$ can be also used.
	we recommend using $l=2, 3$ to make a trade-off between the computation burden and efficiency of model estimation for our proposed BOLT-SSI when the sample size of the data is relatively large.
	
	Furthermore, if $Y$ is a continuous response, similarly we also suggest  to use 2-quantile (median) to split the response $Y$, that is, $m=2$ and $$\widetilde{Y}=\left\{\begin{array}{cc}
	0,\ & Y \leq M_d(Y) \\
	1,\ & Y > M_d(Y)
	\end{array}
	\right.,$$ where $M_d(Y)$ is the median of the response $Y$.
\end{remark}

\subsection{Boolean Representation and Logical Operations }
After discretization, the Boolean operation can be used to speed up the SSI procedure, especially the algorithm to calculate $\widetilde{L}_k^\ast$.  The Boolean Representation and its operations is a classical and fundamental computer computing technique. A standard floating computation that provides a basic operation for many statistical software is composed of hundreds of Boolean operations under a lower level of the computer computing. Hence if the Boolean operation can be directly applied to realize  the proposed algorithm, the computational speed could be much improved.

Assume that the continuous data set $\X$ is one $n\times p$ matrix with $n$ observations and $p$ predictors, $Y$ be the response. After discretizing data set $\X$ and response $Y$, each predictor $\widetilde{X}_i$ has $l$ levels  and $\widetilde{Y}$ has $m$ categories. Here, we take $l=3$ and $m=2$ as an example. Assuming that $\widetilde{Y}$ has two values (0 and 1), then instead of using one row for each predictor $\widetilde{X}_i$, the  new representation uses 3 rows since  3 levels are included in each $\widetilde{X}_i$. Each row consists of two-bit strings, one for samples with $\widetilde{Y}=0$  and the other for them with $\widetilde{Y}=1$, and  each bit can represent one sample in the string. The values (0 and 1) illustrate whether the sample belongs to such a categorical level for each predictor $X_i$.
For instance, we have one discretized data set $\widetilde{\X}$ with 2 predictors and 16 samples, where the first 8 columns represent samples with $\widetilde{Y}=0$ and the others represent samples with $\widetilde{Y}=1$:
$$\widetilde{\X}^T=\begin{array}{c}
\widetilde{Y}\\
\widetilde{X}_1 \\
\widetilde{X}_2
\end{array}
\left[
\begin{array}{ccccccccccccccccc}
0 & 0 & 0 & 0 & 0 & 0 & 0 & 0 & \vdots & 1 & 1 & 1 & 1 & 1 & 1 & 1 & 1 \\
1 & 3 & 2 & 3 & 1 & 2 & 3 & 2 & \vdots & 2 & 2 & 1 & 1 & 3 & 2 & 2 & 1 \\
3 & 2 & 1 & 1 & 3 & 2 & 2 & 1 & \vdots & 2 & 3 & 2 & 3 & 1 & 2 & 3 & 2\\
\end{array}
\right]
$$
and its Boolean representation is
$$\widetilde{\X}_{bit}^T=\begin{array}{c}
\\
\widetilde{X}_1=1 \\
\widetilde{X}_1=2 \\
\widetilde{X}_1=3 \\
\widetilde{X}_2=1 \\
\widetilde{X}_2=2 \\
\widetilde{X}_2=3
\end{array}
\left[
\begin{array}{cc}
\widetilde{Y}=0  & \widetilde{Y}=1     \\
10001000 & 00110001 \\
00100101 & 11000110 \\
01010010 & 00001000 \\
00110001 & 00001000 \\
01000110 & 10100101 \\
10001000 & 01010010 \\
\end{array}
\right]
$$
From the Boolean representation $\widetilde{\X}_{bit}$, we can easily find that the first sample belongs to the first category of $X_1$ and the third category of $X_2$. Further, we can quickly obtain the number of observations that belong to any two categories by taking the logic operation. For example, if we want to calculate the number of  samples with $\widetilde{X}_1=2$ and $\widetilde{X}_2=2$ in the category $\widetilde{Y}=0$, we just conduct the logical {\bf AND} operation:
$$00100101\ \ {\bf AND}\ \  01000110=00000100, $$
and then,  we count the number of 1s in the final string ``00000100'', that is 1. As a result, it is more efficient by using $\widetilde{\X}_{bit}$ to construct the contingency table for any two discretized predictors. Since the fast logic operation with $\widetilde{\X}_{bit}$ is utilized, we can accelerate  our computation for our algorithm.

Obviously, $\widetilde{\X}$ and $\widetilde{\X}_{bit}$ are equivalent and they store the same amount of information.  Because  one byte is composed of 8 bits, $\widetilde{\X}_{bit}$ uses 128 bits to save the data, but  $\widetilde{\X}$ would use $32 \times 64$ bits, 16 times of the space of $\widetilde{\X}_{bit}$,  to save the same data if our computer is a 64-bit computer system.   As a result, the Boolean representation could dramatically reduce the storage space of the data. So all of the large data could be directly uploaded into the RAM, or even be saved in the cache. The transferring amount of time for the data between hard disk and RAM, and between RAM and cache can be largely reduced. This is the other advantage of the Boolean representation or the discretization.

\subsection{New algorithm ``BOLT-SSI'' }
Now, we illustrate our algorithm BOLT-SSI in details. For our ultra-high dimensional generalized linear  model (\ref{model}), instead of calculating the increment $\widetilde{L}_{ij,n}=\widehat{l}_{M_{ij}}-\widehat{l}_{F_{ij}}$ for any pair of $\widetilde{X}_i$ and $\widetilde{X}_j$,   we compute the new increment of the log-likelihood function $\widetilde{L}_{ij,n}'=\widehat{l}_{H_{ij}}-\widehat{l}_{S_{ij}}$ by the IPF method. Then, by taking the thresholding value $\gamma_n$ or choosing the large $d=\left\lfloor\frac{n}{\log n}\right\rfloor$ or $\max(n,p)$,   the selected sure screening set $\widehat{\mathcal{N}}_{\gamma_n}$ is obtained.  Our algorithm BOLT-SSI is summarized as follows:

$\emph {Step 1}.$ For any pair of the continuous variables $X_i$ and $X_j$, $1\leq i<j \leq p$, transform them to the corresponding discretized variables $\widetilde{X}_i$ with level $l_i$ and $\widetilde{X}_j$ with level $l_j$, and change the response $Y$ to a categorical variable $\widetilde{Y}$ if necessary.

$\emph {Step 2}.$  Directly calculate $\widehat{l}_{S_{ij}}$ and  use the IPF algorithm to approximately estimate  $\widetilde{l}_{H_{ij}}$ , and then compute $\widetilde{L}_{ij,n}'=\widehat{l}_{H_{ij}}-\widehat{l}_{S_{ij}}$ for all pairs of $X_i$ and $X_j$.

$\emph {Step 3}. $ Choose the  threshold  $\gamma_n$ and select the following interactions:
$$\mathcal{\widetilde{N}}_{\gamma_n}=\{(i,j): \widetilde{L}_{ij,n}'\geq\gamma_n,\ 1\leq i<j\leq p\}.$$
Usually, we select the $d$ largest $L_{ij,n}$, where $d= \max(n,p)$.

Sometimes, the dimension $p$ is very large and can be in the order of tens of millions. The IPF method may be time-consuming  for computing all $\widehat{l}_{H_{ij}}$. Here, we propose to use an approximation tool to  prune  interaction terms in the second step.  For the homogeneous association regression model (\ref{loglinear1}), Kirkwood Superposition Approximation (KSA), which was firstly proposed by \cite{kirkwood1935statistical}, is  utilized to provide an estimator for $\mu_{ijk}$ in (\ref{loglinear1}).
That is,
$$\widehat{\mu}_{ijk}^{KSA}=\frac{n}{\eta}\frac{\widehat{\pi}_{ij+}\widehat{\pi}_{i+k}\widehat{\pi}_{+jk}}
{\widehat{\pi}_{i++}\widehat{\pi}_{+j+}\widehat{\pi}_{++k}}, $$
where $\eta=\sum_{ijk}\frac{\widehat{\pi}_{ij+}\widehat{\pi}_{i+k}\widehat{\pi}_{+jk}}
{\widehat{\pi}_{i++}\widehat{\pi}_{+j+}\widehat{\pi}_{++k}}$ is a normalization term, $n=\sum_{ijk}n_{ijk}$.
And then, we get the approximation  $\widehat{l}_{KSA}$ for $\widehat{l}_{H_{ij}}$. \cite{wan2010boost} shows that $\widehat{l}_{KSA}-\widehat{l}_S$ is an upper bound of $\widehat{l}_H-\widehat{l}_S$, i.e.,
$$0\leq\widehat{l}_H-\widehat{l}_S\leq \widehat{l}_{KSA}-\widehat{l}_S.$$
Based on this boundary and by setting up one threshold $\gamma_{KSA}$, in the second step, we can filter out many insignificant interaction terms quickly and then reduce the size of a  pool of all interaction effects. The value $\gamma_{KSA}$ can be defined by the conservative Bonferroni correction or specified by user. Obviously, if $\gamma_{KSA}=0$, no interaction term is deleted in this step. In the final step, for the remaining interaction terms, we compute their $\widetilde{L}_{ij,n}'$ by the IPF algorithm. Then select the $d$ largest $\widetilde{L}_{ij,n}'$, where $d= \max(n,p)$ or $\left\lfloor\frac{n}{\log n}\right\rfloor$, or  take  the thresholding value $\gamma_n$  to obtain the sure screening set $\widehat{\mathcal{N}}_{\gamma_n}$.  The term $\gamma_n$ can be taken as the Bonferroni correction $100*(1-0.05*p(p-1)/2)$\% percentile decided by the $\chi^2$ test with degree freedom $(l_i-1)(l_j-1)$ for any one interaction between $\widetilde{X}_i$ and $\widetilde{X}_j$.

In summary, our algorithm BOLT-SSI with KSA is summarized as follows:

$\emph {Step 1}.$ For any pairs of continuous variables $X_i$ and $X_j$, $1\leq i<j \leq p$, transform them to corresponding discretized variables $\widetilde{X}_i$ with level $l_i$ and $\widetilde{X}_j$ with level $l_j$, and change the response $Y$ to a categorical variable $\widetilde{Y}$ if necessary.

$\emph {Step 2}.$  By using the KSA to approximate $\widetilde{l}_{H_{ij}}$ of the IPF algorithm for all pairs of $X_i$ and $X_j$, we  compute $\widehat{l}_{KSA_{ij}}-\widehat{l}_{S_{ij}}$ and set up the threshold $\gamma_{KSA}$ to remove a part of interaction terms.

$\emph {Step 3}.$ For  the remaining interaction effects, we  compute $\widetilde{L}_{ij,n}'=\widehat{l}_{H_{ij}}-\widehat{l}_{S_{ij}}$ and further identify the important interaction effects by $\chi^2$-test with degree freedom $(l_i-1)(l_j-1)$, or directly select the $d$ largest $\widetilde{L}_{ij,n}'$.

So far, we have specified the procedures of our new algorithm ``BOLT-SSI''. Apparently, the new method ``BOLT-SSI'' will be much faster than the original method ``SSI''. Even though BOLT-SSI loses some statistical efficiency by discretizing predictor variables or response variable; its sure screening properties can still be guaranteed for moderate or large sample sizes. Moreover, compared to other screening methods, BOLT-SSI does not rely on hierarchy assumptions but screen significant two-way interactions for all pairs among the predictors.

\section{Sure Screening Properties of BOLT-SSI}
In this section, we derive the sure screening properties of BOLT-SSI by discussing SIS's relationship and discretization SIS. The details of sure screening properties of SSI can be seen in section 1 of the Appendix. And also we demonstrate the efficiency loss by discretization in the last part of this section.

\subsection{Properties of Discretization SIS}
First, without considering interaction effects we investigate the connection between the marginal likelihood and the marginal likelihood after discrization of the predictor variables and response variables, i.e.,  the connection between SIS and Discretized SIS.   As discussed in Section 3.1, after discretization we have such new increment of log-likelihood function
$$\widetilde{L}_k^\star=E\{l(\widetilde{\beta}_0^M,\widetilde{Y})-l(\widetilde{\X}_k^T\widetilde{\Beta}_k^M,\widetilde{Y})\},\ \ \ \ k=1,2,\ldots,p.$$
with $m=2$ and $l \ge 2$.

\indent We need some marginally symmetric conditions for further studies.  Those conditions are used to investigate sure screening properties of a rank robust SIS procedure by \cite{li2012robust}.\\
\indent (M1) Let ($Y_1,X_{1k}$), ($Y_2,X_{2k}$) be the independent copies of ($Y,X_{k}$).Denote $\Delta\varepsilon_k=Y_1-Y_2-\rho_k(X_{1k}-X_{2k})$ and $\Delta X_k=X_{1k}-X_{2k}$, where $\rho_k=corr(Y, X_k)$. The conditional distribution of $\Delta\varepsilon_k$ given $\Delta X_k$ is a symmetric finite mixture distribution, i.e., $f_{\Delta\varepsilon_k|\Delta X_k}(t)=\pi_{0k}f_0(t,\sigma_0^2|\Delta X_k)+(1-\pi_{0k})f_1(t,\sigma_1^2|\Delta X_k)$, where $f_0(t,\sigma_0^2|\Delta X_k)$ is symmetric unimodal probability distribution and $f_1(t,\sigma_1^2|\Delta X_k)$ is a symmetric probability distribution function and $\sigma_0^2$, $\sigma_1^2$ are conditional variances related to $\Delta X_k$, $k\in \mathcal{M}_\star$. Furthermore, there exists a given positive constant $\pi^\star\in (0,1]$ such that $\pi_{0k}\geq\pi^\star$ for any   $k\in \mathcal{M}_\star$.\\
\indent (M2) $c_{\mathcal{M}_\star}=\min_{k\in\mathcal{M}_\star}E|X_{k}|$ is a positive constant and is free of $p$.\\
\indent (M3) The predictors $\X_i=(X_{i1},\ \ldots,\ X_{ip})^T$ and the error term $\varepsilon_i$ are independent, $i=1,\ 2,\ \ldots,\ n$.

\begin{theorem}\label{discret1} Under the marginally symmetric condition (M1)-(M3) and the condition of Theorem 3 in Fan and Song (2010), i.e., for $k \in \mathcal{M}_\star$,
	$$|Cov(b'(\X^T\Beta^\star),\ X_k)|\geq C_1 n^{-\kappa}$$
	where $C_1$ is a positive constant and $\kappa<1/2$. After using 2-quantile and $l-$quantiles to discretize the response $Y$ and the predictor $X_k$, we have\\
	\indent (1) at least one $\widetilde{X}_{k_i}$ such that $$|Cov(\widetilde{Y},\ \widetilde{X}_{k_i})|\geq C_2 n^{-\kappa}$$
	for some positive constant $C_2$.\\
	\indent (2) Furthermore,
	$$\min_{k \in \mathcal{M}_\star}\widetilde{L}_k^\star\geq C_3 n^{-2\kappa}$$
	for some positive constant $C_3$ and $\widetilde{L}_k^\star$ is the corresponding increments of the log-likelihood after discretization.
\end{theorem}

Theorem \ref{discret1} ensures that if predictor variables in the original scale are associated with
the response, they are also related to each other after discretization.
Therefore, as our argument above,  by combining Boolean representation, logical operation, and discretization it could provide us a super-fast way to screen the predictor variables in high dimensional generalized linear models without losing much efficiency. This stimulates us to apply discretization to the interaction pursuit.  Based on the results above, we also get a similar connection between SSI and discretized SSI (BOLT-SSI) as the following.

\subsection{Properties of BOLT-SSI}

\indent Similar to above, we need the following some marginally symmetric conditions to investigate the screening properties of BOLT-SSI.

Let $\zeta_{ij}=Y-b'(\X_{i,j}^T\Beta_{i,j}^M)$, and  $(Y_1,X_{1i},X_{1j},X_{1ij},\zeta_{1ij})$, $(Y_2,X_{2i},X_{2j},X_{2ij},\zeta_{2ij})$ be the independent copies of $(Y,X_i,X_j,X_{ij},\zeta_{ij})$. We further centralize $\zeta_{ij}$ and
denote that $$\rho_{ij}=\frac{\mbox{Cov}(\zeta_{ij},X_{ij})}{\sqrt{\text{Var}(\zeta_{ij})\text{Var}(X_{ij})}}.$$
\indent (M1$'$) Denote $\Delta\varepsilon_{ij}=\zeta_{1ij}-\zeta_{2ij}-\rho_{ij}(X_{1ij}-X_{2ij})$ and $\Delta X_{ij}=X_{1ij}-X_{2ij}$, then the conditional distribution of $\Delta\varepsilon_{ij}$ given $\Delta X_{ij}$ is a symmetric finite mixture distribution, i.e., $f_{\Delta\varepsilon_{ij}|\Delta X_{ij}}(t)=\pi_{0ij}f_0(t,\sigma_0^2|\Delta X_{ij})+(1-\pi_{0ij})f_1(t,\sigma_1^2|\Delta X_{ij})$, where $f_0(t,\sigma_0^2|\Delta X_{ij})$ is symmetric unimodal probability distribution and $f_1(t,\sigma_1^2|\Delta X_{ij})$ is a symmetric probability distribution function and $\sigma_0^2$, $\sigma_1^2$ are conditional variances related to $\Delta X_{ij}$, $i, j\in \mathcal{N}_\star$. Furthermore, there exists a constant $\pi^\star\in (0,1]$ such that $\pi_{0ij}\geq\pi^\star$ for any   $i,j\in \mathcal{N}_\star$.\\
\indent (M2$'$) $c_{\mathcal{N}_\star}=\min_{i,j\in\mathcal{N}_\star}E|X_{ij}|$ is a positive constant and is free of $p$.\\
\indent (M3$'$) The predictors $\X=(X_{1},\ \ldots,\ X_{p})^T$ and the error term $\varepsilon$ are independent.

\begin{remark}  In fact, the marginally symmetric condition (M1)' is also easily satisfied. Denote that $\varepsilon_{ij}=\zeta_{ij}-\rho_{ij}X_{ij}$.
	A special case is that under the  linear model, the conditional distribution of $\varepsilon_{ij}$ given $X_{ij}$ does not depend on $X_{ij}$ and it has $K$ modes, where $K$ is finite. It implies that the conditional distribution $\varepsilon_{ij}|X_{ij}$ is the same as the distribution of $\varepsilon_{ij}$.
	Suppose that  $\varepsilon_{1ij}$, $\varepsilon_{2ij}$  follow a distribution $f_{\varepsilon}(t)$ with $K$ modes, that is, $f_{\varepsilon}(t)=\sum_{k=1}^K\pi_kf_{k}(t)$, where $\pi_k\geq0$ and $\sum_{k=1}^K\pi_k=1$.  Moreover, assume that $f_{lm}^\star(t)$, $1\leq l, m\leq K$, are the distributions of the difference $Z_l-Z_m$, where $Z_l$ and $Z_m$ are independent and follow the distributions $f_l(t)$ and $f_m(t)$, respectively.
	Therefore,  the distribution of $\Delta\varepsilon_{ij}=\varepsilon_{1ij}-\varepsilon_{2ij}$ can be expressed as
	\begin{eqnarray*}
		f_{\Delta\varepsilon}(t) &=& \sum_{l}\sum_{m}\pi_l\pi_mf_{lm}^\star(t)=\sum_{l}\pi_l^2f_{ll}^\star(t)
		+\sum_{l\neq m}\pi_l\pi_mf_{lm}^\star(t)  \\
		&=&\big(\sum_l\pi_l^2\big)\sum_{l}\frac{\pi_l^2}{\sum_l\pi_l^2} f_{ll}^\star(t)
		+(1-\sum_l\pi_l^2)\sum_{l\neq m}\frac{\pi_l\pi_m}{1-\sum_l\pi_l^2}f_{lm}^\star(t)\\
		&\triangleq& \pi_0^\star f_0^\star(t)+(1-\pi_0^\star)f_1^\star(t). \end{eqnarray*}
	Obviously,  $f_{ll}^\star(t)$ are symmetric unimodal distributions because of the unimodal distributions $f_l(t)$, and then $f_0^\star(t)$ is symmetric and unimodal. And  $f_1^\star(t)$ is a symmetric
	and multimodal density function. Moreover, $\pi_0^\star=\sum_l\pi_l^2\geq (\sum_l\pi_l^2)^2/K=1/K.$ \end{remark}

\begin{theorem}\label{discret2} Under the marginally symmetric conditions (M1$'$)$-$(M3$'$) and the condition: for $i, j \in \mathcal{N}_\star$ with
	$$|\text{Cov}_L(Y, X_{ij}|\X_{i,j}^T\Beta_{i,j}^M)|\geq c_1 n^{-\kappa}$$
	where $c_1$ is a positive constant and $\kappa<1/4$. After using 2-quantile, $l_1-$quantiles and $l_2-$quantiles to discretize the response $Y$ and the predictors $X_i$, $X_j$, we have\\
	\indent (1) at least one $\widetilde{X}_{st}^{ij}$ such that $$|\text{Cov}_L(\widetilde{Y},\ \widetilde{X}_{st}^{ij}|\widetilde{\X}_{i,j}^T\widetilde{\Beta}_{i,j}^M)|\geq c_{10} n^{-\kappa}$$
	for some positive constant $c_{10}$.\\
	\indent (2) Furthermore,
	$$\min_{i, j \in \mathcal{N}_\star}\widetilde{L}_{ij}^\star\geq c_{11} n^{-2\kappa}$$
	for some positive constant $c_{11}$ and $\widetilde{L}_{ij}^\star$ is the corresponding increments of the log-likelihood after discretization.
\end{theorem}

Theorem \ref{discret2}  claims that  important interaction terms are still significant after discretization.
Consequently,  similar to sure screening properties of SSI,  we can also show that the sure screening properties of BOLT-SSI, i.e.,  it can  detect significant interaction effects with large probability even when the dimension of the model is ultra-high.

\subsection{Discussion of Efficiency Loss by Discretization}

By Theorem \ref{discret1} and Theorem \ref{discret2}, and following steps in both {Theorem A.5 and A.6 in Appendix,} 
 the sure screening properties of Discretization SIS and BOLT-SIS can be guaranteed as the sample size $n$ tends to infinity. However, there is information loss by discretization, and the efficiency of the proposed screening procedure could be much reduced, especially when the arity $l,m=2$ or 3.

To  simplify  our analysis to obtain the intuition about such efficiency loss by discretization, we just compare the estimation efficiency of the Pearson correlation $\rho$ between the sample correlation estimate and the estimate by our discretization for the bivariate normal random vector
$$\left(\begin{array}{c}
X\\
Y
\end{array} \right) \sim N\left(\left(\begin{array}{c}
0\\
0
\end{array} \right), \left(\begin{array}{cc}
1&\rho\\
\rho&1
\end{array} \right)\right).$$
To discretize $X$ and $Y$, we consider the worst disretization with the largest information loss, i.e. $m=l=2$, and $\widetilde{X}= I(X>M_d(X))$ and $\widetilde{Y}=I(Y>M_d(Y))$. Then based on the proof of Theroem 4.1 in the Appendix, we have
$$\tilde{\rho}=\text{Corr}(\widetilde{X},\widetilde{Y})=4E[I(X_2>X_1)I(Y_2>Y_1)]-1=\tau=\frac{2}{\pi}\arcsin\rho,$$
where $\tau$, in fact, is  the kendall rank correlation  of the bivariate normal random vector $(X,Y)$.  It is well known that $\tau=\frac{2}{\pi}\arcsin\rho$ for the bivariate normal population, and hence if we have the estimate $\hat{\tau}$ of the kendall rank correlation, then the pearson correlation of the bivariate normal random vector can be estimated as $$\hat{\rho}_{\tau}=\sin \frac{\pi}{2} \hat{\tau}.$$

	Let $\hat{\rho}_s$ be the sample Pearson correlation of $X$ and $Y$, which is the optimal estimate of the Pearson correlation $\rho$.  \cite{hotelling1953new} has shown that the  asymptotic property of $\hat{\rho}_s$ under normal assumption should be,
$$\sqrt{n}(\hat{\rho}_s-\rho)\sim N\left(0,(1-\rho^2)^2\right),$$
which implies that $\sqrt{n}\hat{\rho}_s\sim N(0,1)$ when $X$ and $Y$ are independent.

Next let $\hat{\tau}$ be the sample correlation of $\widetilde{X}$ and $\widetilde{Y}$. As discussion above, in fact it is an estimate of the kendall rank correlation $\tau$. By the results of \cite{esscher1924method} and \cite{kendall1949rank} under the normal assumption, and based on the asymptotic normaility of U-statistics (\cite{lee2019u}),    the asymptotic distribution of the estimate  $\hat{\tau}$ is
$$\sqrt{n}(\hat{\tau}-\tau)\sim N\left(0,4\left[\frac{1}{9}-\left(\frac{2}{\pi}\arcsin \frac{\rho}{2}\right)^2\right]\right).$$ Then with Delta method and by simple calculation, the asymptotic normality of $\hat{\rho}_\tau$ should be
$$\sqrt{n}(\hat{\rho}_\tau-\rho)\sim N\left(0,4\left[\frac{1}{9}-\left(\frac{2}{\pi}\arcsin \frac{\rho}{2}\right)^2\right]*\frac{\pi^2}{4}(1-\rho^2)\right),$$  that is,
$\sqrt{n}\hat{\rho}_\tau\sim N(0,\pi^2/9)$ when $\rho=0$.

	Therefore, the relative efficiency of these two procedures is
$$\frac{\text{Var}(\hat{\rho}_\tau)}{\text{Var}(\hat{\rho}_s)}=4\left[\frac{1}{9}-\left(\frac{2}{\pi}\arcsin \frac{\rho}{2}\right)^2\right]*\frac{\pi^2}{4}\frac{1}{1-\rho^2}.$$
\begin{figure}[!htbp]
	\centering
	\includegraphics[width=4 in]{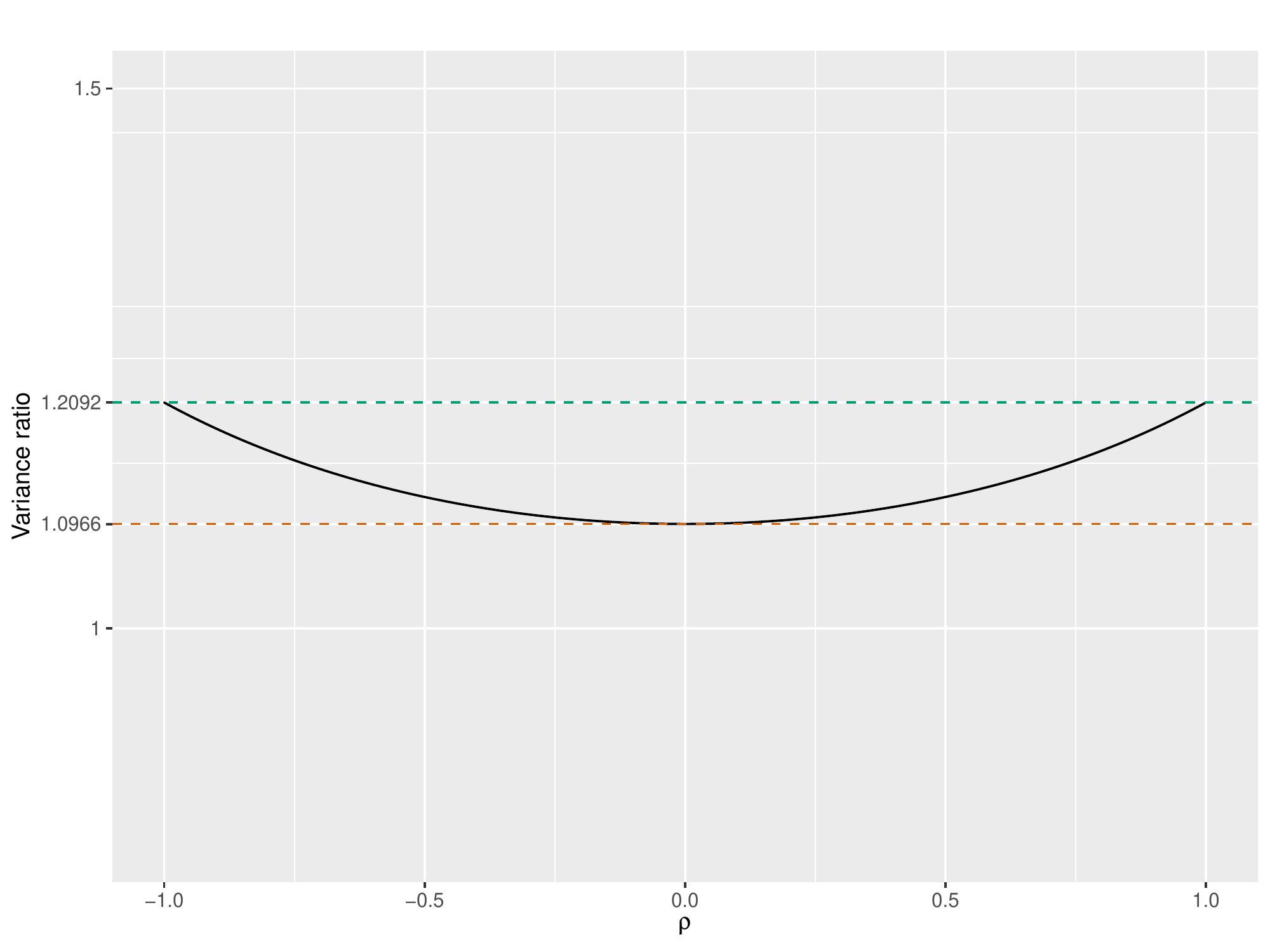}\\
	\caption{Relative efficiency of $\hat{\rho}_{\tau}$ and $\hat{\rho}_s$}\label{fig0}
\end{figure}

	As shown by the Figure \ref{fig0}, such relative efficiency is bounded between $\pi^2/9\approx1.0966$ at $\rho=0$ and $2\sqrt{3}\pi/9 \approx 1.2092$ at $\rho =1$ or $-1$.  It means we do not need much more samples to get the same accurate estimate of $\rho$ as our discretized estimate $\hat{\rho}_\tau$ compared to the sample Pearson correlation estimate $\hat{\rho}_s$ which is the optimal estimate of $\rho$ in some sense.

Though the above discussion is based on the assumption that $(X, Y)$ follows bivariate normal population,  if $(X, Y)$ follows other bivariate distribution, by monotonic transformation, we could transfer $(X, Y)$ to one of bivariate normal random vectors. Usually, under general conditions,  such a monotonic transformation would not change the Pearson correlation between $X$ and $Y$ much under general conditions. Furthermore, the discretized estimate $\hat{\rho}_\tau$ is invariant.  Hence in some sense, as the sample size of data is relatively large,  $\hat{\rho}_\tau$ can be used to screen the relationship between $X$ and $Y$ without losing much efficiency.

The above discussion is based on the worst discretization that the arity $m=l=2$. In such worst case, it has been shown that the statistical  efficiency loss is relatively small, but as shown by our numerical studies, the computational complexity is reduced dramatically.  Hence the discretization approach is an appropriate way to balance the  trade-off between statistical efficiency and computational complexity. The statistical efficiency loss by discretization can be tolerated as long as the sample size of the data is relatively large.

\section{Numerical Studies}
In this section, we  investigate the performance of the proposed SSI and BOLT-SSI by numerical studies. The methods, hierNet (\cite{bien2013lasso}), IP (\cite{fan2016interaction}) RAMP (\cite{hao2018model}) and xyz (\cite{thanei2018xyz})   are employed in  comparisons with respect to the performance on the estimation and prediction.

We consider the linear model (\ref{linear})
\begin{equation}\label{linear}
y=\sum_{i=1}^pX_i\beta_i+\sum_{j<k }X_jX_k\beta_{jk}+\epsilon
\end{equation}
and logistic model (\ref{logistic})
\begin{equation}\label{logistic}
\log(\frac{\pi}{1-\pi})=\sum_{i=1}^pX_i\beta_i+\sum_{j<k }X_jX_k\beta_{jk}.
\end{equation}

 We generate the covariates  $\{x_i\}_{i=1}^n\sim N(0,\Sigma)$ with $\Sigma_{jk}=\rho^{|j-k|}$, where $\rho$ varies in $[0, 0.5]$, and then generate the response $y$ by the linear model (\ref{linear}) and logistic model (\ref{logistic}). For all settings, the set of the important main effects is $S=\{1,2,\ldots,10\}$ with the true coefficients $\beta_S=(1,1,1,1,1,1,1,1,1,1)^T.$  For the linear model,  the error term  $\epsilon\sim N(0,\sigma^2)$ with $\sigma\in\{2,3,4\}$ for different signal-to-noise ratio (SNR) situations. For the logistic model, we change the values of the coefficients of interactions, and let significant interaction effect coefficient $\beta_{ij}=1,2,3$ to obtain the different SNR.  We consider different heredity structures including strong heredity, weak heredity and anti heredity by the following interaction effect settings for linear regression model or logistic model.

  \begin{itemize}
	\item {\bf Example 1} - Linear Model with Strong Heredity.
	The set of 10 important interaction effects is defined as $$T=\{(1,2),(1,3),(2,3),(2,5),(3,4),(6,8),(6,10),(7,8),(7,9),(9,10)\}$$ with corresponding coefficients (2,2,2,2,2,2,2,2,2,2). \\
	
	\item {\bf Example 2} - Linear Model with Weak Heredity.
	The set of 10 important interaction effects is defined as {$$T=\{(1,2),(1,13),(2,3),(2,15),(3,4),(6,10),(6,18),(7,9),(7,18),(10,19)\}$$}
	with corresponding coefficients (2,2,2,2,2,2,2,2,2,2). Here,  for every significant interact effect, only one of its main effects is significant. \\
	
	\item {\bf Example 3} - Linear Model with Anti Heredity.
	The set of 10 important interaction effects is
	{ $$T=\{(11,12),(11,13),(12,13),(12,15),(13,14),(16,18),(16,20),(17,18),(17,19),(19,20)\}$$}
	with corresponding coefficients (2,2,2,2,2,2,2,2,2,2). Here, none of the main effects that have significant interaction effects is included in the linear model (\ref{linear}). \\

	\item {\bf Example 4} - Linear Model with Mixed Heredity.
	Suppose that the set of 10 important interaction effects is
	{$$T=\{(1,2),(1,3),(2,3),(2,15),(6,18),(7,18),(16,20),(17,18),(17,19),(19,20)\}$$} with corresponding coefficients (2,2,2,2,2,2,2,2,2,2). Here, the first three interactions satisfy the strong heredity, next three satisfy  weak heredity assumption, and the last fourth significant interact effects do not have their corresponding main effects in the model. \\

	\item {\bf Example 5} - Logistic  Model with Strong Heredity.
	Consider the set of 10 important interaction effects is
	{$$T=\{(1,2),(1,3),(2,3),(2,5),(3,4),(6,8),(6,10),(7,8),(7,9),(9,10)\}.$$} \\
	
	\item {\bf Example 6} - Logistic  Model with Weak Heredity.
	Denote that the set of 10 important interaction effects is {$$T=\{(1,2),(1,13),(2,3),(2,15),(3,4),(6,10),(6,18),(7,9),(7,18),(10,19)\}.$$}
	
	\item {\bf Example 7} - Logistic  Model with Anti Heredity.
	Assume that the set of 10 important interaction effects is
	{ $$T=\{(11,12),(11,13),(12,13),(12,15),(13,14),(16,18),(16,20),(17,18),(17,19),(19,20)\}.$$}

	\item {\bf Example 8} - Logistic  Model with Mixed Heredity.
	Suppose that the set of 10 important interaction effects is
	{$$T=\{(1,2),(1,3),(2,3),(2,15),(6,18),(7,18),(16,20),(17,18),(17,19),(19,20)\}.$$}
	
\end{itemize}

 We investigate the screening performance and post-screening performance of those interaction effect screening and variable selection methods under different examples.

Let $T$ with cardinality $t=|T|$ denote the significant interaction effects in the model, i.e., $T=\{(j,k):\beta_{j,k}\neq 0\}$. For each scenario, we run $M=100$ Monte-Carlo simulations for each method. For the $m$-th simulation, denote that the estimated interaction subsets as $\widehat{T}_m$. We evaluate the performance on variable selection and model prediction based on the following criteria:
\begin{itemize}
	\item {The average coverage rate (ACR): the percentage of all true interactions included in the selected models.}
	\item Average model size (AMS):  $M^{-1}\sum_{m=1}^M MS_m$,
	where  $MS_m$ is the model size of interaction effect predictors selected by the screening methods or post-model selection method in the $m$-th simulation.
	\item The average  out-of-sample $R^2$ for linear regression model:
	$$R^2=100\% \times \left\{1-\frac{\sum(Y_i^*-\X_i^{*T}\hat{\Beta})^2}{\sum(Y_i^{*}-\bar{Y}^{*})^2}\right\},$$
	where $(\X_i^*,Y_i^*)$ is the testing data and $\hat{\Beta}$ is the estimate of the coefficient based on the training data.
	\item Predictive misclassification rate (PMR) for logistic model:
	$$PMR=I(Y_i^*\neq\hat{Y}),$$
	where $Y_i^*$ is  the true value of the testing data and $\hat{Y}$ is the predictive value of testing data based on the training model.
\end{itemize}

\subsection{Screening Performance}
For the screening procedures, we consider SSI, BOLT-SSI, and the employed  methods, IP and xyz for the linear model and logistic model.  For the method xyz, we choose top 500 interaction terms screened by it (Actually, 500 is the largest number of interactions that the package ``xyz'' can be selected by screening), and let the projection time $L$ of ``xyz" be $10, 100, 1000,$ respectively. For the method IP, we choose the top $n-1$ screened interaction effect predictor variables as the active set. For our method SSI, similarly the top $n-1$ interaction effect terms are selected into the active set. For BOLT-SSI, we consider two cases: keeping the top $n-1$, or the top $\max\{n,p\}$ significant interaction predictors as the screening selected active set.  Since the methods IP and xyz are not available for the logistic model, we only investigate the screening properties of SSI and BOLT-SSI for Example 5-8.

\begin{table}[!htbp]
	\caption{ {\footnotesize Screening results for Linear Models when $p=5000$}}\label{simut2}
	\centering
	\smallskip{\scriptsize \tabcolsep= 3 pt
		
		\begin{tabular}{ccccccccc}
			\hline\hline
			Methods &      $\sigma$ &     SSI &    BOLT-SSI &     BOLT-SSI(p) &    IP &     xyz-L10 &     xyz-L100&    xyz-L1000  \\
			\hline \hline
			\multicolumn{ 9}{c}{($n,p,\rho$)=(500, 5000, 0)} \\
			\hline
			& 2 & 0.98 & 0.03 & 0.64 & 0.73 & 0.00 & 0.01 & 0.76 \\
			Example 1 & 3 & 0.94 & 0.00 & 0.60 & 0.70 & 0.00 & 0.04 & 0.73 \\
			& 4 & 0.80 & 0.00 & 0.48 & 0.59 & 0.00 & 0.01 & 0.55 \\
			\hline
			\multicolumn{9}{c}{($n,p,\rho$)=(500, 5000, 0.5)} \\
			\hline
			& 2 & 1.00 & 0.80 & 0.98 & 0.99 & 0.29 & 0.52 & 0.52 \\
			Example 1  & 3 & 1.00 & 0.58 & 0.94 & 0.99 & 0.22 & 0.51 & 0.52 \\
			& 4 & 1.00 & 0.43 & 0.88 & 0.98 & 0.14 & 0.50 & 0.50 \\
			\hline \hline
			\multicolumn{ 9}{c}{($n,p,\rho$)=(500, 5000, 0)} \\
			\hline
			& 2 & 0.90 & 0.01 & 0.38 & 0.03 & 0.00 & 0.04 & 0.56 \\
			Example 2 & 3 & 0.82 & 0.01 & 0.36 & 0.01 & 0.00 & 0.00 & 0.41 \\
			& 4 & 0.73 & 0.00 & 0.00 & 0.01 & 0.00 & 0.01 & 0.31 \\
			\hline
			\multicolumn{9}{c}{($n,p,\rho$)=(500, 5000, 0.5)} \\
			\hline
			& 2 & 0.73 & 0.03 & 0.60 & 0.00 & 0.00 & 0.00 & 0.00 \\
			Example 2  & 3 & 0.71 & 0.02 & 0.57 & 0.01 & 0.00 & 0.00 & 0.00 \\
			& 4 & 0.67 & 0.00 & 0.45 & 0.00 & 0.00 & 0.00 & 0.00 \\
			\hline \hline
			\multicolumn{ 9}{c}{($n,p,\rho$)=(500, 5000, 0)} \\
			\hline
			& 2 & 0.89 & 0.03 & 0.62 & 0.03 & 0.00 & 0.02 & 0.56 \\
			Example 3 & 3 & 0.82 & 0.03 & 0.44 & 0.02 & 0.00 & 0.01 & 0.53 \\
			& 4 & 0.73 & 0.00 & 0.45 & 0.01 & 0.00 & 0.00 & 0.46 \\
			\hline
			\multicolumn{9}{c}{($n,p,\rho$)=(500, 5000, 0.5)} \\
			\hline
			& 2 & 1.00 & 0.33 & 0.81 & 0.74 & 0.28 & 0.53 & 0.53 \\
			Example 3  & 3 & 1.00 & 0.23 & 0.74 & 0.72 & 0.25 & 0.50 & 0.50 \\
			& 4 & 1.00 & 0.11 & 0.73 & 0.68 & 0.14 & 0.51 & 0.51 \\
			\hline \hline
			\multicolumn{ 9}{c}{($n,p,\rho$)=(500, 5000, 0)} \\
			\hline
			& 2 & 0.91 & 0.00 & 0.44 & 0.06 & 0.00 & 0.03 & 0.47 \\
			Example 4 & 3 & 0.82 & 0.00 & 0.42 & 0.05 & 0.00 & 0.03 & 0.48 \\
			& 4 & 0.69 & 0.00 & 0.23 & 0.03 & 0.00 & 0.00 & 0.34 \\
			\hline
			\multicolumn{9}{c}{($n,p,\rho$)=(500, 5000, 0.5)} \\
			\hline
			& 2 & 0.80 & 0.07 & 0.75 & 0.27 & 0.00 & 0.01 & 0.01 \\
			Example 4  & 3 & 0.78 & 0.05 & 0.73 & 0.28 & 0.00 & 0.01 & 0.01 \\
			& 4 & 0.76 & 0.02 & 0.66 & 0.28 & 0.00 & 0.01 & 0.01 \\
			\hline \hline

		\end{tabular}

	}
\end{table}

\begin{table}[!htbp]
	\caption{ {\footnotesize Screening results for Logistic Models with $n=400$ and $p=2000$ }}\label{simut3}
	\centering
	\smallskip{\scriptsize \tabcolsep= 3 pt
		
		\begin{tabular}{ccccc|cccc}
			\hline\hline
			Methods &      $\beta_{jk}$ &     SSI &    BOLT-SSI &     BOLT-SSI(p) &  &   SSI &    BOLT-SSI &     BOLT-SSI(p)  \\
			\hline
			& & \multicolumn{3}{c}{$\rho=0$} & & \multicolumn{3}{c}{$\rho=0.5$}\\
			\hline
			& 1 & 0.02 & 0.00 & 0.35 & & 0.53 & 0.08 & 0.76  \\
			Example 5 & 2 & 0.40 & 0.04 & 0.56  && 0.84 & 0.30 & 0.86  \\
			& 3 & 0.77 & 0.12 & 0.66 & &0.83 & 0.27 & 0.86  \\
			\hline
			& 1 & 0.02 & 0.00 & 0.28 & & 0.00 & 0.00 & 0.39  \\
			Example 6 & 2 & 0.31 & 0.02 & 0.34 && 0.32 & 0.01 & 0.49  \\
			& 3 & 0.56 & 0.06 & 0.63 & & 0.44 & 0.05 & 0.66\\
			\hline
			& 1 & 0.02 & 0.00 & 0.35 & &   0.53 & 0.08 & 0.76 \\
			Example 7 & 2 & 0.40 & 0.04 & 0.56 & & 0.84 & 0.30 & 0.86\\
			& 3 & 0.77 & 0.12 & 0.66 & & 0.83 & 0.27 & 0.86 \\
			\hline
			& 1 & 0.00 & 0.00 & 0.28 & & 0.04 & 0.00 & 0.43\\
			Example 8 & 2 & 0.33 & 0.05 & 0.57 & & 0.24 & 0.04 & 0.63 \\
			& 3 & 0.52 & 0.05 & 0.70 & & 0.41 & 0.13 & 0.68\\
			\hline \hline
		\end{tabular}

	}
\end{table}

From the results shown by Tables 1-2, the coverage rate will decrease when the signal-noise ratio is relatively small. The proposed SSI has a high coverage percentage in screening interaction effects for different heredity structures. For the methods xyz and IP, they have a lower converge percentage except for the  strong heredity setting compared to SSI. For the proposed BOLT-SSI, though its performance is not better than SSI, its coverage rate is better than the other two methods when the top $p$ significant interaction effects are considered as the screening active set.
By discretization, the data would lose some information, and hence BOLT-SSI would not be as efficient as SSI even though its speed is much faster than SSI. Hence it would increase much probability to keep the true active interaction effect predictors in the screened model by keeping the $p$ top significant interaction effect predictors in the active set after screening.  All in all, the screening performances of SSI and BOLT-SSI($p$) are more stable than the performance of other methods.

\subsection{Post-Screening Performance}
In this subsection, we compare the final model selection and prediction of existing methods (RAMP, xyz, hierNet) with the Lasso after screening by our proposed SSI and BOLT-SSI. For the method RAMP, the tuning  parameter is selected by using EBIC with $\gamma=1$ since the EBIC tends to work the best among most of the settings as shown by \cite{hao2018model}. For the method xyz, we consider the projection time $L$ as 100, 500 and use 5-fold cross-validation (CV) to select the tuning parameter for the post-screening selection.  For our methods SSI and BOLT-SSI, we use 5-folds CV and LASSO to further refine the model selection after screening. All of the simulation settings are the same as the Example 1-8 above. Especially, We set $\rho = 0.5$ for all the studies. To compare the prediction,  for every simulation, we let $n_1=0.75*n$ of the data set as the training data and the remaining data is considered as the testing data. Note that firstly we let $p$  be relatively small so that it is possible to compare the performance of hierNet(\cite{bien2013lasso}) {in Tables 1-2 of Appendix}, where ``w'' stands for weak heredity.

\begin{figure}[!htbp]
	\centering
	\includegraphics[width=4.5 in]{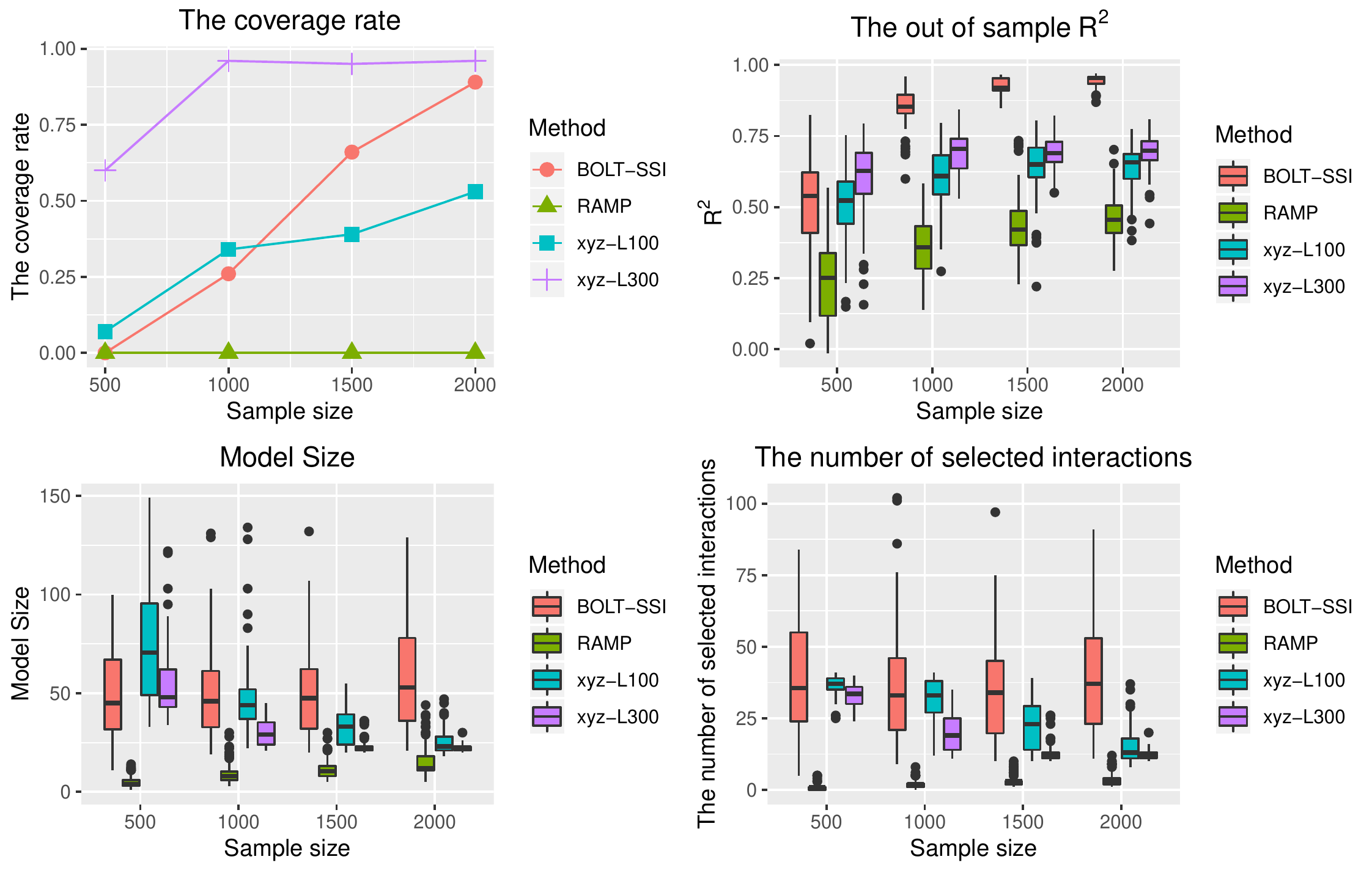}\\
	\caption{Performance of different methods with $p=5000$ and different $n$'s for linear models}\label{fig1}
\end{figure}

\begin{figure}[!htbp]
	\centering
	\includegraphics[width=4.5 in]{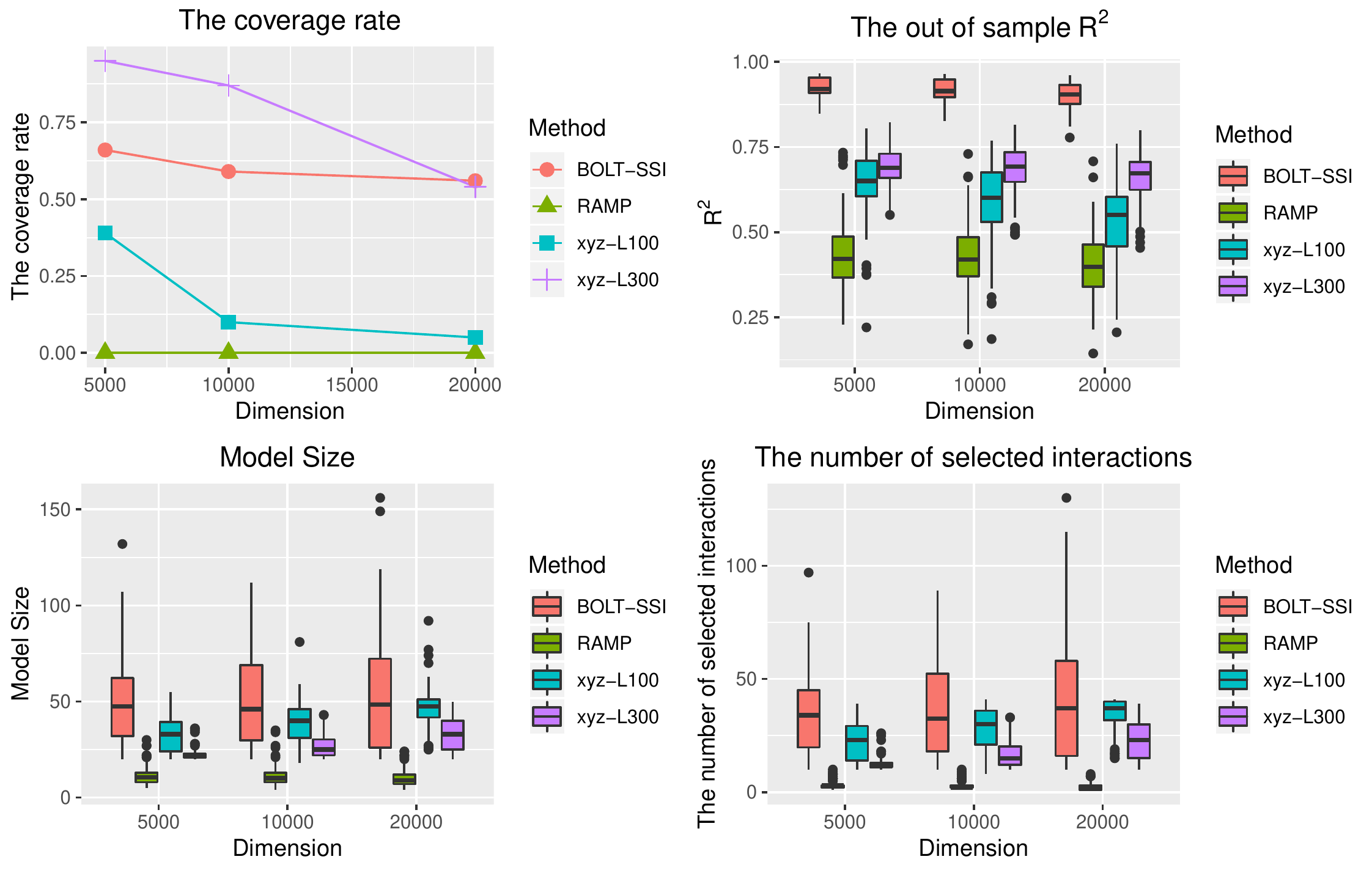}\\
	\caption{Performance of different methods with $n=1500$ and different $p$'s for linear models}\label{fig2}
\end{figure}

\begin{figure}[!htbp]
	\centering
	\includegraphics[width=4.5 in]{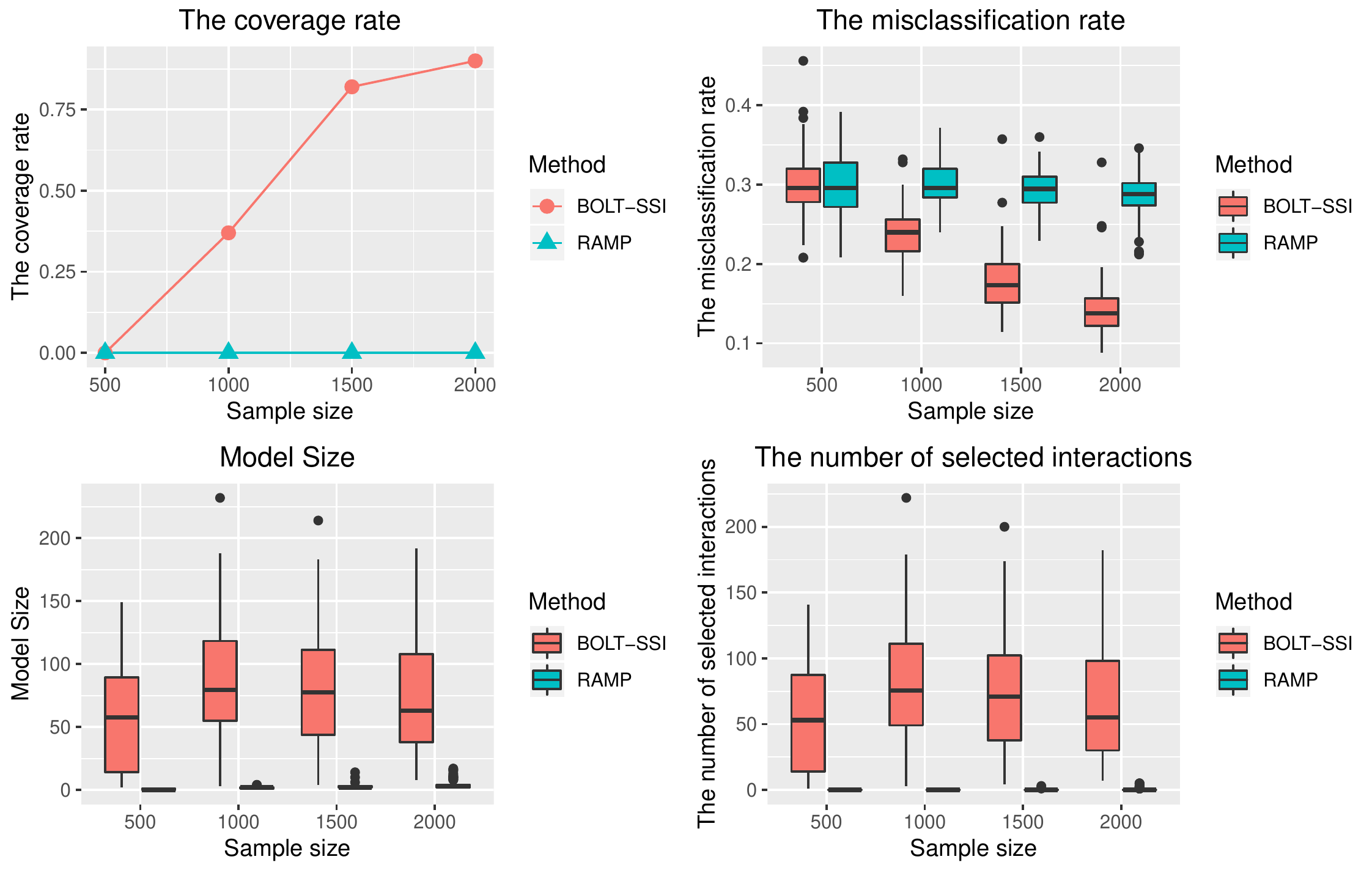}\\
	\caption{Performance of different methods with $p=5000$ and different $n$'s for logistic models}\label{fig3}
\end{figure}

\begin{figure}[!htbp]
	\centering
	\includegraphics[width=4.8 in]{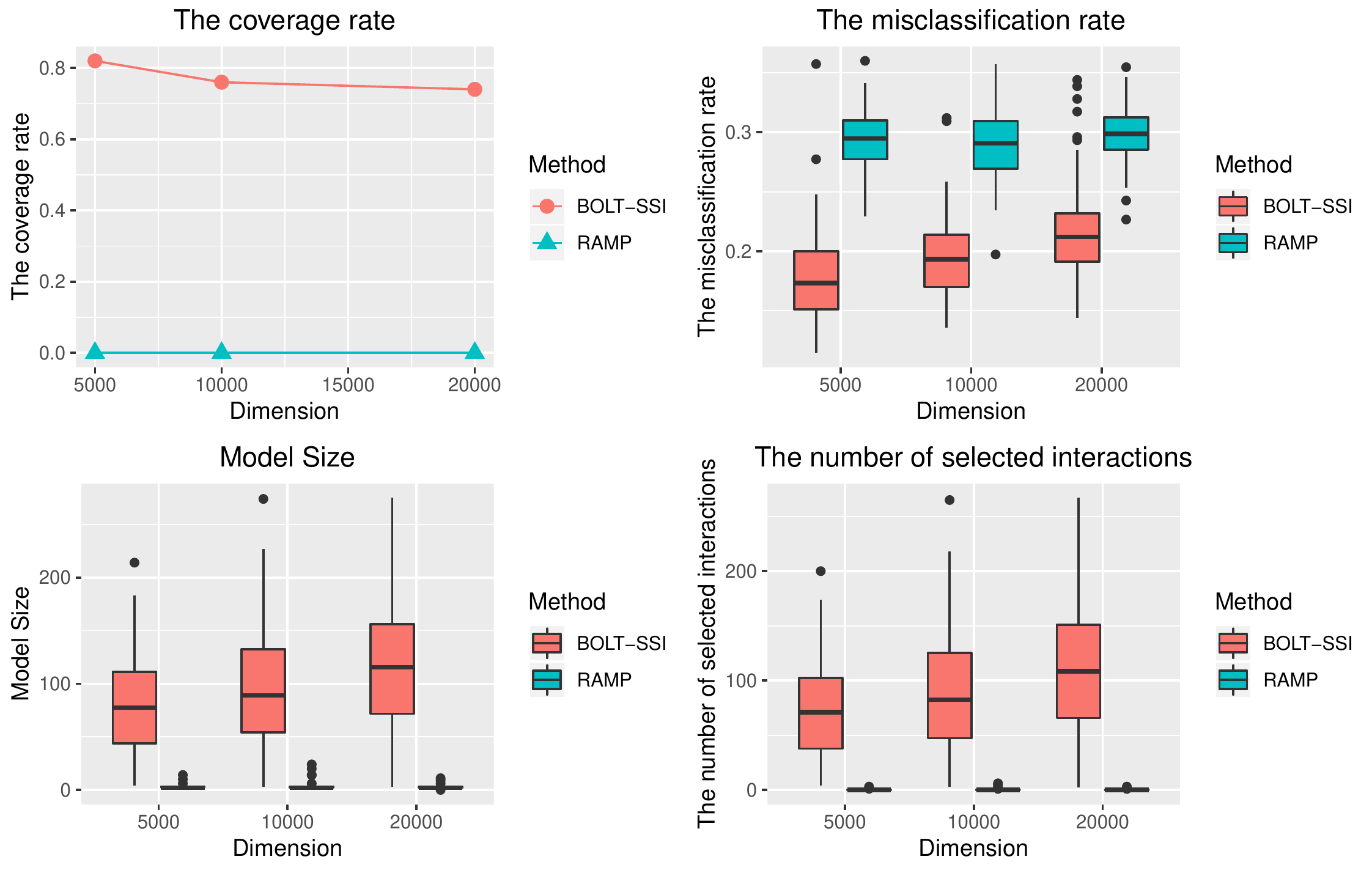}\\
	\caption{Performance of different methods with $n=1500$ and different $p$'s for logistic models}\label{fig4}
\end{figure}

 Note that the computation time for hierNet-s is very large for a single replicate. As a result, we  omit the comparisons with hierNet for the other higher dimensional examples. In the high dimensional settings, we  consider {$(n,p)$=(500,5000),(1000,5000),(1500,5000),\\(2000,5000),(1500,10000),(1500,20000)}
and compare  the performance of BOLT-SSI, RAMP, and xyz. Other methods are very time-consuming, and are not  considered in this setting. {Especially, we set $\sigma=0.5$ for linear models, and $\beta_{ij}=3$ for logistic models.  All results of different methods with $(n,p)=(1000,5000)$ are summarized in Table \ref{simupost9}. It is shown that our method still has a good performance in the high dimensional feature space. Furthermore, we also take Examples 5 and 8 to illustrate the patterns of our method. The results are shown in Figures \ref{fig1}-\ref{fig4}. Obviously, as sample size $n$ increases, the performances of all methods become better, as shown in Figure \ref{fig1} and Figure \ref{fig3}, and our method has the best performance. In Figures \ref{fig2} and \ref{fig4}, though the performance of our method degrades as the dimension $p$ increases, its performance is still much better than others.} The method RAMP is influenced by the heredity assumption, especially if the anti-heredity exists,  the result of RAMP is worst.

\begin{table}[!htbp]
	\caption{ \footnotesize{ Selection and prediction results (standard error) with $(n,p)=(1000,5000)$. The standard errors are in parentheses.}}\label{simupost9}
	\centering
	\smallskip{ \scriptsize \tabcolsep= 2.5 pt
		\begin{tabular}{cccccccccccccccc}
			\hline
			Assumption &    Methods &       {ACR}  &        AMS &          $R^2$  & PMR\\
			\hline
			&    BOLT-SSI &   0.98 & 53.91(2.5) & 94.52(0.22) & --- &\\
			
			Example 1 &       RAMP &    0.16 & 21.67(0.7) & 76.29(1.60)  &--- &\\
			
			&   xyz-L100 &    0.73 & 28.10(0.7) & 58.46(0.95) &--- &\\
			& xyz-L500 &        1 & 23.94(0.2) & 60.07(0.82) & --- &\\
			
			\hline
			&    BOLT-SSI &     0.62 & 45.80(2.3) & 87.16(0.62) & ---  & \\
			
			Example 2 &       RAMP &      1.00 & 20.35(0.1) & 95.34(0.01) & ---  & \\
			
			&   xyz-L100 &        0.23 & 72.70(2.9) & 58.5(1.16) & --- &\\
			& xyz-L500 &          0.97 & 35.64(0.5) & 76.43 (0.56)& --- & \\
			
			\hline
			&    BOLT-SSI &        0.93 & 47.61(1.8) & 90.94(0.33)& --- & \\
			
			Example 3 &       RAMP &          0.00 &   4.5(0.6) & 13.96(0.11)& --- & \\
			
			&   xyz-L100 &          0.80 & 27.85(7.1) & 58.48(1.31)& --- & \\
			&xyz-L500 &             1 & 23.94(0.2) & 59.36(1.20) &---  &\\
			
			\hline
			&    BOLT-SSI &        0.53 & 49.38(1.9) & 88.53(0.50)&---  & \\
			
			Example 4 &       RAMP &        0.00 & 15.54(0.6) & 61.83(0.79) &---  &\\
			
			&   xyz-L100 &       0.34 & 47.26(1.8) & 59.53(1.05)& --- & \\
			& xyz-L500 &           1 & 28.47(0.5) & 68.44(0.89) &---  &\\
			
			\hline
			\hline
			Example 5 &    BOLT-SSI &     0.53 & 36.09(4.0) &- &23.26(0.32) \\
			
			&       RAMP &       0.00 &  0.14(0.1) &- &25.62(0.03) \\
			\hline
			Example 6 &    BOLT-SSI &    0.42 & 47.75(4.9) & - &26.73(0.62) \\
			
			&       RAMP &        0.00 &  6.80(0.5)& -& 28.15(0.60) \\
			\hline
			Example 7 &    BOLT-SSI &      0.62 & 79.80(5.0) &- &20.98(0.31) \\
			
			&       RAMP &       0.00 &  2.97(0.2) &- &28.67(0.31) \\
			\hline
			Example 8 &    BOLT-SSI &        0.53 & 79.26(5.1) & - &22.85(0.41) \\
			
			&       RAMP &        0.00 &  1.69(0.1) &- &25.34(0.24) \\
			\hline
		\end{tabular}
	}
\end{table}

\subsection{Efficiency comparison}
\begin{figure}[!htbp]
	\centering
	\includegraphics[width=4 in]{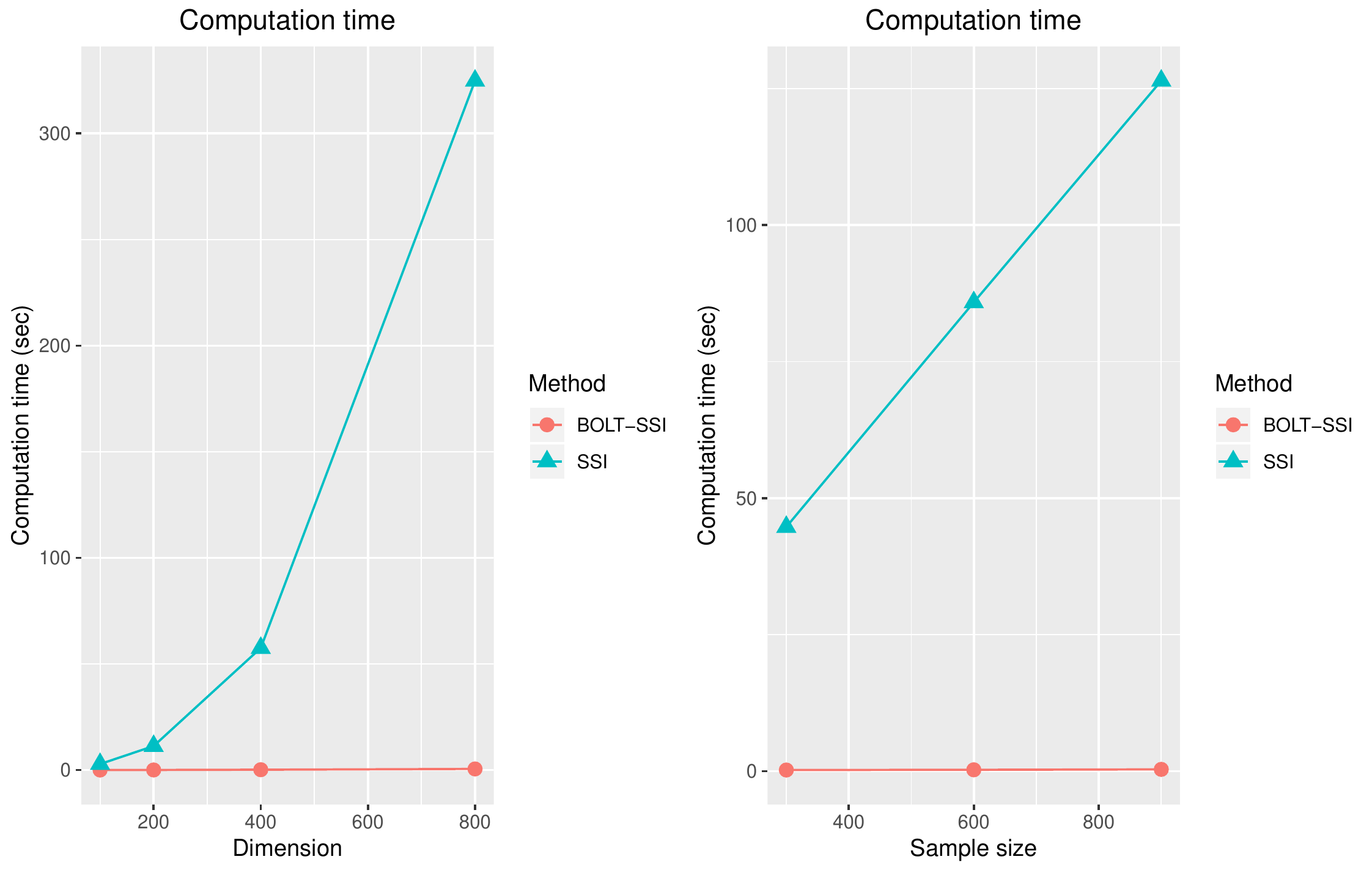}\\
	\caption{ Average computation time of screening procedure for logistic models}\label{fig5}
\end{figure}

\begin{figure}
	\centering
	\includegraphics[width=4.8 in ]{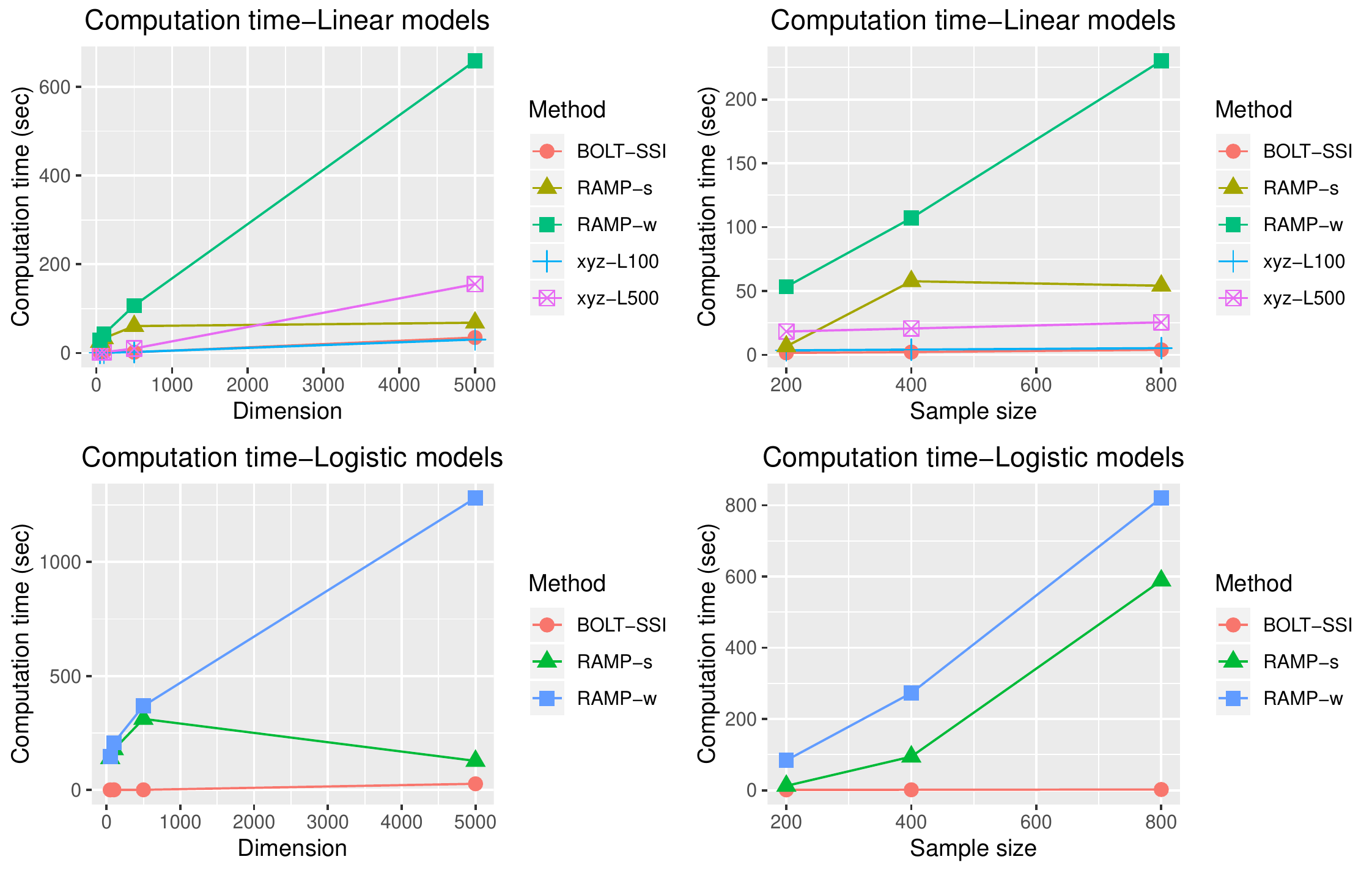}\\
	\caption{Average computation time of post screening procedure }\label{fig6}
\end{figure}

Here, we use Example 1 and Example 5 to study the efficiency of all the above methods. The machine we used equips Intel (R) Xenon(R) CPU E5-1603 v4 @ 2.80GHZ with  8.00 GB RAM. We compare the average computation time of variable selection among the following methods: SSI, BOLT-SSI, xyz, RAMP-s, RAMP-w, hierNet-s, hierNet-w,  based on the 50 simulated data sets by the screening procedure and post-screening procedure, where ``w'' and ``s'' stand for weak heredity and strong heredity respectively. To make fair comparisons, we do not consider the selection of tuning parameters in modeling. Figures \ref{fig5}-\ref{fig6} and Table \ref{efficieny3} summarize the average computation time (seconds per run) for each procedure. Since the differences of computation time are relative small for various $\sigma$ and $\rho$, we only present the results when $\sigma = 2$, $\beta_{jk}=2$ and $\rho=0.5$. It is clear that the method hierNet spends much time on the computation no matter under the strong or weak heredity assumption and the method RAMP with weak heredity is also very slow.  BOLT-SSI is consistently fast and its screening the algorithm does not rely on the heredity assumption of the data structure.

\begin{table}[!htbp]
	\caption{ \footnotesize{ Average computation time of post screening procedure for linear models}}\label{efficieny3}
	\centering
	\smallskip{  \footnotesize \tabcolsep= 2.5 pt
		\begin{tabular}{ccccccccc}
			\hline\hline
			$n$ &          $p$ &   BOLT-SSI &  hierNet-s &  hierNet-w &   xyz-L100 &   xyz-L500 &     RAMP-s &     RAMP-w \\
			\hline
			\multicolumn{9}{c}{Linear Regression Models} \\
			\hline
			
			500 &         50 &       1.13 &      75.26 &       4.92 &       0.22 &       0.86 &      25.00 &      28.85 \\
			
			500 &        100 &       2.55 &     321.88 &      22.43 &       0.39 &       1.61 &      33.11 &      42.44 \\
			
			500 &        500 &       1.66 &      ---   &     669.99 &       2.10 &      10.07 &      60.65 &     106.82 \\
			
			500 &       5000 &      34.75 &      ---   &     ---    &      30.38 &     155.22 &      68.20 &     658.42 \\ \hline
			200 &       1000 &       1.62 &       ---      &      ---      &       3.58 &      18.35 &       6.69 &      53.35 \\
			
			400 &       1000 &       2.26 &       ---     &      ---      &       4.15 &      20.69 &      57.68 &     107.11 \\
			
			800 &       1000 &       4.02 &       ---     &      ---      &       5.32 &      25.52 &      54.18 &     230.20 \\
			\hline\hline
			\multicolumn{9}{c}{Logistic Regression Models}
			\\
			\hline
			500 &         50 &       0.44 &     306.91 &      11.53 &     ---       &  ---        &     139.52 &     147.16 \\
			
			500 &        100 &       0.82 &    1105.96 &      37.16 &     ---       &   ---       &     177.84 &     207.08 \\
			
			500 &        500 &       0.74 &     ---       &     511.21 &     ---       &   ---         &     311.87 &     368.86 \\
			
			500 &       5000 &      27.15 &     ---       &       ---     &     ---       &   ---         &     127.52 &    1281.45 \\ \hline
			200 &       1000 &       1.10 &      ---      &    ---        &    ---        &   ---         &      12.34 &      83.98 \\
			
			400 &       1000 &       1.38 &      ---      &    ---        &   ---         &  ---          &      94.48 &     273.06 \\
			
			800 &       1000 &       2.18 &      ---      &    ---        &   ---         &  ---          &     588.62 &     820.87 \\
			\hline \hline
		\end{tabular}
		
	}
\end{table}

In summary, compared to the other methods,  our proposed SSI and BOLT-SSI($p$) have a stably high coverage rate in terms of the screening performance. When the dimension of data $p$ is not too large, by fine coding, SSI can also finish the screening task in a limited time. After discretization, some data information would be lost, and hence BOLT-SSI can not use all of  the information for screening, and hence it is not as efficient as SSI.  However,  it is much faster than SSI and most of the other screening methods, and can finish screening for huge dimensional data in a relatively small time period. In fact, from our numerical studies, it is shown that BOLT-SSI makes a good trade-off between the computation complexity and the efficiency of screening. Consequently, SSI and BOLT-SSI have absolute competitiveness compared to other interaction screening and variable selection methods.

\section{Real Data}
\subsection{Residential Building Data}
The residential building dataset is available at {{\url{https://archive.ics.uci.edu/ml/datasets/Residential+Building+Data+Set}}}, which contains 8 project physical and financial variables, 19 economic variables and indices in 5-time lag numbers, and two output variables that are construction costs and sale prices. Totally, there are 103 predictors and 372 observations. The total number of interaction terms is  $5.253\times 10^3$. The data set was collected from Tehran, Iran between 1993 and 2008, which is  a city with a metro population of around 8.2 million and much building construction activity. Usually, predicting the price of housing is of paramount importance for economic forecasting in any country. Therefore, our purpose is to use this data set to predict the sale prices. For convenience, the response and all predictors are standardized to have a unit variance before the analysis.

For the method ``LASSO'', we only consider the main effects. For the method ``xyz'', three different projection times $(L=10, 100, 500)$ are studied in this data set. Five-fold cross-validation is used to tune parameter in the method ``xyz'', ``SSI'' and ``BOLT-SSI''. The methods ``RAMP-s'' and ``RAMP-w'' represent the model with strong heredity and weak heredity assumption, respectively. The rule ``EBIC'' is used to select the final model in the method ``RAMP''. For all methods, we randomly select 300 observations as the training set, and use the remaining 72 samples to form the testing data to compute the out-of-sample $R^2$ for the final model. The experiment is repeated 100 times. Time(s) is the average computation time of 100 experiments including variable selection and prediction. The machine still equips Intel (R) Xenon(R) CPU E5-1603 v4 @ 2.80GHZ with  8.00 GB RAM.
The results are listed in the Table \ref{real1}.

\begin{table}[!htbp]\caption{Average results and the standard errors (in parentheses) on the residential building data set }\label{real1}
	\centering
	\smallskip{\footnotesize \tabcolsep= 15 pt
		\begin{tabular}{c|c|c|c|c}
			\multicolumn{ 4}{c}{} \\
			\hline
			& main size & inter size & $R^2(\%)$ &Time(s)\\
			\hline
			SSI     &18.01(0.62) &42.47(1.54) & 98.24(0.16)& 1.15\\
			BOLT-SSI &19.00(0.79) & 27.06(1.35) & 98.11(0.12) &1.48\\
			\hline
			RAMP-s         & 8.40(0.10)  & 7.53(1.81) & 98.54(0.10)&3.79\\
			RAMP-w         & 3.74(0.07)  & 11.33(0.24) & 98.59(0.10)&7.74\\
			\hline
			xyz-L10       & 1(0)    & 1.19(0.04)   & 90.23(0.71)  & 0.97\\
			xyz-L100       &1(0) &  1.27(0.05) & 90.58(0.49)&6.67\\
			xyz-L500       &1(0)    & 1.31(0.06)   & 91.29(0.58)   & 31.17  \\
			\hline
			LASSO-CV5      & 27.33(0.69)  & ------ & 97.77(0.10)& 0.10\\
			LASSO-AIC     &1.14(0.05) &------ & 94.22(0.14)&0.01\\
			LASSO-BIC    & 1(0)    &------& 94.15(0.13)&0.01\\
			\hline
		\end{tabular}
	}
\end{table}

Since the variable ``the price of the unit at the beginning of the project'' has a high correlation with the response ``sales prices'', that is 0.9764 based on the data set,   we find that some methods only select several  variables and the out-of-sample $R^2$ are very high from the results of Table \ref{real1}. Our method BOLT-SSI can easily detect the significant main effect and interaction terms and improve the prediction effect, although the predict performance  is a little worse compared with RAMP. It is reasonable because RAMP is based on the regularization method, which uses all of the  correlation between predictors, not  as the screening methods, which only consider marginal information between the response variable and predict variables. In this sense, our SSI or BOLT-SSI sacrifice some data information to balance the computation complexity.  Because of a relatively small dimension of the data, though our methods are still beneficial and efficient for identifying the important interaction terms in this particular dataset for the prediction, the advantage of such trade-off is not apparent. However,  compared to xyz and LASSO, our methods are more stable, and the sacrifice of the statistical efficiency is much small.

\subsection{Supermarket Data}

The supermarket data was collected from a major supermarket located in northern China and has been analyzed by  \cite{wang2009forward} and \cite{hao2018model}, which includes 6,398 predictors and 464 observations.   The response is the number of customers on a particular day, and each predictor is the corresponding sale volume of the product. The supermarket manager wonders which products would be more associated with the number of customers, which means that he or she wants to select the most informative products to predict the response. Note that here, the total number of interaction terms for the supermarket data in modeling is about $2\times 10^7$, much larger than the number of interaction  effects to model the Residential Building Data.

Here, we randomly select 400 observations as the training data and the remaining 64 observations as the testing data and then use the out-of-sample $R^2$ to evaluate the prediction performance of our methods based on 100 random splits. And the settings of all methods are the same as that of the above example.  The average performance  is summarized in Table \ref{real2}, which includes the average sizes of main effects and interaction effects, the average out-of-sample $R^2$ and their standard errors over 100 experiments. Besides the results of our methods, Table \ref{real2} displays the out-of-sample $R^2$ by other methods,  including RAMP-AIC, RAMP-BIC, RAMP-EBIC, RAMP-GIC,  iFORT \& iFORM, and RAMP. The corresponding results are extracted directly from their papers.
For the results of LASSO-AIC, LASSO-BIC, LASSO-EBIC, LASSO-GIC, we extract them from the paper of \cite{hao2018model} (RAMP). For LASSO-AIC-m, LASSO-BIC-m, we only consider the main effects.

\begin{table}[!htbp]\caption{Average results and the standard errors (in parentheses) on the supermarket data set }\label{real2}
	\centering
	\smallskip{\footnotesize \tabcolsep= 20 pt
		\begin{tabular}{c|c|c|c}
			\multicolumn{ 4}{c}{} \\
			\hline
			& main size & inter size & $R^2(\%)$ \\
			\hline
			BOLT-SSI    &  196.19(3.79) & 42.43(1.13) &93.95(0.15)\\
			SSI      & 107.70(0.73)  & 10.90(0.37) & 92.73(0.14) \\
			\hline
			xyz-L10    &   37.80(0.26)     &    12.61(0.25)           & 87.03(0.26)\\
			xyz-L100      & 35.54(0.24)  & 14.40(0.23)  & 86.94(0.22)\\
			xyz-L500   &    35.26(0.25)  &  14.84(0.24)  & 86.59(0.28)  \\
			\hline
			RAMP-AIC & 229.18(1.68) & 94.53 (1.06) &90.48(0.23)\\
			RAMP-BIC & 101.17(3.25) & 34.36(1.65)  & 91.18(0.20)\\
			RAMP-EBIC& 29.27(1.01) & 3.07(0.29)    &89.67(0.31)\\
			RAMP-GIC & 30.71(0.92) & 3.20(0.30)    &90.08(0.28)\\
			\hline
			iFORT   & --------        & --------      & 88.91(0.17)\\
			iFORM   & --------        & --------      &  88.66(0.18)\\
			\hline
			LASSO-AIC & 264.28 (0.91) & 0(0)          & 92.04(0.18)\\
			LASSO-BIC & 63.47  (0.77) & 0(0)          & 90.76(0.20)\\
			LASSO-EBIC & 15.62(0.46)  & 0(0)          & 72.09(0.53)\\
			LASSO-GIC & 19.19 (0.74)  & 0(0)          & 75.05(0.58)\\
			\hline
			LASSO-AIC-m &  30.72(0.61)   &--------       & 82.65(0.40)\\
			LASSO-BIC-m &  13.21(0.22)   &--------    & 69.58(0.48) \\
			\hline
		\end{tabular}
	}
\end{table}

From the results in Table \ref{real2},  the BOLT-SSI demonstrates the best performance, with the mean out-of-sample $R^2=93.95\%$. Although the products selected by BOLT-SSI are a few more, and it is a  challenging task for the supermarket manager to interpret them, more products can improve the whole supermarket's profit. Therefore,  our method  is helpful for the supermarket manager to make a decision.

To fairly assess the efficiency of the methods ``BOLT-SSI'', ``SSI'', ``xyz'' and ``RAMP'' on this real data set, we still use the machine that  equips Intel (R) Xenon(R) CPU E5-1603 v4 @ 2.80GHZ with  8.00 GB RAM. Time(s) is the average computation time  of 5 experiments,  including variable selection and prediction. The results are listed in Table \ref{real2-1}.

 \begin{table}[!htbp]\caption{Average computation time on the supermarket data set }\label{real2-1}
	\centering
	\footnotesize{ \tabcolsep= 3 pt
		\begin{tabular}{c|ccccccc}
			\hline
			Methods &BOLT-SSI	&SSI	&xyz-L10	&xyz-L100	&xyz-L500	&RAMPs	&RAMPw\\
			\hline
			Time(s)	&98.81	&431.55 &59.09	&463.15	&2252.95	&33.75	&NULL\\
			\hline
		\end{tabular}
	}
\end{table}
Here, the result ``NULL'' means that the error exists. When we only run one time by ``RAMP'' with weak heredity assumption in the above machine, the following error will appear, that is, ``can not allocate vector of size 1.1 Gb'', which implies that the method ``RAMP'' may not be widely used on some ordinary computers when the dimension of the data set is huge. From the above two tables,  at the first step of our screening methods, we only use marginal information of the data or even sacrifice  some information for the method BOLT-SSI. However,  the advantages of computational efficiency are much evident, and especially for BOLT-SSI, the sacrifice  of the data information can be ignored, which is consistent with our theoretical investigation.

\subsection{Northern Finland Birth Cohort (NFBC) Data}
To obtain further insights into the newly developed framework,  we apply it to analyze a real GWAS data set from Finland. The Finland data (NFBC1966) contains
10 quantitative traits, including body mass index (BMI), C-reactive protein (CRP), glucose, insulin, high-density lipoprotein (HDL), low-density lipoprotein (LDL), triglycerides (TG), total cholesterol (TC), systolic blood pressure (SysBP), and diastolic blood pressure (DiaBP). And also, it consists of  5,123 individuals with multiple metabolic traits measured and 319,147 SNPs. We consider BMI as the response,  other 9 phenotypes and all SNPs as the predictors.Hence, totally, the sample size is $n$=5,123,  and the dimension of predictors  is $p=319,156$. The total number of  interaction terms is about $5\times10^{10}$.

Here, we just study the screening performance of our methods. From the BOLT-SSI, we obtain the top $p$ of all the interactions and list the top ten interactions  as follows (Table \ref{real3-1}):
\begin{table}[!htbp]\caption{The top $p$  interactions terms by BOLT-SSI on the NFBC }\label{real3-1}
	\centering\tabcolsep= 25 pt
	\begin{tabular}{c|c|c}
		\hline
		Inter1 & Inter2 & Exact Test Statistic \\
		\hline
		Insulin & HDL & 63.332 \\
		CRP & SysBP & 53.566 \\
		Glucose & Insulin & 53.455 \\
		Insulin & TG & 51.407 \\
		CRP & Insulin & 46.678 \\
		Insulin & SysBP & 42.337 \\
		rs1638742 & rs1958050 & 40.016 \\
		rs2707941 & rs1958050 & 40.010 \\
		rs10217074 & rs7957938 & 39.341 \\
		rs17074280 & rs1890472 & 38.916 \\
		\hline
	\end{tabular}
\end{table}

Base on the Bonferroni correction, $\alpha=0.05/(p(p-1)/2)$, and the critical value are $\chi_{4,\alpha}^2=62.237$ and $\chi_{1,\alpha}^2=50.880$. We can find that the first 4 interactions are significant.

For the Method ``xyz'', we randomly choose $L=500$ and $L=1000$ times of projections and remain 500 significant interactions. There are only 4 same terms between the BOLTSSI's results and xyz(L500)'s results, that is, (Table \ref{real3-2}),
\begin{table}[!htbp]\caption{The 4 same   interactions terms selected by xyz(L500) on the NFBC }\label{real3-2}
	\centering\tabcolsep= 25 pt
	\begin{tabular}{c|c|c}
		\hline
		Inter1 & Inter2 & Exact Test Statistic \\
		\hline
		CRP & TG & 29.662 \\
		rs7631436 & rs7906313 & 7.617 \\
		rs232101 & rs7262267 & 6.479 \\
		rs2250648 & rs7104767 & 4.048 \\
		\hline
	\end{tabular}
\end{table}

The xyz(L1000)'s result has the 5 same terms as the result of our method, that is, (Table \ref{real3-3}),
\begin{table}[!htbp]\caption{The 5 same  interactions terms selected by xyz(L1000) on the NFBC }\label{real3-3}
	\centering\tabcolsep= 25 pt
	\begin{tabular}{c|c|c}
		\hline
		Inter1 & Inter2 & Exact Test Statistic \\
		\hline
		CRP & TG & 29.662 \\
		rs9368950 & rs2474619 & 5.474 \\
		rs1481872 & rs2896268 & 4.532 \\
		rs1355889 & rs969539 & 2.046 \\
		rs11982066&rs1074742&1.519\\
		\hline
	\end{tabular}
\end{table}

Furthermore, the xyz(L500) and xyz(L1000) have 32 same interactions. Based on the  screening results of the method ``xyz'', the interaction terms screened by the method ``xyz'',  cannot pass the Bonferroni's threshold.

From the efficiency,  the screening times of the method ``xyz'' are 0.54 hours (L500) and 0.96 hours (L1000) in the server, respectively; and it takes 5.59 hours to screen the interaction terms by our method when the thread number is 30 in the server.

All in all, the performance of our method is much better than that of the method ``xyz'' on this data set, although the screening time of ``xyz'' is less than that of ``BOLT-SSI''.

\section{Conclusion and Discussion}
In this paper, we study the screening method to detect important significant interaction effects in the generalized high dimensional linear model. A new and straightforward  procedure SSI and its extension BOLT-SSI are proposed. Different from most of the other screening or variable selection methods for the interaction effects detecting, our proposed methods do not depend on the heredity assumption.  The proposed screening methods conduct a  full screening search for all of the  interaction effects among the data. For ultra-high dimensional data, in some sense, such a task seems to be impossible to be completed. Here we show that, by taking advantage of computational structure,  seemly impossible tasks can be done using a standard personal computer. Importantly, the statistical property of the proposed way is guaranteed by our established theory.

Generally speaking, most of the data analysis projects are similar  to engineer projects. Though most of the  theoretical research would be beneficial to projects,  the requirement and expectation of the engineering projects are different from those of theoretical studies. How to combine the advantages of engineering techniques to complete those projects under the requirement and expectation of practice needs further investigation. Our study here attempts to pursuit such a direction by a small step.


%
%
%

%
\bibliographystyle{agsm}

\bibliography{references_boltssi}

\section{Appendix}
\subsection{Properties  of SSI}
Denote that  $\Beta_{ij}=(\beta_{ij0},\beta_i,\beta_j,\beta_{ij})^T$ be the four-dimensional parameter, and  let $\X_{ij}=(1,X_i,X_j,X_{ij})^T$.  Since the log-likelihood function is of the concavity in the generalized linear model with the canonical link function,
the function $\mbox{E}l(\X_{ij}^T\Beta_{ij},Y)$ can arrive at its unique minimum $\mbox{E}l(\X_{ij}^T\Beta_{ij}^M,Y)$ over $\Beta_{ij}\in \mathcal{B}$, in which $\Beta_{ij}^M=(\beta_{ij0}^M,\beta_i^M,\beta_j^M,\beta_{ij}^M)^T$ is an interior point of the set $\mathcal{B}$ and $\mathcal{B}=\{|\beta_{ij,0}^M|\leq B,|\beta_{i}^M|\leq B,|\beta_{j}^M|\leq B,|\beta_{ij}^M|\leq B\}$ is an area with the large width $B$ where the marginal likelihood is maximized. The following conditions are needed:
\fl (A) The marginal Fisher information: $\I_{ij}(\Beta_{ij})=\mbox{E}\{b''(\X_{ij}^T\Beta_{ij})\X_{ij}\X_{ij}^T\}$ is finite and positive definite at $\Beta_{ij}=\Beta_{ij}^M$, for $1\leq i<j \leq p_n.$ Moreover, $\|\I_{ij}(\Beta_{ij})\|_{\mathcal{B}}=\sup\limits_{\Beta_{ij}\in\mathcal{B},\|\x\|=1}\|\I_{ij}(\Beta_{ij})^{1/2}\x\|$ is bounded.

\fl (B) (i) Let $\X_{i,j}=(1,X_i,X_j)^T$. As the definition of the conditional linear expectation, provided by \cite{barut2016conditional},  is the best linearly fitted regression within the class of linear functions, we denote that
$$E_L(Y|\X_{ij}^T\Beta_{ij}^M)=b'(\X_{ij}^T\Beta_{ij}^M)\ \ \ \ \mbox{and}\ \ \ E_L(Y|\X_{i,j}^T\Beta_{i,j}^M)=b'(\X_{i,j}^T\Beta_{i,j}^M);$$ and
$$\text{Cov}_L(Y,X_{ij}|\X_{i,j}^T\Beta_{i,j}^M)\equiv E(X_{ij}-E_L(X_{ij}|\X_{i,j}^T\Beta_{i,j}^M))(Y-E_L(Y|\X_{i,j}^T\Beta_{i,j}^M)).$$For $(i,j)\in \mathcal{N}_\star$, there exists a constant $c_1>0$ such that $|\text{Cov}_L(Y,X_{ij}|\X_{i,j}^T\Beta_{i,j}^M)|\geq c_1n^{-\kappa}$ for some $0<\kappa<1/4$.

(ii) Denote
$$m_{ij}=\frac{b'(\X_{ij}^T\Beta_{ij}^M)-b'(\X_{i,j}^T\Beta_{i,j}^M)}{\X_{ij}^T\Beta_{ij}^M-\X_{i,j}^T\Beta_{i,j}^M},$$
and $\mbox{E}(m_{ij}X_{ij}^2)=E(m_{ij}X_i^2X_j^2)\leq c_2$ for some constant $c_2$, in which $1\leq i<j \leq p$.

\fl (C) For all $\Beta_{ij}\in\mathcal{B}$, $\mbox{E}(l(\X_{ij}^T\Beta_{ij},Y)-l(\X_{ij}^T\Beta_{ij}^M,Y))\geq V\|\Beta_{ij}-\Beta_{ij}^M\|^2$, for some constant $V>0$, for all $1\leq i<j\leq p$.

\fl (D) There exist constants $m_0$, $m_1$, $s_0$, $s_1>0$ and $\alpha>0$, such that for  sufficiently large $t>0$,
$$P(|X_{i}|>t)\leq m_1\exp\{-m_0t^\alpha\}\ \ \  \ \ \ \  \mbox{for}\ \ 1\leq i\leq p,$$
and that
$$\mbox{E}\exp(b(\X^T\Beta^\star+s_0)-b(\X^T\Beta^\star))+E\exp(b(\X^T\Beta^\star-s_0)-b(\X^T\Beta^\star))\leq s_1.$$

\fl (E) For the function $b(\theta)$, the second derivative $b''(\theta)$ is continuous and $b''(\theta)>0$. There exists  $\varepsilon_1>0$ such that for all $1\leq i<j\leq p$,
$$\sup_{\Beta\in\mathcal{B}, \|\Beta-\Beta_{ij}^M\|\leq\varepsilon_1}|\mbox{E}\{b(\X_{ij}^T\Beta)I(|X_{ij}|>K_n)\}|\leq o(n^{-1}),$$
where $I(\cdot)$ is the indicator function and $K_n$ is an arbitrarily large constant such that for a given $\Beta$ in $\mathcal{B}$, the function $l(\x^T\Beta,y)$ satisfies the Lipschitz property with positive constant $k_n$ for all $(\x,y)$ in the set $\Omega_n=\{(\x,y): \|\x\|_\infty\leq K_n, |y|\leq K_n^\star\}$ with $K_n^\star=m_0K_n^\alpha/s_0$, where $\|\cdot\|_\infty$ is the supremum norm.

\fl (F) The variance $\text{Var}(\X^T\Beta^\star)=\Beta^{\star T}\bfm \Sigma \Beta^\star$ and $b''(\cdot)$ are bounded,
where $\bfm \Sigma=\text{diag}(0,\bfm \Sigma_1)$ with $\bfm \Sigma_1=\text{Var}(\X)$.

\fl (G) The minimum eigenvalue of the matrix $\mbox{E}[m_{ij}\X_{ij}\X_{ij}^T]$ is larger than a positive constant for any $i, j$, where $m_{ij}$ is defined in Condition B(ii).

\fl (H) Denote that $\Beta_{ij-}^M=(\beta_{i,j0}^M,0,\ldots,0,\beta_{i,}^M,0,\ldots,0,\beta_{j,}^M,0,\ldots,0)^T$, $\Delta\Beta_{ij}=\Beta_{\mathcal{C}}^\star-\Beta_{ij-}^M$. Let $R_{ij}=\mbox{E}[X_{ij}\X_{\mathcal{C}}^T\Delta\Beta_{ij}]$ and $\R=(R_{12},R_{13},\ldots,R_{(p-1)p})^T$, then $\|\R\|_2^2=o(\lambda_{\mbox{max}}(\bfm \Sigma_{\mathcal{I}}))$, where $\lambda_{\mbox{max}}(\bfm \Sigma_{\mathcal{I}})$ is the largest eigenvalue of the matrix $\bfm \Sigma_{\mathcal{I}}.$

\

All conditions are similar to the ones proposed by \cite{fan2010sure} and \cite{barut2016conditional}, and are satisfied by most of the generalized linear models such as linear regression and logistic regression. By the strict convexity property of $b(\theta)$, $m_{ij}$ is almost surely larger than 0. If $b(\theta)=\theta^2/2$, then $m_{ij}=1$ and Condition B(ii) is automatically satisfied by the uniform bounded property of $\mbox{E}(X_{ij}^2)$ since  $X_{i}$ and $X_j$ are normalized. The first part of Condition (D) builds an exponential bound on the tails of $X_j$. Actually, since the event $\{\omega: |X_{ij}(\omega)|>t\}$ is a subset of the union of $\{\omega: |X_{i}(\omega)|>\sqrt{t}\}$ and $\{\omega: |X_{j}(\omega)|>\sqrt{t}\}$, when $P(|X_{i}|>\sqrt{t})\leq m_1'\exp\{-m_0t^{\alpha/2}\}$ and $P(|X_{j}|>\sqrt{t})\leq m_1'\exp\{-m_0t^{\alpha/2}\}$, we  have that
$$P(|X_{ij}|>t)\leq 2m_1'\exp\{-m_0t^{\alpha/2}\}\ \ \ \ \mbox{for}\ \ 1\leq i<j\leq p.$$
Then, we can take $m_1=2m_1'$ and by $\exp\{-m_0t^{\alpha}\}<\exp\{-m_0t^{\alpha/2}\}$,  the exponential bound on the tails is simultaneously available for main effect and interaction terms. Hence, the first part of Condition (D) also implies an exponential bound for the tails of $X_{ij}$.
The second part of Condition (D) also guarantees that the response variable $Y$ possesses the exponentially light tail, as shown by Lemma 1 of \cite{fan2010sure}.

To detect  the important interactions in our model, one critical question would be: at what level the interactions of variables should be preserved by the screening procedure? If one interaction $X_{ij}$ is jointly important i.e. $\beta_{ij}^\star\neq 0$, will it still be marginally important, i.e. $\beta_{ij}^M\neq 0$? On the other hand,  when one interaction is jointly unimportant i.e., $\beta_{ij}^\star= 0$, will it still be marginally unimportant, i.e. $\beta_{ij}^M=0$? The following theorems are trying to give the answers to these questions.

\

\fl{\bf Theorem A.1} \label{ssith1} For $1\leq i<j\leq p$, the marginal likelihood increment $L_{ij}^\star=0$ if and only if $\beta_{ij}^M=0.$

\

\fl{\bf Theorem A.2} \label{ssith2} For $1\leq i<j\leq p$, the marginal regression parameters $\beta_{ij}^M=0$ if and only if $\text{Cov}_L(Y,X_{ij}|\X_{i,j}^T\Beta_{i,j}^M)=0$.

\

\fl{\bf Corollary A.1}  \l
abel{ssico1} For $1\leq i<j\leq p$, the marginal likelihood increment $L_{ij}^\star=0$ if and only if $\text{Cov}_L(Y,X_{ij}|\X_{i,j}^T\Beta_{i,j}^M)=0$.

\

The above theorems and corollary reveal that both the increment of the log-likelihood and the marginal regression parameters are measurements of the relationship between the interaction and the mean response function. They are equivalent under mild conditions.

To distinguish the active interactions $\{X_{ij}: (i,j)\in\mathcal{N}_\star\}$ and inactive interactions $\{X_{ij}: (i,j)\not\in\mathcal{N}_\star\}$, we need to set up one appropriate threshold value $\gamma_n$, so that the minimum marginal signal strength is stronger than the stochastic noise and the sure screening property will be guaranteed.  Theorem A.3 and Theorem A.4 show that active interaction set and inactive interaction set can be separated by the marginal coefficient $\beta_{ij}$ of $X_{ij}$ or the increment of marginal likelihood functions.

\

\fl{\bf Theorem A.3} \label{ssith3} If Condition (B) holds, then there exists a positive constant $c_3$ such that
$$\min_{(i,j)\in \mathcal{N}_\star}|\beta_{ij}^M|\geq c_3n^{-\kappa}.$$

\

\fl{\bf Theorem A.4} \label{ssith4} Under the conditions (B) and (C), we have
$$\min_{(i,j)\in \mathcal{N}_\star}L_{ij}^\star\geq c_4n^{-2\kappa}$$
for some positive constant $c_4$.

\

By the uniform convergence of the marginal likelihood ratio, we obtain the uniform convergence rate and sure screening properties of the proposed SSI by the following theorems. The convergence rate will help control the size of the selected set.

\

\fl{\bf Theorem A.5} \label{ssith5} Assume that Conditions (A), (B), (C), (D) and (E) hold. Let $k_n=b'(3K_n B+B)+m_0K_n^\alpha/s_0$, with $K_n$ given in Condition (E).

(i) If $n^{1-2\kappa}/(k_nK_n)^2\rightarrow\infty$, then for any $c_5>0$, there exists a constant $c_6>0$ such that
\begin{eqnarray*}
	&&P\left(\max_{1\leq i<j\leq p}|\hat{\beta}_{ij}^M-\beta_{ij}^M|\geq c_5n^{-\kappa}\right)\\
	&\leq& q\left(\exp(-c_6n^{1-2\kappa}/(k_nK_n)^2)+nm_2\exp(-m_0K_n^{\alpha/2})\right),
\end{eqnarray*}
where $q=\frac{p(p-1)}{2}$ and $m_2=3m_1+s_1$.

(ii) If $n^{1-2\kappa}/(k_nK_n)^2\rightarrow\infty$, then for any $c_7>0$, there exist constants $c_8>0$ and $c_9>0$ such that
\begin{eqnarray*}
	&&P\left(\max_{1\leq i<j\leq p}|L_{ij,n} -L_{ij}^\star|\geq c_7n^{-2\kappa}\right)\\
	&\leq& q\left(2\exp(-c_8n^{1-2\kappa}/(k_nK_n)^2)+4\exp(-c_9n^{1-4\kappa})+4nm_2\exp(-m_0K_n^{\alpha/2})\right),
\end{eqnarray*}

(iii) In addition, by taking $\gamma_n=c_{10}n^{-2\kappa}$ with $c_{10}\leq c_4/2$ , we have
\begin{eqnarray*}
	&&P(\mathcal{N}_\star\subset\widehat{\mathcal{N}}_{\gamma_n})\\
	&\geq & 1-s_n\left(2\exp(-c_8n^{1-2\kappa}/(k_nK_n)^2)+4\exp(-c_9n^{1-4\kappa})+4nm_2\exp(-m_0K_n^{\alpha/2})\right),
\end{eqnarray*}
where $s_n=|\mathcal{N}_\star|$ is the size of active interactions.

\

Note that the sure screening property given in Theorem A.5(iii) only relates to the size $s_n$ of the active interaction effects. The dimensionality $p$ or $q$ does not affect the sure screening. For generalized linear model, such as logistic regression, $b(\theta)=\ln(1+\exp(\theta))$ , and $b'(\theta)=\frac{1}{1+\exp(-\theta)}$ is bounded. By Theorem \ref{ssith5}(ii), the optimal order of $K_n$ is $n^{(1-4\kappa)/(\alpha+2)}$, and
$$P\left(\max_{1\leq i<j\leq p}|L_{ij,n} -L_{ij}^\star|\geq c_7n^{-2\kappa}\right)=O\left\{p^2\exp(-c_9n^{(1-4\kappa)\alpha/(\alpha+2)})\right\}.$$
Thus, the tail probability will be exponentially small. That is, we can deal with the NP-dimensionality
$$\ln p=o\left(n^{(1-4\kappa)\alpha/(\alpha+2)}\right)$$
with $\alpha=\infty$ for the special case of the bounded covariates and $\alpha=2$ for normal covariates. Similar results for unconditional screening and conditional screening are shown in \cite{fan2010sure} and \cite{barut2016conditional}, respectively.

For SSI,  the following theorem shows that the false selection rate can be controlled absolutely.  In other words, the size of the set $\widehat{\mathcal{N}}_{\gamma_n}$ can be controlled, and hence the number of interactions would be significantly reduced for the final model estimation.

\

\fl{\bf Theorem A.6} \label{ssith6} Under Conditions (A)-(H), we have
\begin{eqnarray*}
	&&P\left(|\widehat{\mathcal{N}}_{\gamma_n}|\leq O(n^{2\kappa}\lambda_{max}(\bfm \Sigma_{\mathcal{I}}))\right)\\
	&\geq & 1-q\left(2\exp(-c_8n^{1-2\kappa}/(k_nK_n)^2)+4\exp(-c_9n^{1-4\kappa})+4nm_2\exp(-m_0K_n^{\alpha/2})\right).
\end{eqnarray*}
where $q=p(p-1)/2$ and $m_2=3m_1+s_1$.

\

From the proof of Theorem A.6, without Condition (H), Theorem \ref{ssith6} still holds with $\bfm \Sigma_{\mathcal{I}}$ replaced by $\bfm \Sigma_{\mathcal{I}}+\R\R^T$.   If $\lambda_{\text{max}}(\bfm \Sigma_{\mathcal{I}})=O(n^\uptau)$, the size of the selected set has order $O(n^{2\kappa+\uptau})$, the same order as in the approach of \cite{fan2008sure}. Our result is an extension of the work of \cite{fan2008sure}. Similar results have been shown in  \cite{fan2010sure}, \cite{fan2011nonparametric}, \cite{li2012robust}, and \cite{barut2016conditional}.

\subsection{Proofs}
In this section, we will provide the proofs of the main theorems in this paper.\\

\fl{\bf Proof of Theorem A.1:} 
If $\beta_{ij}^M=0$, by the model identifiability, $\beta_{i,j0}^M=\beta_{ij0}^M$, $\beta_{i,}^M=\beta_{i}^M$ and  $\beta_{j,}^M=\beta_{j}^M$. Hence, $L_{ij}^\star=0$. On the other hand, if $L_{ij}^\star=0$, by Condition (C), $\Beta_{i,j}^M=\Beta_{ij}^M$. It follows that $\beta_{i,j0}^M=\beta_{ij0}^M$, $\beta_{i,}^M=\beta_{i}^M$, $\beta_{j,}^M=\beta_{j}^M$ and $\beta_{ij}^M=0$.

\

\fl{\bf Proof of Theorem A.2:} 
Note that the condition $\text{Cov}_L(Y,X_{ij}|\X_{i,j}^T\Beta_{i,j}^M)=0$ is equivalent to $E\{(Y-b'(\X_{i,j}^T\Beta_{i,j}^M))X_{ij}\}=0.$ We prove the necessarity first. The marginal regression coefficients $\Beta_{ij}^M$ satisfy the score equation
\begin{equation}\label{s1}
E\{b'(\X_{ij}^T\Beta_{ij}^M)\X_{ij}\}=E(Y\X_{ij})=E(E(Y|\X)\X_{ij})=E(b'(\X^T\Beta^\star)\X_{ij}),
\end{equation}
i.e.,
\begin{equation}\label{s2}
E\{b'(\X_{ij}^T\Beta_{ij}^M)\X_{i,j}\}=E(Y\X_{i,j})=E(E(Y|\X)\X_{i,j})=E(b'(\X^T\Beta^\star)\X_{i,j}),
\end{equation}
and the coefficients $\Beta_{i,j}^M$ satisfy the score equation
\begin{equation}\label{s3}
E\{b'(\X_{i,j}^T\Beta_{i,j}^M)\X_{i,j}\}=E(Y\X_{i,j})=E(b'(\X^T\Beta^\star)\X_{i,j}).
\end{equation}
If $\beta_{ij}^M=0$, by the equation (\ref{s2}), the first three components of $\Beta_{ij}^M$, should be equal to $\Beta_{i,j}^M$ by the uniqueness of the solution to the score equation (\ref{s3}). Therefore, the score equation (\ref{s1}) on the component $X_{ij}$ gives
\begin{equation}\label{s4}
E\{b'(\X_{i,j}^T\Beta_{i,j}^M)X_{ij}\}=E(YX_{ij}).
\end{equation}
It follows that  $E\{(Y-b'(\X_{i,j}^T\Beta_{i,j}^M))X_{ij}\}=0$, i.e., $\text{Cov}_L(Y,X_{ij}|\X_{i,j}^T\Beta_{i,j}^M)=0$.\\
For the sufficiency, if $E\{(Y-b'(\X_{i,j}^T\Beta_{i,j}^M))X_{ij}\}=0$,  we have the equation (\ref{s4}). By equation (\ref{s3}), $((\Beta_{i,j}^M)^T,0)^T$ is a solution to the equation (\ref{s1}). By the property of solution's uniqueness, it follows that $\Beta_{ij}^M=((\Beta_{i,j}^M)^T,0)^T$, so $\beta_{ij}^M=0$.

\

\fl{\bf Proof of Corollary  A.1:} 
By Theorem A.1 and A.2, we can easily obtain this Corollary.

\

\fl{\bf Proof of Theorem A.3:} 
Denote that the matrix $\A=E(m_{ij}\X_{ij}\X_{ij}^T)$ and partition it as
\[\A=
\begin{bmatrix}
E(m_{ij}\X_{i,j}\X_{i,j}^T) & E(m_{ij}\X_{i,j}X_{ij}) \\
E(m_{ij}X_{ij}\X_{i,j}^T) & E(m_{ij}X_{ij}^2)\\
\end{bmatrix}=
\begin{bmatrix}
\A_{11} & \A_{12} \\
\A_{21} & \A_{22}\\
\end{bmatrix}.
\]
Hence, the matrix $\A$ is a positive definite matrix. By the convexity of the function $b(\cdot)$, $m_{ij}>0$ almost surely. Therefore, for any nonzero constant vector $ a$, $ a^T\A a=E(m_{ij} a^T\X_{ij}\X_{ij}^T a)=E(m_{ij} a^T\X_{ij}\X_{ij}^T a)=E(m_{ij}(a^T\X_{ij})^2)>0$ and the inverse matrix $A_{11}^{-1}$ exists.

Based on the  equation (\ref{s2}) and (\ref{s3}), we have $$E\{b'(\X_{ij}^T\Beta_{ij}^M)\X_{i,j}\}=E\{b'(\X_{i,j}^T\Beta_{i,j}^M)\X_{i,j}\},$$ i.e.,
$E\{[b'(\X_{ij}^T\Beta_{ij}^M)-b'(\X_{i,j}^T\Beta_{i,j}^M)]\X_{i,j}\}=0$. Let $\Delta_{\Beta_{ij}}=(\beta_{ij0}^M,\beta_{i}^M,\beta_{j}^M)^T-\Beta_{i,j}^M$ and by the definition of $m_{ij}$, we have
$$E\{m_{ij}(\X_{i,j}^T\Delta_{\Beta_{ij}}+X_{ij}\beta_{ij}^M)\X_{i,j}\}=0,$$
that is, $\Delta_{\Beta_{ij}}=-A_{11}^{-1}A_{12}\beta_{ij}^M$.
Moreover,
\begin{eqnarray*}
	\text{Cov}_L(Y,X_{ij}|\X_{i,j}^T\Beta_{i,j}^M)&=& E\{(Y-E_L(Y|\X_{i,j}^T\Beta_{i,j}^M))X_{ij}\}\\
	&=& E\{[b'(\X_{ij}^T\Beta_{ij}^M)-b'(\X_{i,j}^T\Beta_{i,j}^M)]X_{ij}\} \\
	&=& E\{m_{ij}[\X_{i,j}^T\Delta_{\Beta_{ij}}+X_{ij}\beta_{ij}^M]X_{ij}\} \\
	&=&A_{21} \Delta_{\Beta_{ij}}+A_{22}\beta_{ij}^M\\
	&=&(A_{22}-A_{21}A_{11}^{-1}A_{12})\beta_{ij}^M.
\end{eqnarray*}
By the positive definiteness of Matrix $A$, $A_{22}-A_{21}A_{11}^{-1}A_{12}>0$. Hence, by Condition (B),
$$|\text{Cov}_L(Y,X_{ij}|\X_{i,j}^T\Beta_{i,j}^M)|=|A_{22}-A_{21}A_{11}^{-1}A_{12}||\beta_{ij}^M|\leq A_{22}|\beta_{ij}^M|\leq c_2 |\beta_{ij}^M|.$$
Letting $c_3=\frac{c_1}{c_2}$, we have $|\beta_{ij}^M|\geq c_3n^{-\kappa}$. The conclusion follows.\ \ \ $\Box$

\

\fl{\bf Proof of Theorem A.4:}
If Condition (B) holds, 
by Theorem A.3, 
we have $|\beta_{ij}^M|\geq c_3n^{-\kappa}$ for $1\leq i<j\leq p$. Then by Condition (C), we have
\begin{eqnarray*}
	L_{ij}^\star &=&E\{l(\X_{i,j}^T\Beta_{i,j}^M,Y)-l(\X_{ij}^T\Beta_{ij}^M)\}\\
	&=&E\{l(\X_{ij}^T((\Beta_{i,j}^M)^T,0)^T,Y)-l(\X_{ij}^T\Beta_{ij}^M)\}\\
	&\geq& V\|(\Beta_{i,j}^M)^T,0)^T-\Beta_{ij}^M\|^2\\
	&\geq& V |\beta_{ij}^M|^2\\
	&\geq& Vc_3^2 n^{-2\kappa}.
\end{eqnarray*}
Letting $c_4=Vc_3^2 $, we have $\min_{(i,j)\in \mathcal{N}^\star}L_{ij}^\star\geq c_4n^{-2\kappa}$.\ \ \ $\Box$

\

\


In order to prove Theorem A.5, we need some results in Fan and Song (2010) and Barut $et\ al.$ (2016), which are listed below.

Let $\Beta_0= \argmin\limits_\Beta El(\X^T\Beta,Y)$ be the population parameter. Assume that $\Beta_0$ is an interior point of a sufficiently large, compact and convex set $\mathbf{B}\subseteq \mathbf{R}^p$, and assume the conditions below.

\fl (A1) The Fisher information
$$\I(\Beta)=E\left\{\left[\frac{\partial}{\partial\Beta}l(\X^T\Beta,Y)\right]\left[\frac{\partial}{\partial\Beta}l(\X^T\Beta,Y)\right]^T\right\}$$
is finite and positive definite at $\Beta=\Beta_0$. Furthermore, $\|\I(\Beta)\|_\mathbf{B}=\sup\limits_{\Beta\in\mathbf{B},\|\x\|=1}\|\I(\Beta)^{1/2}\x\|$ exists.

\

\fl (B1) The function $l(\x^T\Beta,Y)$ satisfies the Lipschitz property with positive constant $k_n$: For any $\Beta$,$\Beta'\in\mathbf{B}$ and $(\x,y)\in \Omega_n=\{(\x,y): \|\x\|_\infty\leq K_n, |y|\leq K_n^\star\}$,
$$|l(\x^T\Beta,Y)-l(\x^T\Beta',Y)|\leq k_n|\x^T\Beta-\x^T\Beta'|$$
for some sufficiently large positive constants $K_n$ and $K_n^\star$.
Furthermore, there exists a sufficiently large constant $C$ such that
$$\sup_{\Beta\in\mathbf{B},\ \|\Beta-\Beta_0\|\leq b_n}|E\{[l(\X^T\Beta,Y)-l(\X^T\Beta_0,Y)](1-I_n(\X,Y))\}|\leq o(p/n),$$
where $b_n=Ck_nV_n^{-1}(p/n)^{1/2}$, $V_n$ is defined in Condition (C1) and $I_n(\x,y)=I((\x,y)\in\Omega_n).$

\

\fl (C1) The function $l(\X^T\Beta,Y)$ is convex in $\Beta$ and
$$E[l(\X^T\Beta,Y)-l(\X^T\Beta_0,Y)]\geq V_n\|\Beta-\Beta_0\|^2,$$
for some positive constant $V_n$, where $\|\Beta-\Beta_0\|\leq b_n$, $b_n$ is defined in Condition (B1).

The proof of Theorem A.5
needs an exponential bound for the tail probability of the quasi maximum likelihood estimator $\hat{\Beta}=\argmin\limits_{\Beta}\mathbb{P}_n l(\X^T\Beta,Y)$.


\fl{\bf Lemma A.1} \emph{(Fan and Song (2010)) Under conditions (A1)-(C1),  for any $t>0$,
	$$P\left(\sqrt{n}\|\hat{\Beta}-\Beta_0\|\geq 16K_n(1+t)/V_n\right)\leq \exp(-2t^2/K_n^2)+nP(\Omega_n^c).$$}

\


\fl{\bf Lemma A.2} \emph{(Fan and Song (2010)) Under condition (D), for any $t>0$,
	$$P(|Y|\geq m_0t^\alpha/s_0)\leq s_1\exp(-m_0t^\alpha).$$
}

\fl{\bf Lemma A.3}  \emph{Under conditions (A1)-(C1), there exist positive constants $c_6$, $c_7$, $c_9$ and $\kappa$, such that
	\begin{eqnarray*}
		&&P\left(|\mathbb{P}_n\{l(\X^T\hat{\Beta},Y)\}-E\{l(\X^T\Beta_0,Y)\}|\geq c_7n^{-2\kappa}\right)\\
		&\leq& \exp(-c_6n^{1-2\kappa}/(k_nK_n)^2)+2\exp\left(-c_9n^{1-4\kappa}\right)+2nP(\Omega_n^c).
	\end{eqnarray*}
}
\proof
\begin{eqnarray*}
	&& |\mathbb{P}_n\{l(\X^T\hat{\Beta},Y)\}-E\{l(\X^T\Beta_{0},Y)\}| \\
	&=& |\mathbb{P}_n\{l(\X^T\hat{\Beta},Y)\}-\mathbb{P}_n\{l(\X^T\Beta_{0},Y)\}
	+\mathbb{P}_n\{l(\X^T\Beta_0,Y)\}-E\{l(\X^T\Beta_{0},Y)\}| \\
	&\leq& |\mathbb{P}_n\{l(\X^T\hat{\Beta},Y)\}-\mathbb{P}_n\{l(\X^T\Beta_{0},Y)\}|+ |\mathbb{P}_n\{l(\X^T\Beta_0,Y)\}-E\{l(\X^T\Beta_{0},Y)\}|\\
	&\triangleq& S_1+S_2
\end{eqnarray*}
For the terms $S_1$, by Taylor's expansion and $\mathbb{P}_n l'(\X^T\hat{\Beta},Y)=0$, we have
$$S_1=\frac{1}{2}(\hat{\Beta}-\Beta_0)^Tg(\bfm \xi_n)(\hat{\Beta}-\Beta_0)\leq\frac{1}{2}D_0\lambda_{max}(\mathbb{P}_n\X\X^T)\|\hat{\Beta}-\Beta_0\|^2,$$
where $D_0=\mbox{sup}_xb''(x)$, $\lambda_{max}(\mathbb{P}_n\X\X^T)$ is the maximum eigenvalue of the sample variance matrix $\mathbb{P}_nb''(\bfm \xi_n^T\X)\X\X^T$, and $\bfm \xi_n$ lies between $\hat{\Beta}$ and $\Beta_0$.
By Lemma A.1  and taking $1+t=c_5Vn^{1/2-\kappa}/(16k_n)$,
$$P(\|\hat{\Beta}-\Beta_0\|^2\geq c_5^2n^{-2\kappa})=P(\|\hat{\Beta}-\Beta_0\|\geq c_5n^{-\kappa})
\leq \exp(-c_6n^{1-2\kappa}/(k_nK_n)^2)+nP(\Omega_n^c).$$
Furthermore, by the Hoeffding inequality (Hoeffding(1963)), for a random variable $X$ and any given $K>0$, we have
$$P\left((\mathbb{P}_n-E)X^k\I(|X|\leq K)>\epsilon\right)\leq \exp\left(-2n\varepsilon^2/(4K^{2k})\right)$$
for any $k\geq0$ and $\epsilon>0$.
Therefore, with exception on a set with negligible probability and a constant $\delta>1$, it follows that
$$\lambda_{max}(\mathbb{P}_n\X\X^T)\leq \delta\lambda_{max}(E\X\X^T)=O(1).$$
Consequently, $S_1\leq D_1\|\hat{\Beta}-\Beta_0\|^2$ for some $D_1>0$ and then taking $c_7=D_1c_5^2$,
\begin{eqnarray*}
	&&P(S_1\geq c_7 n^{-2\kappa})\leq P(\|\hat{\Beta}-\Beta_0\|^2\geq c_5^2n^{-2\kappa}) \\
	&\leq&  \exp(-c_6n^{1-2\kappa}/(k_nK_n)^2)+nP(\Omega_n^c).
\end{eqnarray*}
For the term $S_2$, for any $\varepsilon>0$,
\begin{eqnarray*}
	&&P\left(|\mathbb{P}_n\{l(\X^T\Beta_0,Y)\}-E\{l(\X^T\Beta_{0},Y)\}|>\varepsilon\right)  \\
	&\leq& P\left(|(\mathbb{P}_n-E)\{l(\X^T\Beta_0,Y)\}|>\varepsilon,\Omega_n \right)+nP(\Omega_n^c).
\end{eqnarray*}
Since $l(\x^T\Beta_0,y)$ satisfies the Lipschitz property, it can be bounded by some interval with length $C>0$ on the set $\Omega_n$ for each $1\leq i\leq n$. Using Hoeffding inequality again, we have
$$P\left(|(\mathbb{P}_n-E)\{l(\X^T\Beta_0,Y)\}|>\varepsilon,\Omega_n \right)\leq 2\exp\left(-2n\varepsilon^2/C^2\right).$$
Taking $\varepsilon=c_7n^{-2\kappa}$, we have
\begin{eqnarray*}
	&&P\left(|\mathbb{P}_n\{l(\X^T\Beta_0,Y)\}-E\{l(\X^T\Beta_{0},Y)\}|>c_7n^{-2\kappa}\right)  \\
	&\leq& P\left(|(\mathbb{P}_n-E)\{l(\X^T\Beta_0,Y)\}|>c_7n^{-2\kappa},\Omega_n \right)+nP(\Omega_n^c)\\
	&\leq& 2\exp\left(-2c_7^2n^{1-4\kappa}/C^2\right)+nP(\Omega_n^c) .
\end{eqnarray*}
Taking $c_9=2c_7^2/C^2$, we have
\begin{eqnarray*}
	&&P\left( |\mathbb{P}_n\{l(\X^T\hat{\Beta},Y)\}-E\{l(\X^T\Beta_{0},Y)\}|\geq c_7n^{-2\kappa} \right)\\
	&\leq& P(S_1\geq c_7n^{-2\kappa}) +P(S_2\geq  c_7n^{-2\kappa})\\
	&\leq& \exp(-c_6n^{1-2\kappa}/(k_nK_n)^2)+2\exp\left(-c_9n^{1-4\kappa}\right)+2nP(\Omega_n^c).
\end{eqnarray*}

\

\fl{\bf Proof of Theorem A.5:} 
(i) We want to use Lemma A.1 to get the exponential bound for the tail property, hence we need to check the conditions (A1)-(C1). By the conditions (A)-(E), we can easily find that most of the conditions are satisfied except for the second part of condition (B1). Now we check it. In our case,
\begin{eqnarray*}
	&& |E\{[l(\X_{ij}^T\Beta_{ij},Y)-l(\X_{ij}^T\Beta_{ij}^M,Y)](1-I_n(\X_{ij},Y))\}|\\
	&=& |E\{[b(\X_{ij}^T\Beta_{ij})-Y\X_{ij}^T\Beta_{ij})-b(\X_{ij}^T\Beta_{ij}^M)+Y\X_{ij}^T\Beta_{ij}^M](1-I_n(\X_{ij},Y))\}| \\
	&\leq& |E\{b(\X_{ij}^T\Beta_{ij})I(|X_{ij}>K_n|)\}|+|E\{b(\X_{ij}^T\Beta_{ij}^M)I(|X_{ij}>K_n|)\}|\\
	&&+|E\{Y\X_{ij}^T\Beta_{ij}(1-I_n(\X_{ij},Y))\}|+|E\{Y\X_{ij}^T\Beta_{ij}^M(1-I_n(\X_{ij},Y))\}|
\end{eqnarray*}
By condition (E), the first two terms are of order $o(1/n)$. For the last two terms, using the Cauchy-Schwarz inequality and Lemma A.2  with $K_n^\star=m_0K_n^\alpha/s_0$,
\begin{eqnarray*}
	&& |E\{Y\X_{ij}^T\Beta_{ij}(1-I_n(\X_{ij},Y))\}| \\
	&\leq& (E|Y\X_{ij}^T\Beta_{ij}|^2)^{1/2}(E|1-I_n(\X_{ij},Y)|^2)^{1/2}  \\
	&\leq&C(P(1-I_n(\X_{ij},Y)))^{1/2}
	\leq C[P(|X_{ij}|>K_n)+P(|Y|>K_n^\star)]^{1/2}\\
	&\leq& C[m_1\exp(-m_0K_n^{\alpha/2})+s_1\exp(-m_0K_n^\alpha)]^{1/2}\\
	&\leq&C[(m_1+s_1)\exp(-m_0K_n^{\alpha/2})]^{1/2}
\end{eqnarray*}
When $n$ tends to infinity, the last two terms can be very small. In summary, the second part of  condition (B1) are satisfied. As a result, we have
$$P\left(\sqrt{n}\|\hat{\Beta}_{ij}^M-\Beta_{ij}^M\|\geq 16K_n(1+t)/V\right)\leq \exp(-2t^2/K_n^2)+nP(\Omega_n^c).$$
And then, using condition (D) and Lemma A.2  with $m_2=3m_1+s_1$, we have
\begin{eqnarray*}
	P(\Omega_n^c) &\leq& P(\|\X_{ij}\|_\infty>K_n)+P(|Y|>K_n^\star)\\
	&\leq& 3m_1\exp(-m_0K_n^{\alpha/2})+ s_1\exp(-m_0K_n^\alpha) \\
	&\leq& m_2\exp(-m_0K_n^{\alpha/2}).
\end{eqnarray*}
Next, by using the above inequalities and taking $1+t=c_5Vn^{1/2-\kappa}/(16k_n)$, it follows that
\begin{eqnarray*}
	&& P(|\hat{\beta}_{ij}^M-\beta_{ij}^M|\geq c_5n^{-\kappa}) \\
	&\leq& P(\|\hat{\Beta}_{ij}^M-\Beta_{ij}^M\|\geq c_5n^{-\kappa})\\
	&\leq& \exp(-c_6n^{1-2\kappa}/(k_nK_n)^2)+nm_2\exp(-m_0K_n^{\alpha/2})
\end{eqnarray*}
for some positive constant $c_6$. Consequently, by Bonferroni's inequality with $q=\frac{p(p-1)}{2}$, we have
$$P\left(\max_{1\leq i<j\leq p}|\hat{\beta}_{ij}^M-\beta_{ij}^M|\geq c_5n^{-\kappa}\right)
\leq q\left(\exp(-c_6n^{1-2\kappa}/(k_nK_n)^2)+nm_2\exp(-m_0K_n^{\alpha/2})\right).$$
(ii) By definition of $L_{ij,n}$ and $L_{ij}^\star$,
\begin{eqnarray*}
	L_{ij,n} &=& \mathbb{P}_n\{l(\hat{\beta}_{i,j0}^M+\hat{\beta}_{i,}^M X_i+\hat{\beta}_{j,}^M X_j,Y)-l(\hat{\beta}_{ij0}^M+\hat{\beta}_i^M X_i+\hat{\beta}_j^M X_j+\hat{\beta}_{ij}^M X_{ij},Y)\} \\
	&=& \mathbb{P}_n\{l(\X_{i,j}^T\hat{\Beta}_{i,j}^M,Y)-l(\X_{ij}^T\hat{\Beta}_{ij}^M,Y) \}
\end{eqnarray*}
and 
\begin{eqnarray*}
	L_{ij}^\star&=&E\{l(\beta_{i,j0}^M+\beta_{i,}^M X_i+\beta_{j,}^M X_j,Y)-l(\beta_{ij0}^M+\beta_i^M X_i+\beta_j^M X_j+\beta_{ij}^M X_{ij},Y)\}\\
	&=&E\{l(\X_{i,j}^T\Beta_{i,j}^M,Y)-l(\X_{ij}^T\Beta_{ij}^M,Y)\}
\end{eqnarray*}
Hence,
\begin{eqnarray*}
	&&|L_{ij,n} -L_{ij}^\star|\\
	&=& |\mathbb{P}_n\{l(\X_{i,j}^T\hat{\Beta}_{i,j}^M,Y)\}-E\{l(\X_{i,j}^T\Beta_{i,j}^M,Y)\}-\mathbb{P}_n\{l(\X_{ij}^T\hat{\Beta}_{ij}^M,Y)\}+E\{l(\X_{ij}^T\Beta_{ij}^M,Y)\}|\\
	&\leq& |\mathbb{P}_n\{l(\X_{i,j}^T\hat{\Beta}_{i,j}^M,Y)\}-E\{l(\X_{i,j}^T\Beta_{i,j}^M,Y)\}|+|\mathbb{P}_n\{l(\X_{ij}^T\hat{\Beta}_{ij}^M,Y)\}-E\{l(\X_{ij}^T\Beta_{ij}^M,Y)\}|\\
	&\triangleq& T_1+T_2
\end{eqnarray*}
Using Lemma A.3, we have
\begin{eqnarray*}
	&& P(|L_{ij,n} -L_{ij}^\star|\geq c_7n^{-2\kappa})\\
	&\leq&P(T_1\geq c_7n^{-2\kappa}) + P(T_2\geq c_7n^{-2\kappa})  \\
	&\leq&  2\exp(-c_6n^{1-2\kappa}/(k_nK_n)^2)+4\exp\left(-c_9n^{1-4\kappa}\right)+4nP(\Omega_n^c)\\
	&\leq&  2\exp(-c_6n^{1-2\kappa}/(k_nK_n)^2)+4\exp\left(-c_9n^{1-4\kappa}\right)+4nm_2\exp(-m_0K_n^{\alpha/2}).
\end{eqnarray*}
Consequently, by Bonferroni's inequality with $q=\frac{p(p-1)}{2}$ and $c_8=c_6$, we have
\begin{eqnarray*}
	&&P\left(\max_{1\leq i<j\leq p}|L_{ij,n} -L_{ij}^\star|\geq c_7n^{-2\kappa}\right)\\
	&\leq& q\left(2\exp(-c_8n^{1-2\kappa}/(k_nK_n)^2)+4\exp\left(-c_9n^{1-4\kappa}\right)+4nm_2\exp(-m_0K_n^{\alpha/2})\right).
\end{eqnarray*}
(iii) Define the event
$$A_n=\left\{\max_{(i,j)\in \mathcal{N}^\star}|L_{ij,n}-L_{ij}^\star|\leq c_4n^{-2\kappa}/2\right\}.$$
By Theorem A.4, 
we have
$\min\limits_{(i,j)\in \mathcal{N}_\star}|L_{ij}^\star|\geq c_4n^{-2\kappa}$, and then for all $(i,j)\in \mathcal{N}_\star$,
$$L_{ij,n}=|L_{ij,n}-L_{ij}^\star+L_{ij}^\star|\geq L_{ij}^\star-|L_{ij,n}-L_{ij}^\star|\geq c_4n^{-2\kappa}/2. $$
Taking $\gamma_n=c_{10}n^{-2\kappa}$ with $c_{10}\leq c_4/2$, we have $\mathcal{N}_\star\subset\widehat{\mathcal{N}}_{\gamma_n}$. Furthermore, $P(A_n)\leq P(\mathcal{N}_\star\subset\widehat{\mathcal{N}}_{\gamma_n})$. And then, by Theorem A.5(ii),
we have
$$P(A_n^c)
\leq s_n\left(2\exp(-c_8n^{1-2\kappa}/(k_nK_n)^2)+4\exp\left(-c_9n^{1-4\kappa}\right)+4nm_2\exp(-m_0K_n^{\alpha/2})\right).$$
Finally,
\begin{eqnarray*}
	&&P(\mathcal{N}_\star\subset\widehat{\mathcal{N}}_{\gamma_n})\\
	&\geq & 1-s_n\left(2\exp(-c_8n^{1-2\kappa}/(k_nK_n)^2)+4\exp\left(-c_9n^{1-4\kappa}\right)+4nm_2\exp(-m_0K_n^{\alpha/2})\right).
\end{eqnarray*}

\

\fl{\bf Proof of Theorem A.6:} 
The key idea of the proof is similar to that of Theorem 5 of Fan and Song (2010).
The idea of this proof is to show that
\begin{equation}\label{s5}
\|\Beta_{\mathcal{I}}^M\|^2=O(\lambda_{max}(\bfm \Sigma_{\mathcal{I}})).
\end{equation}
If so,  by definition,  we have
\begin{eqnarray*}
	0 \leq L_{ij}^\star &=& E\{l(\X_{i,j}^T\Beta_{i,j}^M,Y)-l(\X_{ij}^T\Beta_{ij}^M,Y)\} \\
	&\leq& E\{l(\beta_{ij0}^M+\beta_{i}^MX_i+\beta_{j}^MX_j,Y)-l(\X_{ij}^T\Beta_{ij}^M,Y)\}
\end{eqnarray*}
Using Taylor's expansion, for some $D_2>0$, we have
$$E\{l(\beta_{ij0}^M+\beta_{i}^MX_i+\beta_{j}^MX_j,Y)-l(\X_{ij}^T\Beta_{ij}^M,Y)\} \leq D_2 (\beta_{ij}^M)^2.$$
As a result, with vector form, we have
$$\|\L^\star\|\leq O(\|\Beta_{\mathcal{I}}^M\|^2)=O(\lambda_{max}(\bfm \Sigma_{\mathcal{I}})).$$
Therefore, for any $\varepsilon>0$, the number of $\{(i,j): L_{ij}^\star >\varepsilon n^{-2\kappa} , 1\leq i<j \leq p\}$ cannot exceed $O(\lambda_{max}(\Sigma_{\mathcal{I}}))$. Thus, on the set
$$B_n=\left\{\max_{1\leq i<j \leq p}|L_{ij,n}-L_{ij}^\star|\leq \varepsilon n^{-2\kappa}\right\},$$
the number of $\{(i,j): L_{ij,n} >2\varepsilon n^{-2\kappa} , 1\leq i<j \leq p\}$ cannot exceed the number of $\{(i,j): L_{ij}^\star >\varepsilon n^{-2\kappa} , 1\leq i<j \leq p\}$, which is bounded by $O(n^{2\kappa}\lambda_{max}(\Sigma_{\mathcal{I}}))$. By taking $\varepsilon=c_7/2$, we have
$$P\left(|\widehat{\mathcal{N}}_{\gamma_n}|\leq O(n^{2\kappa}\lambda_{max})\right)\geq P(B_n).$$
Consequently, the conclusion can follow from Theorem A.5(ii).

Now we prove the equation (\ref{s5}). By condition (B) and the proof of Theorem A.3, 
$A_{22}-A_{21}A_{11}^{-1}A_{12}$ is uniformly bounded from below, we have
$$|\beta_{ij}^M|\leq D_3|\text{Cov}_L(Y,X_{ij}|\X_{i,j}^T\Beta_{i,j}^M)|$$
for a positive constant $D_3$. Using the  Lipschitz continuity of $b'(\cdot)$, we have
\begin{eqnarray*}
	|\beta_{ij}^M|&\leq& D_3|\text{Cov}_L(Y,X_{ij}|\X_{i,j}^T\Beta_{i,j}^M)|\\
	&= & D_3|E(b'(\X^T\Beta^\star)-b'(\X_{i,j}^{T}\Beta_{i,j}^{M}))X_{ij}|\\
	&\leq& D_4 |EX_{ij}(\X^T\Beta^\star-\X_{i,j}^{T}\Beta_{i,j}^{M})|\\
	&=& D_4 |[EX_{ij}(\X_{\mathcal{I}}^T\Beta_{\mathcal{I}}^\star+\X_{\mathcal{C}}^T\Delta\Beta_{ij})]|
\end{eqnarray*}
for some constant $D_4>0$, where $\Beta_{ij-}^M=(\beta_{i,j0}^M,0,\ldots,0,\beta_{i,}^M,0,\ldots,0,\beta_{j,}^M,0,\ldots,0)^T$, $\Delta\Beta_{ij}=\Beta_{\mathcal{C}}^\star-\Beta_{ij-}^M$. Let $R_{ij}=E[X_{ij}\X_{\mathcal{C}}^T\Delta\Beta_{ij}]$ and $\R=(R_{12},R_{13},\ldots,R_{(p-1)p})^T$. Therefore,
$$|\beta_{ij}^M|^2\leq D_4^2|E[X_{ij}\X_{\mathcal{I}}^T\Beta_{\mathcal{I}}^\star]+R_{ij}|^2$$
and
\begin{eqnarray*}
	\|\Beta_{\mathcal{I}}^M\|^2 &\leq& D_4^2D_5^2\|E[\X_{\mathcal{I}}\X_{\mathcal{I}}^T\Beta_{\mathcal{I}}^\star]+\R\|^2.
\end{eqnarray*}
Now
\begin{eqnarray*}
	&& \|E[\X_{\mathcal{I}}\X_{\mathcal{I}}^T\Beta_{\mathcal{I}}^\star]+\R\|^2 = \|\bfm \Sigma_{\mathcal{I}}\Beta_{\mathcal{I}}^\star+\R\|^2 \\
	&=& {\Beta_{\mathcal{I}}^\star}^T \bfm \Sigma_{\mathcal{I}}^2\Beta_{\mathcal{I}}^\star+2\R^T\bfm\Sigma_{\mathcal{I}}\Beta_{\mathcal{I}}^\star+\R^T\R\\
	&\leq& \lambda_{\max}(\bfm \Sigma_{\mathcal{I}}){\Beta_{\mathcal{I}}^\star}^T \bfm\Sigma_{\mathcal{I}}\Beta_{\mathcal{I}}^\star+2\R^T\bfm\Sigma_{\mathcal{I}}\Beta_{\mathcal{I}}^\star+\R^T\R\\
	&\leq& \lambda_{\max}(\bfm \Sigma_{\mathcal{I}})Var(\X^T\Beta^\star)+2\R^T\bfm\Sigma_{\mathcal{I}}\Beta_{\mathcal{I}}^\star+\R^T\R.
\end{eqnarray*}
Since $Var(\X^T\Beta^\star)=O(1)$, and by Condition (H), we have $\|\Beta_{\mathcal{I}}^M\|^2=O(\lambda_{max}(\bfm \Sigma_{\mathcal{I}})).$

\

\noindent{\bf Proof of Theorem 4.1:} 
(1)The idea is similar to the proof of Theorem 1 of Li et al. (2012). 
Denote that $X_k^\star=\Phi^{-1}[F_k^X(X_k)]$ and $Y^\star=\Phi^{-1}[F^Y(Y)]$, where $F_k^X$ and $F^Y$ are the cumulative  distribution functions  of $X_k$ and $Y$, respectively; $\Phi^{-1}$ is the standard normal distribution function.\\
The condition is equivalent to $$|\text{Cov}(Y,\ X_k)|\geq C_1 n^{-\kappa}.$$
Note that $\text{Cov}(b'(\X^T\Beta^\star),\ X_k)=E(b'(\X^T\Beta^\star)X_k)=E(E(Y|\X)X_k)=E(YX_k)=\text{Cov}(Y,X_k).$
Since $Y$ and $X_k$ are standardized, we have $|\rho_k|\geq C_1 n^{-\kappa}$. \\
Firstly, we consider the special case $l=m=2$ and then $\widetilde{Y}=I(Y>M_d(Y))$ and $\widetilde{X}_{k}=\widetilde{X}_{k_2}=I(X_{k}>M_d(X_k))$. We only need to prove that $|\text{Cov}(\widetilde{Y},\ \widetilde{X}_{k})|\geq C_2 n^{-\kappa}$ for some positive constant $C_2$. \\
Furthermore, assume that $\rho_k\geq C_1 n^{-\kappa}$ and let $X_{1k}^\star=\Phi^{-1}[F_k^X(X_{1k})]$, $X_{2k}^\star=\Phi^{-1}[F_k^X(X_{2k})]$ and $Y_1^\star=\Phi^{-1}[F^Y(Y_1)]$, $Y_2^\star=\Phi^{-1}[F^Y(Y_2)]$, thus, $\frac{1}{\sqrt{2}}(X_{2k}^\star-X_{1k}^\star)$ and $\frac{1}{\sqrt{2}}(Y_{2}^\star-Y_{1}^\star)$ follow the standard normal distribution. Consequently,
\begin{eqnarray*}
	\text{Cov}(\widetilde{Y},\ \widetilde{X}_{k}) &=& \text{Cov}(I(Y>M_d(Y),I(X_{k}>M_d(X_k))) \\
	&=&  \text{Cov}(I(Y^\star>0),I(X_{k}^\star>0))\\
	&=& E(I(Y^\star>0)I(X_{k}^\star>0))-\frac{1}{4}\\
	&=& E\left\{I\left(\frac{1}{\sqrt{2}}(Y_{2}^\star-Y_{1}^\star)>0\right)
	I\left(\frac{1}{\sqrt{2}}(X_{2k}^\star-X_{1k}^\star)>0\right)\right\}-\frac{1}{4}\\
	&=&E\left\{I(X_{2k}^\star>X_{1k}^\star)I(Y_{2}^\star>Y_{1}^\star)\right\}-\frac{1}{4}
\end{eqnarray*}
Since the function $\Phi^{-1}\cdot F_k^X$ and $\Phi^{-1}\cdot F^Y$ are two increasing functions, their inverse functions are also increasing. Therefore, we have
\begin{eqnarray*}
	\text{Cov}(\widetilde{Y},\ \widetilde{X}_{k}) &=& E\left\{I(X_{2k}>X_{1k})I(Y_{2}>Y_{1})\right\}-\frac{1}{4} \\
	&=& E\left\{I(X_{2k}-X_{1k}>0)I(Y_{1}-Y_{2}<0)\right\}-\frac{1}{4} \\
	&=& E\left\{I(X_{2k}-X_{1k}>0)I(\Delta \varepsilon_k<\rho_k(X_{1k}-X_{2k}))\right\}-\frac{1}{4}
\end{eqnarray*}
Taking into account the symmetry of $f_{\Delta\varepsilon_k|\Delta X_k}(t)$,
$$1-F_{\Delta\varepsilon_k|\Delta X_k}(-t)=F_{\Delta\varepsilon_k|\Delta X_k}(t)$$
and $$F_{\Delta\varepsilon_k|\Delta X_k}(0)=\frac{1}{2},$$
where $F_{\Delta\varepsilon_k|\Delta X_k}(\cdot)$ is the cumulative distribution function of $\Delta \varepsilon_k$ given $\Delta X_k$.
Hence,
\begin{eqnarray*}
	&&\text{Cov}(\widetilde{Y},\ \widetilde{X}_{k}) \\
	&=& E\left\{I(X_{2k}>X_{1k})F_{\Delta\varepsilon_k|\Delta X_k}(\rho_k(X_{1k}-X_{2k}))\right\}-E\left\{I(X_{2k}>X_{1k})\right\}F\left\{\Delta\varepsilon_k<0|\Delta X_k\right\} \\
	&=& E\left\{I(X_{2k}-X_{1k}>0)[F_{\Delta\varepsilon_k|\Delta X_k}(\rho_k(X_{2k}-X_{1k}))-F_{\Delta\varepsilon_k|\Delta X_k}(0)]\right\} \\
	&=& E\left\{I(X_{2k}-X_{1k}>0)\int_{0}^{\rho_k(X_{2k}-X_{1k})}f_{\Delta\varepsilon_k|\Delta X_k}(t)dt\right\}
\end{eqnarray*}
According to Condition (M1),
\begin{eqnarray*}
	\text{Cov}(\widetilde{Y},\ \widetilde{X}_{k})&=& E\bigg\{I(X_{2k}-X_{1k}>0)\\
	&\times&\int_{0}^{\rho_k(X_{2k}-X_{1k})}\left[\pi_{0k}f_0(t,\sigma_0^2|\Delta X_k)+(1-\pi_{0k})f_1(t,\sigma_1^2|\Delta X_k)\right]dt\bigg\}\\
	&\geq& \pi_{0k}E\left\{I(X_{2k}-X_{1k}>0)\int_{0}^{\rho_k(X_{2k}-X_{1k})}f_0(t,\sigma_0^2|\Delta X_k)dt\right\}.
\end{eqnarray*}
By the Gaussian inequality for the symmetric unimodal distribution (See Pukelshemim (1994), and Sellke (1997)),
$$P(|X|\geq k\sigma)\leq \left\{\begin{array}{cc}
1-\frac{k}{\sqrt{3}}, & k\leq \frac{2}{\sqrt{3}},  \\
\frac{4}{9K^2}, &   k\leq \frac{2}{\sqrt{3}},
\end{array}
\right.$$
therefore,
$$P(|X|\geq k\sigma)\leq \frac{1}{1+k/\sqrt{3}},$$
where $X$ is a unimodal random variable with a mode at the origin zero and variance $\sigma^2$. Using this Gaussian inequality, we have
\begin{eqnarray*}
	\int_{0}^{\rho_k(X_{2k}-X_{1k})}f_0(t,\sigma_0^2|\Delta X_k)dt &=&\left\{\int_{0}^{\infty}-\int_{\rho_k(X_{2k}-X_{1k})}^{\infty}\right\}f_0(t,\sigma_0^2|\Delta X_k)dt  \\
	&=&\frac{1}{2}-P(\Delta \varepsilon_k > \rho_k(X_{2k}-X_{1k}))\\
	&\geq& \frac{1}{2}-\frac{1}{2}\frac{1}{1+\frac{\rho_k(X_{2k}-X_{1k})}{\sqrt{3\sigma_0^2}}}\\
	&=& \frac{\rho_k(X_{2k}-X_{1k})}{\sqrt{12\sigma_0^2}+2\rho_k(X_{2k}-X_{1k})}.
\end{eqnarray*}
Since
$$Var(\Delta \varepsilon_k |\Delta X_k)=\pi_{0k}\sigma_0^2+(1-\pi_{0k})\sigma_1^2\geq \pi_{0k}\sigma_0^2\geq \pi^\star\sigma_0^2,$$
we have
$$ \int_{0}^{\rho_k(X_{2k}-X_{1k})}f_0(t,\sigma_0^2|\Delta X_k)dt\geq \frac{\rho_k(X_{2k}-X_{1k})}{\sqrt{12Var(\Delta \varepsilon_k |\Delta X_k)/\pi^\star}+2\rho_k(X_{2k}-X_{1k})}.$$
Define the variable $Z_k=\sqrt{Var(\Delta \varepsilon_k |\Delta X_k)}=\sqrt{Var(\Delta Y-\rho_k\Delta X_k|\Delta X_k)}$ and note that $Var(\Delta \varepsilon_k)=Var(\Delta Y-\rho_k\Delta X_k)=2(1-\rho_k^2)$. By Condition (M1),  $$E(\Delta \varepsilon_k |\Delta X_k)=0,$$ and then by the law of total variance,
$$Var(\Delta \varepsilon_k)=E(Var(\Delta \varepsilon_k |\Delta X_k))+Var(E(\Delta \varepsilon_k |\Delta X_k))=E(Var(\Delta \varepsilon_k |\Delta X_k)).$$
Hence, for a given large positive constant $T$, by Markov inequality,
$$P(Z_k>T)\leq \frac{E(Z_k^2)}{T^2}=\frac{E(Var(\Delta \varepsilon_k |\Delta X_k))}{T^2}=\frac{Var(\Delta \varepsilon_k)}{T^2}\leq \frac{2}{T^2},$$
that is, $$P(Var(\Delta \varepsilon_k |\Delta X_k)>T^2)\leq \frac{2}{T^2},$$
which means that with at least probability $1-\frac{2}{T^2}$, we have
\begin{eqnarray*}
	\int_{0}^{\rho_k(X_{2k}-X_{1k})}f_0(t,\sigma_0^2|\Delta X_k)dt &\geq& \frac{\rho_k(X_{2k}-X_{1k})}{\sqrt{12T^2/\pi^\star}+2\rho_k(X_{2k}-X_{1k})} \\
	&\geq& \frac{\rho_k(X_{2k}-X_{1k})}{4T/\sqrt{\pi^\star}+2\rho_k(X_{2k}-X_{1k})}.
\end{eqnarray*}
Consequently, by $\pi_{0k}\geq\pi^\star$,
\begin{eqnarray*}
	\text{Cov}(\widetilde{Y},\ \widetilde{X}_{k}) &\geq& \pi^\star E\left\{I(X_{2k}-X_{1k}>0)\frac{\rho_k(X_{2k}-X_{1k})}{4T/\sqrt{\pi^\star}+2\rho_k(X_{2k}-X_{1k})}\right\}I(Z_k\leq T) \\
	&=&\pi^\star E\left\{I(X_{2k}-X_{1k}>0)\frac{\rho_k(X_{2k}-X_{1k})}{4T/\sqrt{\pi^\star}+2\rho_k(X_{2k}-X_{1k})}\right\}\\
	&&-\pi^\star E\left\{I(X_{2k}-X_{1k}>0)\frac{\rho_k(X_{2k}-X_{1k})}{4T/\sqrt{\pi^\star}+2\rho_k(X_{2k}-X_{1k})}\right\}I(Z_k>T)\\
	&\triangleq& I_1+I_2.
\end{eqnarray*}
For the term $I_1$,
\begin{eqnarray*}
	I_1 &\geq& \pi^\star E\left\{I(T/\rho_k>X_{2k}-X_{1k}>0)\frac{\rho_k(X_{2k}-X_{1k})}{4T/\sqrt{\pi^\star}+2\rho_k(X_{2k}-X_{1k})}\right\}  \\
	&\geq& \frac{\pi^\star \rho_k}{4T/\sqrt{\pi^\star}+2T} E\left\{(X_{2k}-X_{1k})I(T/\rho_k>X_{2k}-X_{1k}>0)\right\}\\
	&\geq& \frac{\pi^\star \rho_k}{4T/\sqrt{\pi^\star}+2T}\\
	&&\times E\left\{(X_{2k}-X_{1k})I(X_{2k}-X_{1k}>0)-(X_{2k}-X_{1k})I(X_{2k}-X_{1k}>T)\right\}
\end{eqnarray*}
Using the inequality $E|X-Y|\geq E|X|$, where  $X$ and $Y$ are i.i.d. random variables with $E(X)=E(Y)=0$, and by Condition (M2), we have
$$E\left\{|X_{2k}-X_{1k}|\right\}\geq E|X_{1k}|\geq c_{\mathcal{M}_\star},$$
and then, by the symmetry property of the distribution of $X_{2k}-X_{1k}$, $$E\left\{(X_{2k}-X_{1k})I(X_{2k}-X_{1k}>0)\right\}\geq \frac{1}{2}c_{\mathcal{M}_\star}.$$
On the other hand, according to Cauchy-Schwarz inequality,
$$E\left\{(X_{2k}-X_{1k})I(X_{2k}-X_{1k}>T)\right\}\leq \sqrt{P(X_{2k}-X_{1k}>T)E(X_{2k}-X_{1k})^2}\leq\frac{2}{T}.$$
Consequently,  $$I_1\geq \frac{\pi^\star \rho_k c_{\mathcal{M}_\star} }{8T/\sqrt{\pi^\star}+4T}-\frac{2\pi^\star\rho_k}{4T^2/\sqrt{\pi^\star}+2T^2}=\frac{\pi^\star \rho_k c_{\mathcal{M}_\star} }{8T/\sqrt{\pi^\star}+4T}-\frac{\pi^\star\rho_k}{2T^2/\sqrt{\pi^\star}+T^2}.$$
As for the term $I_2$, using Cauchy-Schwarz inequality again,
\begin{eqnarray*}
	I_2 &=& -\pi^\star E\left\{I(X_{2k}-X_{1k}>0)\frac{\rho_k(X_{2k}-X_{1k})}{4T/\sqrt{\pi^\star}+2\rho_k(X_{2k}-X_{1k})}\right\}I(Z_k>T)\\
	&\geq& -\frac{\pi^\star\rho_k }{4T/\sqrt{\pi^\star}}E\left\{(X_{2k}-X_{1k})I(X_{2k}-X_{1k}>0)\right\}I(Z_k>T)\\
	&\geq& -\frac{\pi^\star\rho_k }{4T/\sqrt{\pi^\star}}\sqrt{E\left\{(X_{2k}-X_{1k})^2I(X_{2k}-X_{1k}>0)\right\}}\sqrt{P(Z_k>T)}\\
	&\geq& -\frac{\pi^\star\rho_k }{4T/\sqrt{\pi^\star}}\cdot \sqrt{2}\cdot \frac{\sqrt{2}}{T}=-\frac{\pi^\star\rho_k }{2T^2/\sqrt{\pi^\star}}.
\end{eqnarray*}
Combing the above two inequalities for the terms $I_1$ and $I_2$,
\begin{eqnarray*}
	\text{Cov}(\widetilde{Y},\ \widetilde{X}_{k}) \geq I_1 + I_2 &\geq& \frac{\pi^\star \rho_k c_{\mathcal{M}_\star} }{8T/\sqrt{\pi^\star}+4T}-\frac{\pi^\star\rho_k}{2T^2/\sqrt{\pi^\star}+T^2}-\frac{\pi^\star\rho_k }{2T^2/\sqrt{\pi^\star}} \\
	&\geq&\frac{(\pi^\star)^2 \rho_k c_{\mathcal{M}_\star} }{12T}-\frac{5\pi^\star\rho_k}{6T^2}.
\end{eqnarray*}
Taking the large positive value $T=\frac{15}{c_{\mathcal{M}_\star}\pi^\star}$,  it follows that
$$\text{Cov}(\widetilde{Y},\ \widetilde{X}_{k}) \geq \left[\frac{(\pi^\star)^2 c_{\mathcal{M}_\star}}{12}\frac{c_{\mathcal{M}_\star}\pi^\star}{15}-
\frac{5\pi^\star}{6}\frac{(c_{\mathcal{M}_\star}\pi^\star)^2}{15^2}\right]\rho_k
=\frac{(\pi^\star)^3c_{\mathcal{M}_\star}^2}{540}\rho_k\geq C_1'n^{-\kappa},$$
where the positive constant $C_1'=C_1(\pi^\star)^3c_{\mathcal{M}_\star}^2/540$.\\
If $\rho_k\leq -C_1 n^{-\kappa}$, by the similar steps as above, we also have $\text{Cov}(\widetilde{Y},\ \widetilde{X}_{k}) \leq -C_1'n^{-\kappa}$.\\
In summary, if  the condition (M1)-(M3) hold and $|\text{Cov}(b'(\X^T\Beta^\star),\ X_k)|\geq C_1 n^{-\kappa}$ for any $k\in \mathcal{M}_\star$ with a positive constant $C_1$, and after discretizing the response and predictor, there exists a positive constant $C_2=C_1'$ such that  $|\text{Cov}(\widetilde{Y},\ \widetilde{X}_{k_2})| \geq C_2n^{-\kappa}$ in the special case $l=m=2$.
Furthermore, following the above same steps, we also have that  $|\text{Cov}(\widetilde{Y},\ \widetilde{X}_{k_1})| \geq C_2n^{-\kappa}$ when $l=m=2$. \\
\indent In the following, we will consider the general case: $m=2$ and $l\geq 3$. By the above proof for the special case $l=2$, if we divide the predictor into two parts, we have shown that for some positive constant $C_1'$,
$$|\text{Cov}(I(Y>M_d(Y), I(X>M_d(X_k))| \geq C_1'n^{-\kappa},$$
and $$|\text{Cov}(I(Y>M_d(Y), I(X< M_d(X_k))| \geq C_1'n^{-\kappa}.$$
As for case $l>2$, when $l$ is a even number,
$$I(X_k>M_d(X_k))=\bigcup_{i=\frac{l}{2}+1}^{l}I\left(X_k\in P_i^{X_k}\right),$$
and
$$I(X_k<M_d(X_k))=\bigcup_{i=1}^{\frac{l}{2}}I\left(X_k\in P_i^{X_k}\right).$$
Hence, \begin{eqnarray*}
	\text{Cov}(I(Y>M_d(Y), I(X>M_d(X_k)) &=& \text{Cov}\left(I(Y>M_d(Y),\bigcup_{i=\frac{l}{2}+1}^{l}I\left(X_k\in P_i^{X_k}\right)\right) \\
	&= & \sum_{i=\frac{l}{2}+1}^{l} \text{Cov}\left(I(Y>M_d(Y),\ I\left(X_k\in P_i^{X_k}\right)\right)
\end{eqnarray*}
and
\begin{eqnarray*}
	\text{Cov}(I(Y>M_d(Y), I(X<M_d(X_k)) &=& \text{Cov}\left(I(Y>M_d(Y),\bigcup_{i=1}^{\frac{l}{2}}I\left(X_k\in P_i^{X_k}\right)\right) \\
	&= & \sum_{i=1}^{\frac{l}{2}} \text{Cov}\left(I(Y>M_d(Y),\ I\left(X_k\in P_i^{X_k}\right)\right)
\end{eqnarray*}

which means that there exists at least one term   $i=1,\ldots, \frac{l}{2}$  or $i=\frac{l}{2}+1,\ldots, l $,  such that $$\left|\text{Cov}\left(I(Y>M_d(Y),\ I\left(X_k\in P_i^{X_k}\right)\right)\right|\geq C_2n^{-\kappa},$$
where $C_2=\frac{2}{l}C_1'$ is a positive constant number, that is,
$$|\text{Cov}(\widetilde{Y},\ \widetilde{X}_{k_i})|\geq C_2 n^{-\kappa}.$$
When $l$ is odd, the support set of $X_k$ is divided into $l$ parts and denote that $\{Q_i\}_{i=1}^{l-1}$ are a series of cutting points ($l-$quantiles), and then $P_1^{X_k}=(-\infty,\ Q_1)$, $P_l^{X_k}=[Q_{l-1},\ \infty)$, $P_i^{X_k}=[Q_i,\ Q_{i+1})$ , for $1<i<l$. Therefore,
\begin{eqnarray*}
	\{X_k>M_d(X_k)\} &=& \left[M_d(X_k),Q_{\frac{l+1}{2}}\right)\ \bigcup\ \left[Q_{\frac{l+1}{2}}, Q_{\frac{l+3}{2}}\right)\ \bigcup \cdots \bigcup\ \left[Q_{l-1},\ \infty\right) \\
	&=&\left[M_d(X_k),Q_{\frac{l+1}{2}}\right)\ \bigcup\  \bigcup_{i=\frac{l+3}{2}}^{l}\left(X_k\in P_i^{X_k}\right),
\end{eqnarray*}
and
\begin{eqnarray*}
	\{X_k<M_d(X_k)\} &=& (-\infty,\ Q_1)\ \bigcup\cdots \bigcup\  \left[Q_{\frac{l-3}{2}}, Q_{\frac{l-1}{2}}\right)\ \bigcup \left[Q_{\frac{l-1}{2}},\ M_d(X_k)\right) \\
	&=&\bigcup_{i=1}^{\frac{l-1}{2}}\left(X_k\in P_i^{X_k}\right)\ \bigcup\ \left[Q_{\frac{l-1}{2}},\ M_d(X_k)\right).
\end{eqnarray*}
Based on two results of the case $l=m=2$,  we conclude that two cases will happen.\\
Case (i): There exists at least one term $i=1,\ldots, \frac{l-1}{2}$  or $i=\frac{l+3}{2},\ldots, l $,  such that $$\left|\text{Cov}\left(I(Y>M_d(Y),\ I\left(X_k\in P_i^{X_k}\right)\right)\right|\geq \frac{2}{l+1}C_1'n^{-\kappa}.$$
Take $C_2=\frac{2}{l+1}C_1'$, our proof will be completed.\\
Case (ii): For all $i=1,\ldots, \frac{l-1}{2}$  and $i=\frac{l+3}{2},\ldots, l $, we have
$$\left|\text{Cov}\left(I(Y>M_d(Y),\ I\left(X_k\in P_i^{X_k}\right)\right)\right|< \frac{2}{l+1}C_1'n^{-\kappa};$$
but
$$\left|\text{Cov}\left\{I(Y>M_d(Y),\ I\left(X_k\in \left[M_d(X_k),Q_{\frac{l+1}{2}}\right)\right)\right\}\right|\geq \frac{2}{l+1}C_1'n^{-\kappa}$$
and
$$\left|\text{Cov}\left\{I(Y>M_d(Y),\ I\left(X_k\in \left[Q_{\frac{l-1}{2}},\ M_d(X_k)\right)\right)\right\}\right|\geq \frac{2}{l+1}C_1'n^{-\kappa}.$$
In this case, $P_{\frac{l+1}{2}}^{X_k}=\left[Q_{\frac{l-1}{2}},\ M_d(X_k)\right)\bigcup \left[M_d(X_k),Q_{\frac{l+1}{2}}\right)$, and
\begin{eqnarray*}
	&&\text{Cov}\left(I(Y>M_d(Y),\ I\left(X_k\in P_{\frac{l+1}{2}}^{X_k}\right)\right)  \\
	&=& \text{Cov}\left\{I(Y>M_d(Y),\ I\left(X_k\in \left[M_d(X_k),Q_{\frac{l+1}{2}}\right)\right)\right\}\\
	&+&\text{Cov}\left\{I(Y>M_d(Y),\ I\left(X_k\in \left[Q_{\frac{l-1}{2}},\ M_d(X_k)\right)\right)\right\}
\end{eqnarray*}
It follows that $$\left|\text{Cov}\left(I(Y>M_d(Y),\ I\left(X_k\in P_{\frac{l+1}{2}}^{X_k}\right)\right)\right|> \frac{4}{l+2}C_1'n^{-\kappa}.$$
If the condition (M1)-(M2) hold and $|\text{Cov}(b'(\X^T\Beta^\star),\ X_k)|\geq C_1 n^{-\kappa}$ for any $k\in \mathcal{M}_\star$ with a positive constant $C_1$, after using 2-quantile and $l-$quantiles to discretize the response $Y$ and the predictor $X_k$, there exists at least one $\widetilde{X}_{k_i}$  such that  $|\text{Cov}(\widetilde{Y},\ \widetilde{X}_{k_i})| \geq C_2n^{-\kappa}$ for some positive constant $C_2$, which is dependent on $l$.

\

(2) Assume that $\widetilde{X}_{k_i}$ satisfies $|\text{Cov}(\widetilde{Y},\ \widetilde{X}_{k_i})| \geq C_2n^{-\kappa}$ for some positive constant $C_2$ and $\widetilde{\X}_{k_i}=(1,\widetilde{X}_{k_i})^T$. The coefficient $\widetilde{\Beta}_{k_i}^M$ is defined as the minimizer of the componentwise regression
$$\widetilde{\Beta}_{k_i}^M=(\widetilde{\beta}_{k_i,0}^M,\ \widetilde{\beta}_{k_i}^M)=\argmin_{\widetilde{\beta}_{0},\ \widetilde{\beta}_{k_i}}El(\widetilde{\beta}_{0}+\widetilde{\beta}_{k_i}\widetilde{X}_{k_i},\ \widetilde{Y}).$$
Define $\widetilde{\mathcal{M}}_\star=\{1\leq j \leq \tilde{p},\ \widetilde{\Beta}_j^\star\neq0\}$, where $$\widetilde{\Beta}^\star=(\widetilde{\Beta}_0^\star,\ \widetilde{\Beta}_1^\star,\ \ldots,\ \widetilde{\Beta}_{\tilde{p}}^\star).$$
Consider the new categorical response $\widetilde{Y}$ and predictor $\widetilde{\X}=\{1,\ \widetilde{X}_1,\ \widetilde{X}_2,\ \ldots,\ \widetilde{X}_{\tilde{p}}\}$, we have
$$\left|\text{Cov}(b'(\widetilde{\X}^T\widetilde{\Beta}^\star),\ \widetilde{X}_{k_i})\right|=\left|\text{Cov}(\widetilde{Y},\ \widetilde{X}_{k_i})\right| \geq C_2n^{-\kappa}.$$
By Theorem 3 in Fan and Song (2010), we have $\left|\widetilde{\beta}_{k_i}\right|\geq C_2'n^{-\kappa}$ for some positive constant $C_2'$, and
\begin{eqnarray*}
	\widetilde{L}_k^\star &=& E\left\{l(\widetilde{\beta}_0^M,\widetilde{Y})-l(\widetilde{\X}_k^T\widetilde{\Beta}_k^M,\widetilde{Y})\right\} \\
	&\geq& E\left\{l(\widetilde{\beta}_0^M,\widetilde{Y})-l(\widetilde{\X}_{k_i}^T\widetilde{\Beta}_{k_i}^M,\widetilde{Y})\right\}\\
	&\geq& V \left|\widetilde{\beta}_{k_i}\right|^2\geq C_3n^{-2\kappa}
\end{eqnarray*}
where $V$ is some positive constant and $C_3=V(C_2')^2$.

\

\noindent{\bf Proof of Theorem 4.2:}
$$\text{Cov}_L(Y,X_{ij}|\X_{i,j}^T\Beta_{i,j}^M)=E\{(Y-b'(\X_{i,j}^T\Beta_{i,j}^M))X_{ij}\}=\text{Cov}(\zeta_{ij},X_{ij}).$$
After discretizing $Y$, $X_i$ and $X_j$, that is, $Y=\sum_{k=1}^{2}Y I(Y\in P_k^Y)$, $X_i=\sum_{s=1}^{l_1}X_i I(X_i\in P_{s}^{X_i})$ and
$X_t=\sum_{t=1}^{l_2}X_j I(X_j\in P_{t}^{X_j})$, $X_{ij}$ is transformed into
$$X_{ij}=\sum_{s,t}X_{ij}I\left(\left\{X_i\in P_s^{X_i}\right\}\ \bigcap\ \left\{X_j\in
P_t^{X_j}\right\}\right),\ \ \ 1\leq s\leq l_1,\ \ 1\leq t\leq l_2.$$
Hence, the support set of $\zeta_{ij}$  becomes the union of several intervals.
Suppose that $\zeta_{ij}=\sum_{k'}\zeta_{ij}I(\zeta_{ij}\in \Omega_{k'} )$, where $1\leq k'\leq 2l_1l_2$.
By taking $\zeta_{ij}$ as the  response $Y$ of Theorem 4.1 
and $X_{ij}$ as
the predictor $X_k$ in Theorem 4.1,
there exists at least one term such that
$$\Big|\text{Cov}\Big(I(\zeta_{ij}\in \Omega_{k'}),I\left(\left\{X_i\in P_s^{X_i}\right\}\ \bigcap\ \left\{X_j\in P_t^{X_j}\right\}\right)\Big)\Big|\geq a_1
n^{-\kappa},$$
for some positive constant $a_1$, where $a_1$ is related to $l_1$ and $l_2$. Therefore,
$$|\text{Cov}(\widetilde{Y}-b'(\widetilde{\X}_{i,j}^T\widetilde{\Beta}_{i,j}^M),\widetilde{X}_{st}^{ij})|\geq a_2 n^{-\kappa}.$$
By taking $c_{10}=a_2$,
$$|\text{Cov}_L(\widetilde{Y},\ \widetilde{X}_{st}^{ij}|\widetilde{\X}_{i,j}^T\widetilde{\Beta}_{i,j}^M)|\geq c_{10} n^{-\kappa}.$$
(2) By the proof of Theorem A.3 and A.4, we directly have the conclusion.

\subsection{Simulation Study}
In this section,  we   provide some  simulation results as the supplement to  our paper. Table \ref{simut1} provides the screening results with $p=2000$ in the linear models by using different methods.  Table \ref{post1}-\ref{post2} compare the post-screening performance of diffenent methods  with realtively small $p$,  where ``w'' stands for weak heredity. In the setting with a small $p$, it is clear that our methods SSI and BOLT-SSI outperform other methods, with respect to the coverage rate, the out-of-sample $R^2$ and the predictive misclassification rate for most examples.
Sometimes, the method hierNet has a perfect coverage rate and the out-of-sample $R^2$, but its average model size is much larger than the model size of our methods, especially in the linear models.

\begin{table}[!htbp]
	\caption{ {\footnotesize Screening results for Linear Models when $p=2000$}}\label{simut1}
	\centering
	\smallskip{\scriptsize \tabcolsep= 3 pt
		
		\begin{tabular}{ccccccccc}
			\hline\hline
			Methods &      $\sigma$ &     SSI &    BOLT-SSI &     BOLT-SSI(p) &    IP &     xyz-L10 &     xyz-L100&    xyz-L1000  \\
			\hline \hline
			\multicolumn{ 9}{c}{($n,p,\rho$)=(400, 2000, 0)} \\
			\hline
			& 2 & 0.96 & 0.07 & 0.56 & 0.76 & 0.00 & 0.29 & 0.85 \\
			Example 1 & 3 & 0.92 & 0.02 & 0.48 & 0.70 & 0.02 & 0.21 & 0.84 \\
			& 4 & 0.81 & 0.03 & 0.45 & 0.64 & 0.00 & 0.14 & 0.71 \\
			\hline
			\multicolumn{9}{c}{($n,p,\rho$)=(400, 2000, 0.5)} \\
			\hline
			& 2 & 1.00 & 0.60 & 0.95 & 1.00 & 0.51 & 0.61 & 0.61 \\
			Example 1  & 3 & 1.00 & 0.56 & 0.91 & 1.00 & 0.48 & 0.61 & 0.61 \\
			& 4 & 1.00 & 0.36 & 0.82 & 1.00 & 0.43 & 0.60 & 0.60 \\
			\hline \hline
			\multicolumn{ 9}{c}{($n,p,\rho$)=(400, 2000, 0)} \\
			\hline
			& 2 & 0.90 & 0.03 & 0.45 & 0.08 & 0.01 & 0.11 & 0.69 \\
			Example 2 & 3 & 0.78 & 0.00 & 0.38 & 0.06 & 0.00 & 0.06 & 0.64 \\
			& 4 & 0.63 & 0.00 & 0.41 & 0.05 & 0.00 & 0.05 & 0.48 \\
			\hline
			\multicolumn{9}{c}{($n,p,\rho$)=(400, 2000, 0.5)} \\
			\hline
			& 2 & 0.77 & 0.04 & 0.58 & 0.11 & 0.01 & 0.01 & 0.01 \\
			Example 2  & 3 & 0.71 & 0.03 & 0.49 & 0.08 & 0.01 & 0.01 & 0.01 \\
			& 4 & 0.64 & 0.01 & 0.43 & 0.06 & 0.01 & 0.03 & 0.01 \\
			\hline \hline
			\multicolumn{ 9}{c}{($n,p,\rho$)=(400, 2000, 0)} \\
			\hline
			& 2 & 0.90 & 0.01 & 0.56 & 0.12 & 0.01 & 0.21 & 0.82 \\
			Example 3 & 3 & 0.80 & 0.03 & 0.50 & 0.08 & 0.00 & 0.14 & 0.72 \\
			& 4 & 0.69 & 0.01 & 0.36 & 0.04 & 0.01 & 0.08 & 0.55 \\
			\hline
			\multicolumn{9}{c}{($n,p,\rho$)=(400, 2000, 0.5)} \\
			\hline
			& 2 & 1.00 & 0.27 & 0.75 & 0.72 & 0.62 & 0.69 & 0.69 \\
			Example 3  & 3 & 1.00 & 0.21 & 0.78 & 0.71 & 0.55 & 0.68 & 0.68 \\
			& 4 & 0.99 & 0.20 & 0.82 & 0.68 & 0.47 & 0.65 & 0.65 \\
			\hline \hline
			\multicolumn{ 9}{c}{($n,p,\rho$)=(400, 2000, 0)} \\
			\hline
			& 2 & 0.90 & 0.04 & 0.47 & 0.11 & 0.00 & 0.14 & 0.70 \\
			Example 4 & 3 & 0.85 & 0.01 & 0.42 & 0.13 & 0.01 & 0.08 & 0.61 \\
			& 4 & 0.64 & 0.00 & 0.32 & 0.09 & 0.00 & 0.08 & 0.51 \\
			\hline
			\multicolumn{9}{c}{($n,p,\rho$)=(400, 2000, 0.5)} \\
			\hline
			& 2 & 0.84 & 0.09 & 0.70 & 0.45 & 0.02 & 0.03 & 0.03 \\
			Example 4  & 3 & 0.77 & 0.07 & 0.65 & 0.46 & 0.03 & 0.04 & 0.03 \\
			& 4 & 0.72 & 0.03 & 0.56 & 0.42 & 0.04 & 0.04 & 0.04 \\
			\hline \hline

		\end{tabular}

	}
\end{table}


\begin{table}[!htbp]
	\caption{ \footnotesize{ Selection and prediction results (standard errors) with $(n,p)=(400,100)$. }}\label{post1}
	\centering
	\smallskip{ \scriptsize \tabcolsep= 2 pt

		\begin{tabular}{ccccccccc}
			\hline\hline
			Methods &   &   $\sigma$ &     SSI &    BOLT-SSI &     RAMP &       xyz-L100 &     xyz-L500&   hierNet-w  \\
			\hline \hline
			&         & 2 & 1 & 0.95 & 0.00 & 1    &  1   & 1\\
			&  {ACR}     & 3 & 1 & 0.95 & 0.00 & 1    &  1   & 1 \\
			&         & 4 & 1 & 0.90 & 0.00 & 0.99 & 0.98 & 1 \\
			\hline
			
			&      & 2 & 54.4(1.0) & 52.0(1.0) & 12.1(0.3) & 27.3(0.3) & 27.5(3.5) &  230.6(4.4)\\
			Example 1 & AMS & 3 & 56.8(1.0) & 53.2(1.0) & 11.2(0.4) & 31.0(0.6) & 29.2(0.5)& 216.6(4.4) \\
			&      & 4 & 58.4(1.0) & 53.9(1.0) & 10.6(0.4)& 31.7(0.6)& 31.2(0.7) & 195.8(4.8) \\
			\hline
			&       & 2 & 95.4(0.15) & 91.3(0.37)& 58.5(1.34)& 53.3(2.02) & 53.3(1.59) & 94.1(0.18)\\
			& $R^2$  & 3 & 90.0(0.27)& 86.4(0.47) & 54.3(1.62) & 50.5(1.64) & 48.7(1.02) & 87.2(0.39) \\
			&      & 4 & 83.1(0.44) & 79.5(0.63) & 48.8(1.64) & 48.9(1.72) & 47.9(1.02) &  78.4(0.60) \\
			\hline \hline
			
			&         & 2 & 1     & 0.68 & 0.00 & 0.49 &  1   & 1\\
			&  {ACR}    & 3 & 0.99  & 0.58 & 0.00 & 0.48 &  1   & 0.99 \\
			&         & 4 & 0.97  & 0.47 & 0.00 & 0.30 & 0.99 & 0.81 \\
			\hline
			&      & 2 & 60.7(1.1) & 52.1(1.5) & 15.5(0.2) & 27.7(0.3) & 34.6(0.4) &  301.3(3.9)\\
			Example 2 & AMS & 3 & 65.9(1.1) & 54.7(1.5) & 14.7(0.3) & 31.1(0.4) & 39.0(0.4)& 262.8(5.4) \\
			&      & 4 & 68.2(1.1) & 55.6(1.5) & 13.7(0.3)& 35.3(0.4)& 40.4(0.4) & 167.7(7.3) \\
			\hline
			&       & 2 & 92.4(0.24) & 81.6(0.86)& 77.2(1.14)& 85.4(0.48) & 72.9(1.06) & 89.2(0.31)\\
			& $R^2$  & 3 & 84.5(0.35)& 74.2(0.82) & 71.0(1.20) & 83.6(0.42) & 64.8(1.33) & 76.1(0.77) \\
			&      & 4 & 74.6(0.51) & 63.5(1.04) & 63.2(1.34) & 79.5(0.50) & 58.9(1.12) &  58.6(1.08) \\
			\hline \hline
			&         & 2 & 1 & 0.84 & 0.00 & 1    &  1   & 1\\
			&  {ACR}    & 3 & 1 & 0.82 & 0.00 & 1    &  1   & 1 \\
			&         & 4 & 1 & 0.78 & 0.00 & 1    &  1   & 1 \\
			\hline
			
			&         & 2 & 51.3(0.9) & 47.7(1.3) & 4.3(0.4) & 28.1(0.4) &
			28.1(0.3) &  383.5(3.6)\\
			Example 3& AMS     & 3 & 53.8(1.0) & 50.8(1.6) & 4.2(0.5) & 31.4(0.6) &
			30.4(0.6)& 340.1(3.8) \\
			&         & 4 & 56.6(1.1) & 49.1(1.3) & 3.9(0.4)& 31.7(0.6)&
			32.7(0.6) & 297.6(4.3) \\
			\hline
			&       & 2 & 95.3(0.17) & 87.0(0.66)& 14.6(1.82)& 58.4(1.45) &
			57.8(1.41) & 92.3(0.20)\\
			& $R^2$ & 3 & 89.9(0.27) & 81.6(0.81)& 13.4(1.85)& 51.6(1.65) &
			50.7(1.67) & 84.4(0.40) \\
			&        & 4 & 83.0(0.42) & 75.5(0.81)& 12.4(1.80) & 48.7(1.69)
			& 48.4(1.71) & 74.2(0.63) \\
			\hline \hline
			&         & 2 & 0.94 & 0.69 & 0.00 & 1    &  1   & 1\\
			&  {ACR}      & 3 & 0.97 & 0.60 & 0.00 & 0.99 &  1   & 1 \\
			&           & 4 & 0.95 & 0.61 & 0.00 & 0.97 &  1   & 0.98 \\
			\hline
			&         & 2 & 54.7(0.9) & 49.1(1.4) & 10.2(0.4) & 31.7(0.4) &
			31.1(0.3) &  332.8(3.6)\\
			Example 4& AMS     & 3 & 58.3(1.0) & 50.5(1.2) & 9.6(0.4) & 35.9(0.5) &
			35.2(0.4)&  292.1(3.7) \\
			&         & 4 & 61.6(1.2) & 52.1(1.5) & 8.7(0.4)& 36.9(0.4)&
			37.0(0.5) & 240.6(5.8) \\
			\hline
			&       & 2 & 93.4(0.24) & 84.0(0.72)& 51.9(1.52)& 63.6(1.15) &
			63.5(1.41) & 90.8(0.24)\\
			& $R^2$ & 3 & 86.6(0.37) & 77.2(0.80)& 47.7(1.65)& 58.3(1.12) &
			60.0(1.31) & 80.5(0.51) \\
			&        & 4 & 78.2(0.52) & 70.0(0.96)& 41.9(1.62) & 54.8(1.09)
			& 55.2(1.37) & 67.1(0.85) \\
			\hline \hline

		\end{tabular}

	}
\end{table}

\begin{table}[!htbp]
	\caption{ \footnotesize{ Selection and prediction results (standard errors) with $(n,p)=(400,100)$. }}\label{post2}
	\centering
	\smallskip{ \scriptsize \tabcolsep= 2 pt

		\begin{tabular}{ccccccccc}
			\hline\hline
			Methods &   &   $\beta_{ij}$ &     SSI &    BOLT-SSI &     RAMP &         hierNet-w  \\
			\hline \hline
			&      & 1 & 0.34 & 0.38 & 0.00 & 0.17 \\
			& {ACR}  & 2 & 0.64 & 0.18 & 0.00 &  0.28 \\
			&      & 3 & 0.74 & 0.26 & 0.00&   0.30 \\
			\hline
			&      & 1 & 63.1(3.8) & 56.9(3.1) & 6.7(0.4) & 36.9(3.3) \\
			Example 5 & AMS & 2 & 69.2(3.4) & 15.7(1.8) & 0.7(0.2) & 83.6(7.4) \\
			&      & 3 & 61.3(3.7) & 15.8(1.3) & 0.2(0.1)& 103.8(9.2) \\
			\hline
			&     & 1 & 28.2(0.54) & 25.5(0.46) & 26.8(0.62) & 25.7(0.47) \\
			&PMR  & 2 & 24.6(0.44) & 27.1(0.49) & 28.6(0.45) & 38.5(0.52) \\
			&     & 3 & 22.7(0.46) & 23.5(0.46) & 26.2(0.39)& 41.7(0.55) \\
			\hline \hline
			&      & 1 & 0.07 & 0.11 & 0.00 & 0.21 \\
			& {ACR}  & 2 & 0.73& 0.31 & 0.00 &  0.34 \\
			&      & 3 & 0.88 & 0.34 & 0.00&   0.28 \\
			\hline
			&      & 1 & 39.0(3.0) & 41.9(2.1) & 3.9(0.2) & 53.4(3.2) \\
			Example 6 & AMS & 2 & 74.7(3.5) & 49.0(2.6) & 4.3(0.4) & 82.3(4.9) \\
			&      & 3 & 76.2(3.1) & 35.1(2.5) & 2.3(0.3) & 82.3(7.1) \\
			\hline
			&     & 1 & 24.3(0.51) & 23.1(0.45) & 26.8(0.62) & 21.0(0.48) \\
			&PMR  & 2 & 24.7(0.48) & 25.1(0.54) & 31.1(0.76) & 29.0(0.47) \\
			&     & 3 & 22.8(0.52) & 26.5(0.67) & 33.6(0.58)& 32.3(0.46) \\
			\hline \hline
			&      & 1 & 0.53 & 0.14 & 0.00 & 0.09 \\
			& {ACR}  & 2 & 0.77 & 0.44 & 0.00 &  0.18 \\
			&      & 3 & 0.85 & 0.46 & 0.00&   0.13 \\
			\hline
			
			&      & 1 & 58.9(3.0) & 41.8(1.6) & 3.4(0.1) & 93.7(6.2) \\
			Example 7 & AMS & 2 & 69.8(2.5) & 46.1(1.9) & 1.8(0.1) & 94.1(7.9) \\
			&      & 3 & 76.7(3.3) & 37.8(1.8) & 1.1(0.1)&  65.6(8.0) \\
			\hline
			&     & 1 & 18.2(0.37) & 18.4(0.39) & 25.3(0.44) & 20.8(0.37) \\
			&PMR  & 2 & 19.6(0.45) & 19.0(0.47) & 27.4(0.51) & 24.3(0.54) \\
			&     & 3 & 19.2(0.42) & 19.6(0.46) & 25.7(0.42)& 23.4(0.43) \\
			\hline \hline
			&      & 1 & 0.20 & 0.10 & 0.00 & 0.19 \\
			& {ACR}  & 2 & 0.68 & 0.32 & 0.00 &  0.28 \\
			&      & 3 & 0.81 & 0.42 & 0.00&   0.14 \\
			\hline
			
			&      & 1 & 47.5(2.8) & 44.7(2.3) & 3.4(0.1) & 80.8(5.1) \\
			Example 8 & AMS & 2 & 70.1(2.7) & 41.1(2.2) & 2.5(0.1) & 94.7(5.1) \\
			&      & 3 & 76.3(3.4) & 31.4(1.9) & 1.5(0.2)&  62.9(6.0) \\
			\hline
			&     & 1 & 20.6(0.43) & 20.5(0.47) & 26.2(0.45) & 20.3(0.42) \\
			&PMR  & 2 & 21.6(0.51) & 22.3(0.49) & 30.0(0.55) & 25.1(0.44) \\
			&     & 3 & 21.1(0.44) & 23.8(0.52) & 29.7(0.48)& 28.1(0.48) \\
			\hline \hline

		\end{tabular}

	}
\end{table}

%

\end{document}